\documentclass[aap]{imsart}

\usepackage{amsthm}
\usepackage{amsmath}
\usepackage{marvosym}
\usepackage{amsfonts}
\usepackage{booktabs}
\usepackage{array}
\usepackage{paralist}
\usepackage{verbatim}
\usepackage{mathtools}
\usepackage{booktabs}
\usepackage{dsfont}
\usepackage{mathtools}
\usepackage{graphicx}
\usepackage{subcaption}
\usepackage{color}
\usepackage{url}
\usepackage{hyperref}
\usepackage[T1]{fontenc}

\graphicspath{{}{figure/}}

\DeclareMathOperator*{\argmin}{argmin}

\newcommand{\pp}{\mathbf{P}}
\newcommand{\rr}{\mathbf{R}}
\newcommand{\bQ}{\mathbf{Q}}
\newcommand{\ii}{\mathbf{I}}

\newcommand{\mm}{\mathbf{m}}
\newcommand{\mmm}{\mathbf{M}}

\newcommand{\bb}{\mathbf{B}}
\newcommand{\kk}{\mathbf{K}}
\newcommand{\kkm}{\kk_{s(\mm^*)}}
\newcommand{\zzm}{\zz_{s(\mm^*)}}
\newcommand{\xx}{\mathbf{x}}
\newcommand{\ee}{\mathbf{e}}
\newcommand{\xxx}{\mathbf{X}}
\newcommand{\zz}{\mathcal{Z}}
\newcommand{\bZ}{\mathbf{Z}}
\newcommand{\bY}{\mathbf{Y}}
\newcommand{\bz}{\mathbf{z}}
\newcommand{\by}{\mathbf{y}}
\newcommand{\bu}{\mathbf{U}}
\newcommand{\bv}{\mathbf{V}}
\newcommand{\calZ}{\mathcal{Z}}

\providecommand{\norm}[1]{\| #1\|}

\newcommand{\R}{\mathbb{R}}

\newcommand{\n}{\mathcal{N}}

\newcommand{\e}{\mathbb{E}}
\newcommand{\ind}{V^{(N)}_{\mathrm{WIP}} (\alpha)}
\newcommand{\cycle}{V_{\mathrm{cycle}} (\alpha)}
\newcommand{\indcontinuous}{V^{(N)}_{\mathrm{WIP-async.}} (\alpha)}
\newcommand{\opt}{V^{(N)}_{\mathrm{opt}} (\alpha)}
\newcommand{\rel}[1][N]{V^{(#1)}_{\mathrm{rel}}(\alpha)}

\newcommand{\sss}{\mathbf{S}}
\newcommand{\floor}{V^{(N)}_{\mathrm{WIP}}(\lfloor N\alpha\rfloor/N)}
\newcommand{\ceil}{V^{(N)}_{\mathrm{WIP}}(\lceil N\alpha\rceil/N)}
\newcommand{\continue}{V^{(N)}_{\mathrm{WIP}}(\overline{\alpha})}

\newcommand\newalpha{\tilde{\alpha}}

\newcommand\bS{\mathbf{S}}

\newcommand\bMN{\mathbf{M}^{(N)}}
\newcommand\MN{M^{(N)}}
\newcommand\ba{\mathbf{a}}
\newcommand\N{\mathbb{N}}
\newcommand\calO{\mathcal{O}}
\newcommand{\red}[1]{{\color{red}#1}}
\newcommand{\expect}[1]{\mathbb{E}\left[#1\right]}
\newcommand{\proba}[1]{\mathbb{P}\left[#1\right]}

\usepackage{amsthm}					
\theoremstyle{definition} 			
\newtheorem{definition}{Definition}[section]
\newtheorem{lem}[definition]{Lemma} 

\newtheorem{thm}[definition]{Theorem}
\newtheorem{rem}[definition]{Remark}

\newtheorem{theorem}{Theorem}[section]
\newtheorem{lemma}[theorem]{Lemma}

\begin{document}


\begin{frontmatter}
  \title{Exponential Convergence Rate for the Asymptotic Optimality of
    Whittle Index Policy}


\begin{abstract}

  We evaluate the performance of Whittle index policy for restless Markovian bandits, when the number of bandits grows. It is proven in \cite{WeberWeiss1990} that this performance is asymptotically optimal if the bandits are indexable and the associated deterministic system has a global attractor fixed point.  In this paper  we show  that, under the same conditions, the  convergence rate is exponential in the number of bandits, unless the fixed point is \emph{singular} (to be defined later). Our proof is based on the nature of the deterministic equation governing the stochastic system: We show that it  is a piecewise affine continuous dynamical system inside the simplex of the empirical measure of the bandits. Using simulations and numerical solvers, we also investigate  the cases where  the conditions for the exponential rate theorem are violated, notably when attracting limit cycles appear, or when the fixed point is singular. We illustrate our theorem on a Markovian fading channel model, which has been well studied in the literature. Finally, we extend our synchronous model results to the asynchronous model.
\end{abstract}


\begin{aug}
\author[A]{\fnms{Nicolas} \snm{Gast}\ead[label=e1]{nicolas.gast@inria.fr}},
\author[A]{\fnms{Bruno} \snm{Gaujal}\ead[label=e2]{bruno.gaujal@inria.fr}}
\and
\author[A]{\fnms{Chen} \snm{Yan}\ead[label=e3]{chen.yan@univ-grenoble-alpes.fr}}
\address[A]{ Univ. Grenoble Alpes, Inria, CNRS, Grenoble INP, LIG, 38000 Grenoble, France,
\printead{e1,e2,e3}.}

\end{aug}



\end{frontmatter}


\section{Introduction}

A multi-armed bandit (MAB) problem is a sequential allocation problem: At each decision epoch, one or several bandits are activated and some observable rewards are obtained. The goal is to maximize the total reward obtained by  a sequence of activations. There are at least three fundamental formalizations of the bandit problem depending on the assumed nature of the reward process: stochastic, adversarial, and Markovian. Each bandit model has its own specific playing strategies and uses distinct techniques of analysis. We focus here on the Markovian bandits (for a thorough analysis of the other two types of bandit models, see \emph{e.g.} \cite{DBLP:journals/corr/abs-1204-5721,lattimore2020bandit}). Each time, a subset of bandits are chosen to be activated. All bandits generate an instantaneous reward that depend on their state and their activation. The state of each activated bandit then changes in a Markovian fashion, based on an underlying transition matrix (or a rate matrix in the asynchronous case). Both the reward and the new state are revealed to the decision maker for its next decision. The bandits that are not activated change state according to a different transition matrix. When the underlying stochastic transition laws are  assumed to be known (see  \cite{Duff:1995:QBP:3091622.3091648} for a treatment of the case where the transition matrices are unknown), the optimal policy can be computed via dynamic programming, and the problem is essentially of computational nature.

The above  Markovian MAB problem has been solved in the {\it restful case} (non activated bandits do not change their states) with one active bandit at each decision epoch in \cite{Gittins79banditprocesses} by the Gittins index policy, which is a greedy policy that can be computed efficiently. In \cite{whittle-restless} Whittle generalizes the model in two aspects. Firstly, at each decision epoch more than one bandit can be activated, and secondly, the bandits that are not activated can also change states ({\it restless bandits}), according to a different stochastic transition matrix, as mentioned before. Under these generalizations, the problem can no longer be solved  by a similar efficient index-type greedy policy, and indeed it has been proven in \cite{Papadimitriou99thecomplexity} that this problem  is PSPACE-hard. In \cite{whittle-restless}, however, Whittle conjectures that, under some conditions, the so-called "Whittle index policy" (WIP) should be optimal asymptotically, \emph{i.e.} when the number of  bandits goes to infinity with a fixed proportion of active bandits.

This conjecture has been proven in the famous paper \cite{WeberWeiss1990} for the asynchronous model, under several technical conditions (further discussed in \cite{weber1991addendum}), namely when the bandits are indexable and the drift of the Markov system has a fixed point that is a global attractor.
These results further reinforce the interest of Whittle index, as restless bandit models have been used in a wide range of applications and Whittle index policies turn to be efficient solutions. Among them one can cite wireless scheduling \cite{aalto2015whittle,hsu2018age,raghunathan2008index}, queuing systems \cite{ansell2003whittle},  crawling optimal content on the web \cite{avrachenkov2016whittle}, load-balancing \cite{larrnaaga2016dynamic} and sensors \cite{nino2011sensor}. In particular, partially observable Markov decision process (POMDP) falls into the category of restless Markovian bandits by using a Bayesian approach to construct the transition matrices. One concrete example is the multi-channel wireless scheduling problem of \cite{liu2010indexability,meshram2018whittle}. In this system, there are $N$ Gilbert-Elliott channels and the state of a channel is only observed when a transmission is scheduled on this channel. This example will be further analyzed in Section~\ref{sec:channels} to illustrate our result.

There are several extensions to the restless bandit model. In \cite{hodge2015}, individual bandits have multiple levels of activation (instead of only two) and are subject to an overall resource constraint; in \cite{oatao22728}, the environment in which lives the bandits also changes along time; in \cite{Ve2016.6}, the author considers the larger set of all priority policies (to which WIP belongs) and aims at finding best policies among them. Of all those generalizations, some similar index policies have been proposed and asymptotic optimality results have been proved therein. However, to the best of our knowledge, nobody has  considered the question of evaluating the convergence rate of the performance of the
Whittle index policy  to the optimal one.

\subsection*{Contributions}

Despite the well-known asymptotic optimality of WIP (under some conditions) and its empirically good performance on numerous models listed above, as well as its many extensions, there is very limited research on how fast WIP becomes optimal. In this paper we show that the convergence of the performance of WIP to the performance of an optimal policy is exponentially fast with the number $N$ of bandits, giving a theoretical explanation  for the good performance of WIP in practice, even when the number of bandits is small. This result holds in the synchronous as well as the asynchronous cases, under  the same conditions as the asymptotic optimality proven in \cite{WeberWeiss1990}, namely the bandits are indexable and that the ordinary differential equation driving the dynamics of the mean field approximation has a fixed point that is a global attractor, plus the additional condition that the fixed point is {\it non-singular}. This last condition will be discussed in length in the rest of the paper.
The exponential convergence rate  extends to the multi-action case developed in \cite{hodge2015}. This extension is direct (the piece-wise affine structure of the dynamics  is preserved in the multi-action case) and will not be discussed further in the paper.

The proof of our main result ({\it i.e.} exponential convergence rate in the general case) relies on two main ingredients. The first one comes by noticing that the dynamics of the mean field approximation of the $N$ bandits, each with $d$ states, under WIP is piecewise affine and continuous over a finite number of polytopes partitioning the configuration space (the simplex in dimension $d$). This piecewise linearity of the mean field approximation comes as  a mixed blessing when one tries to compute the convergence rate: On the one hand the  dynamics is  not differentiable at the interface between the polytopes. Therefore,  previous approaches based on the smoothness of the drift such as \cite{gast:hal-01891636,Kolokoltsov2011,gast:hal-01622054,Ying2017} collapse here. On the other hand, when the global attracting fixed point  falls into the interior of  a polytope ({\it i.e.} it is non-singular), the dynamics in a small neighborhood around the fixed point is affine and the expected behavior of the system is relatively simple to analyze.

The second ingredient is to divide the analysis of the behavior of the stochastic system into two parts: before it enters a small neighborhood of the fixed point and after it does. The Stein's method is used to compare its  behavior with its mean field approximation inside the neighborhood. Hoeffding's inequality (in the synchronous case) or an exponential martingale concentration inequality (in the asynchronous case) is used to control its behavior outside the neighborhood.\ \\

To be more precise,  we show that under indexability, global attraction of the fixed point of the mean field dynamics and non-singularity of this fixed point,  the  average performance of a  stochastic Markovian bandit system under WIP converges  to its mean field limit  as $b \cdot \exp(-cN)$ where $N$ is the number of bandits and $b,c$ are positive constants independent of $N$. Our result comes with  several novelties.

\begin{itemize}
\item
Firstly, we believe that this is the first example where  an exponential convergence to a  mean field limit has been obtained. This exponential rate relies crucially on the piecewise affine nature of the deterministic dynamical system, as opposed to most other mean field approximation results that prove convergence rates that are polynomial in $1/\sqrt{N}$ and for which  the deterministic  dynamics is  smooth everywhere.

\item
Secondly, although a part of our proof has a large deviation flavor, our result concerns the expected behavior of
the stochastic bandits and not its deviations so that our result cannot be obtained by simply using general results on
dynamical systems in the presence of random perturbations, such as the large deviation bounds presented in Section 1.5 in \cite{kifer1988random}.
As for the part of our proof  on concentration bounds that might have been obtained using large deviation principles, we believe that our  direct proof, based on concentration inequalities, is simple enough and provides a clearer understanding of what is actually going on.

\item The contrast between singular and non-singular attractors has gone unnoticed so far.
Our theoretical results (exponential convergence in the non-singular case and possibly only in $1/\sqrt{N}$ in the singular case) are backed by numerical experiments showing that  for a moderate number of bandits ($N$ ranging from 10 to 50),
the relative performance of WIP w.r.t. the optimal policy can be almost perfect (less than 0.1 \% difference) in the non-singular case to simply good (around 4 \%) in the singular case.

\item Finally,  we also investigate the behavior of the restless bandits under WIP when the fixed point of the deterministic dynamical system is \emph{not} a global attractor. In the synchronous case, the system can become  periodic, with a stable periodic cycle of variant length (the fixed point being unstable). In such cases, the performance of WIP converges to the average performance over the cycle and in general is not asymptotically optimal.
\end{itemize}

\subsection*{Organization of the paper}
In Section~\ref{sec:model}, we introduce the synchronous restless bandit model: all bandits change their state simultaneously in discrete time, according to transition matrix ${\bf P^1}$ when being activated  and  ${\bf P^0}$ when not being activated. We also define the Whittle indices and the main notations used in the paper. We then present the main result of the paper in Section \ref{sec:main}, namely exponential convergence for the performance of WIP to the optimal one in the general situation. In Section~\ref{sec:numerical}, we illustrate our results with several examples. We provide simulation and numerical estimations for the performance of WIP in different cases: singular, non-singular, cyclic. In Section~\ref{sec:channels}, we present an application of our result to the Markovian fading channel problem, where we check numerically with parameters that fall into the general case framework (non-singular global attracting fixed point). Finally, in Section~\ref{sec:continuous}, we extend our result to the classical asynchronous bandit model (bandits are continuous-time Markov chains, and decisions are made every time when one bandit changes its state). We show that exponential convergence rate also holds in the asynchronous case, and highlight the differences between the synchronous and asynchronous models.

\section{The synchronous restless bandit model}
\label{sec:model}

We first describe the restless bandit model in Section~\ref{model}.  We then recall the definition of Whittle index in Section~\ref{ssec:whittle} and its relation with a linear relaxation in Section~\ref{ssec:relax}. Note that in our model, all bandits are synchronous. This is a discrete-time version of the classical continuous-time model studied in \cite{WeberWeiss1990}. We will discuss an extension of our results to the latter model in Section~\ref{sec:continuous}.

\subsection{Model description}\label{model}

The synchronous restless bandit model with parameters

$\big\{ (\pp^0, \pp^1, \rr^0, \rr^1); \alpha, N\big\}$ is a Markov decision process (MDP) defined as follows:
\begin{enumerate}
\item The model is composed of $N$ bandits. Each bandit evolves in an identical finite state space $\{1\dots d\}$ and the state of bandit $n$ at time $t$ is denoted by $S_n(t)\in\{1\dots d\}$. The state space of the whole process is denoted at time $t$ by $\bS(t) = \big( S_1(t), S_2(t), ..., S_N(t) \big)$.

\item Decisions are taken at times $t \in \N$. At each decision epoch, a decision maker observes $\bS(t)$ and chooses $\alpha N$ of the $N$ bandits to be activated, where we assume that $\alpha$ and $N$ are such that $\alpha N$ is an integer. We set $a_n(t)=1$ if bandit $n$ is activated at time $t$ and $a_n(t)=0$ otherwise. The action vector at time $t$ is $\ba(t) = \big( a_1(t),a_2(t), ..., a_N(t) \big)$. It satisfies $\sum_{n=1}^N a_n(t)=\alpha N$.

\item Bandit $n$ evolves according to Markovian laws: for all states $i,j$, action $a\in\{0,1\}$ and $t\in\N$:
    \begin{equation}
      \mathbb{P} ( S_n (t+1) = j \mid S_n (t) = i, a_n (t) = a ) = P^a_{ij}.
      \label{eq:markov}
    \end{equation}
Given $\ba(t)$, the $N$ bandits make their transitions independently.
\item If bandit $n$ is in state $i$ and the decision maker takes action $a \in \{0,1\}$, a bounded reward $R^a_{i}\in\R$ is earned.
\end{enumerate}

The goal of the decision maker is to compute a decision rule in order to maximize the long-term expected average reward. The theory of stochastic dynamic programming \cite{Puterman:1994:MDP:528623} shows that there exists an optimal policy which is Markovian and stationary (\emph{i.e.} $\ba(t)$ can be chosen as a time-independent function of $\bS(t)$). Denote by $\Pi$ the set of such Markovian stationary policies, the optimization problem of the decision maker can be formalized as
\begin{align}
  \opt & := \sup_{\ba \in \Pi} \lim_{T \rightarrow \infty} \frac{1}{T} \mathbb{E} \Big[ \sum_{t=0}^{T-1} \sum_{n=1}^N R^{a_n(t)}_{S_n(t)} \Big] \label{eq1}\\
  & \mbox{subject to} \sum_{n=1}^N a_n(t) = \alpha N, \ \mbox{for all} \  t \in \mathbb{N}. \label{eq2}
\end{align}

In the rest of the paper, we assume that matrices $\pp^0$ and $\pp^1$ are such that the states form a single aperiodic closed class, regardless of the policy employed. This assumption was also used in \cite{WeberWeiss1990} and guarantees that neither the value of the optimization problem \eqref{eq1} nor the optimal policy depend on the initial state $\bS(0)$ of the system at time $0$. We call such a bandit an \emph{aperiodic unichain} bandit.

\subsection{Indexability and Whittle index}
\label{ssec:whittle}

In theory, a dynamic programming approach can be used to solve Equations ~\eqref{eq1}-\eqref{eq2}, but this approach is computationally intractable, as the numbers of possible states and actions grow exponentially with $N$. In fact, such problems have been proven to be PSPACE-hard in \cite{Papadimitriou99thecomplexity}. To overcome this difficulty, Whittle introduces in \cite{whittle-restless} a very efficient heuristic known as Whittle index policy (WIP). This heuristic is obtained by computing an index $\nu_i$ for each state $i$. At a given decision epoch, WIP activates the $\alpha N$ bandits having currently the highest indices. We describe below how these indices are defined.


The index of a bandit can be computed by considering each individual bandit in isolation. For a given $\nu \in \mathbb{R}$, we define the subsidy-$\nu$ problem as the following MDP. The state space is the one of a single bandit. At each time $t$, the decision maker chooses whether or not to activate this bandit. As in the original problem, the bandit evolves at time $t$ according to \eqref{eq:markov}. The difference lies in the passive action that is subsidized: If the bandit is in state $i$ and action $1$ is taken, then as before, a reward $R^1_i$ is earned; if the bandit is in state $i$ and action $0$ is taken, then a reward $R^0_i + \nu$ is earned.

The goal of the decision maker is to maximize the total expected reward (including passive subsidies) earned over an infinite horizon. For a given $\nu \in \mathbb{R}$, let us denote by $\omega(\nu)$ the set of states for which there exists an optimal policy of the $\nu$-subsidized MDP such that the passive action is optimal in these states. Whittle indices are defined as follows:


\begin{definition}[Indexability and Whittle index] A bandit $(\pp^0, \pp^1, \rr^0, \rr^1)$ is $\emph{indexable}$ if $\omega(\nu)$ is increasing in $\nu$, namely if  for all $\nu \le \nu'$, we have $\omega(\nu) \subseteq \omega(\nu')$.  In this case, the Whittle index of a state $i$, that we denote by $\nu_i$, is defined as the smallest subsidy such that the passive action is optimal in this state:
\begin{equation*}
  \nu_i := \inf_{\nu \in \mathbb{R}} \big\{ \nu \ \big| \ i \in \omega(\nu) \big\}.
\end{equation*}
\end{definition}
It should be emphasized that there exist restless bandit problems that are \emph{not} indexable, although this is relatively rare (we discuss this in more detail in Section~\ref{sec:numerical_statistics}). Note that when $\pp^{0}$ is the identity matrix, bandits are restful, \emph{i.e.} the states of the bandits that are not activated do not change. In such a case, a bandit is always indexable and Whittle index coincides with the classical definition of Gittins index \cite{gittins2011multi}. An algorithm to test indexability and calculate the index, as well as a geometric interpretation can be found in \cite{nino2007dynamic}. For given parameters $(\pp^0,\pp^1,\rr^0,\rr^1)$, the complexity of this algorithm is $\calO(2^d)$. However, if a restless bandit is known to be indexable beforehand, then its indices can be computed by a greedy algorithm in time $\mathcal{O}(d^3)$.



\subsection{Whittle relaxation and the asymptotic optimality theorem}
\label{ssec:relax}

An intuition behind the definition of Whittle index is that it is related to a relaxation of the original $N$ bandits problem (\ref{eq1}) where the constraint \eqref{eq2} is replaced by $\lim_{T \rightarrow \infty} \frac{1}{T} \sum_{t=0}^{T-1} \sum_{n=1}^N a_n(t) = \alpha N$. While the constraint \eqref{eq2} imposes that exactly $\alpha N$ bandits are activated at each time step, the relaxed constraint only imposes the time-averaged number of activated bandits to be equal to $\alpha N$:
\begin{align}
  \rel & := \sup_{\ba \in \Pi} \lim_{T \rightarrow \infty} \frac{1}{T} \mathbb{E} \Big[ \sum_{t=0}^{T-1} \sum_{n=1}^N R^{a_n(t)}_{S_n(t)} \Big] \label{eq:relax_synchro}\\
  & \mbox{subject to} \lim_{T \rightarrow \infty} \frac{1}{T} \sum_{t=0}^{T-1} \sum_{n=1}^N a_n(t)  = \alpha N. \label{eq3}
\end{align}
By using $\nu$ as a Lagrange multiplier of the constraint $\lim_{T \rightarrow \infty} \frac{1}{T} \sum_{t=0}^{T-1} \sum_{n=1}^N a_n(t)  = \alpha N$, the Lagrangian of the problem \eqref{eq:relax_synchro}-\eqref{eq3} is
\begin{align*}
    &\lim_{T \rightarrow \infty} \frac{1}{T} \mathbb{E} \Big[ \sum_{t=0}^{T-1} \sum_{n=1}^N R^{a_n(t)}_{S_n(t)} \Big]
  + \nu \Big(\alpha N -  \lim_{T \rightarrow \infty} \frac{1}{T} \sum_{t=0}^{T-1} \sum_{n=1}^N a_n(t)\Big)  \\
  =& \ \nu \alpha N + \sum_{n=1}^N \lim_{T \rightarrow \infty} \frac{1}{T} \mathbb{E} \Big[ \sum_{t=0}^{T-1} \big(R^{a_n(t)}_{S_n(t)} - \nu a_n(t)\big)\Big].
\end{align*}

Note that, for a fixed $\nu$, finding a policy that maximizes the above Lagrangian can be done by solving $N$ independent optimization problems (one for each bandit), and each problem is a $\nu$-subsidized MDP.

It should be clear that the constraint \eqref{eq3} is weaker than the constraint \eqref{eq2}. This shows that $ \opt \leq \rel$. Hence $\rel$ is an upper bound on the value of the original optimization problem \eqref{eq1}. In fact, the next result shows that, as the number of bandits grows\footnote{In the rest of the paper, unless otherwise specified, we restrict our attention to the values of $N$ such that $\alpha N$ is an integer. The notation $\lim_{N\to\infty}f(N)=\ell$ is to be understood as $\lim_{k\to\infty}f(k N_0)=\ell$ where $N_0$ is the smallest positive integer such that $\alpha N_0 \in\N$. We will discuss non-integer values of $\alpha N$ in Section~\ref{ssec:non-integer-alphaN}.}, the value of the original problem converges to this value:
\begin{theorem} \label{th:rel}
  Consider a synchronous restless bandit model with $N$ identical aperiodic unichain bandits and such that the matrices $\pp^0$ and $\pp^1$ are rational. Then $\rel = N \rel[1]$ and
  \begin{equation}
    \label{eq:th_rel}
    \limsup_{N\to\infty}\sqrt{N}\left(\frac{\opt}{N} - \rel[1]\right) < \infty.
  \end{equation}
\end{theorem}
The above theorem justifies the relaxation \eqref{eq3} by showing that when the number of bandit is large, the value of the optimization problem \eqref{eq1} is close to $\rel$: $\lim_{N\to\infty}\frac{\opt}{N}=\rel[1]$. In Theorem~1 of \cite{WeberWeiss1990}, the result $\lim_{N\to\infty}\frac{\opt}{N}=\rel[1]$ was proved for the asynchronous bandit model that we will discuss in Section~\ref{sec:continuous}. To the best of our knowledge, the statement of this theorem in our setting of synchronous bandit model is new. Moreover, our result shows that the convergence is \emph{at least} in $\calO(1/\sqrt{N})$.  For completeness, we provide a proof of Theorem~\ref{th:rel} in Appendix~\ref{apx:proof_rel}. It is an adaptation of the proof of \cite[Theorem~1]{WeberWeiss1990}: we use a similar coupling argument, although the coupling has to be adapted to our synchronous setting, and we also need the additional aperiodic assumption on the model.

While Theorem~\ref{th:rel} guarantees that the original optimization problem converges to the relaxation, it does not guarantee any result on the performance of WIP. This leaves two important questions: \emph{(1) Is indexability a sufficient condition for WIP to be asymptotically optimal (for our synchronous bandit model)? (2) If WIP is asymptotically optimal, then, at which speed does it become optimal?}  In the remainder of the paper we will see that, similarly to the asynchronous bandit model, there exist examples for which WIP is \emph{not} asymptotically optimal. We will also exhibit sufficient conditions to guarantee asymptotic optimality that is similar to the conditions of \cite{WeberWeiss1990}. Our main result concerns the rate of convergence. We will show that, except in rare cases, when WIP is asymptotically optimal, it does so at exponential speed with the number of bandits $N$. This complements Theorem~\ref{th:rel} by proving that, under the same conditions, the convergence in \eqref{eq:th_rel} occurs at exponential rate.

\section{Main Results}
\label{sec:main}

We first show in Section~\ref{ssec:phi} that, when $N$ is large, the stochastic system governed by WIP behaves like a piecewise affine deterministic system. We then present the exponential convergence result in Section~\ref{ssec:expo}. Later in Section~\ref{sec:continuous} we will see how to extend this result to the classical model of asynchronous bandits of \cite{WeberWeiss1990}.

\subsection{Piecewise affine dynamics and definition of a singular point}
\label{ssec:phi}

To avoid ambiguity in the definition of WIP, we assume that the problem is \emph{strictly} indexable. By this, we mean that there does not exist two states that have the same Whittle index. This is mostly a technical assumption that guarantees that there is a unique\footnote{If two states or more had the same index, to specify an index policy, one would need a tie-breaking rule. Our proof could be easily adapted if either the tie-breaking rule defines a strict order of the states or if ties are broken at random.} WIP.

Recall that the state space of a single bandit is $\{ 1\dots d \}$, and assume without loss of generality that the states are already sorted according to their Whittle indices in deceasing order: $\nu_1>\nu_2>\dots>\nu_d$. We shall call a  \emph{configuration} of an $N$ bandits system  the vector representing the proportion of bandits being in each state. Let $\Delta^{d} \in \mathbb{R}^{d}_{\geq 0}$ be the unit $d$-simplex, that is $\Delta^d := \{ \mm \in [0,1]^d \mid m_1 + m_2 + ... + m_{d} = 1 \}$. A possible configuration of the system at a given time step can be represented by a point $\mm$ in $\Delta^{d}$, where $m_i$ is the proportion of bandits in state $i\in\{1\dots d\}$.

Our result on the rate at which WIP becomes asymptotically optimal depends on the property of the iterations of a deterministic map that we define below. Denote by $\bMN(t)$ the $N$ bandits system configuration at time $t$ under WIP. The bandits being time homogeneous Markov chains, we can define a map $\phi:\Delta^d\to\Delta^d$ as
\begin{equation*}
  \phi_i(\mm):=\expect{\MN_i(t+1)\mid \bMN(t)=\mm}
\end{equation*}
for all $i\in\{1\dots d\}$ and $\mm\in\Delta^d$. It is the expected proportion of bandits going to state $i$ at time $t+1$ under WIP, knowing that the system was in configuration $\mm$ at time $t$. This map has the following properties:
\begin{lemma}\label{lem:phi}
  Assume that the bandits are unichain and indexable. The map $\phi$ satisfies:
  \begin{enumerate}[(i)]
    \item The definition of $\phi$ does not depend on $N$ (as long as $\alpha N$ is an integer) nor on $t$.
    \item $\phi$ is a piecewise affine function, with $d$ affine pieces, and $\phi$ is Lipschitz-continuous.
    \item $\phi$ has a unique fixed point, \emph{i.e.} there exists a unique $\mm\in\Delta^d$ such that $\phi(\mm)=\mm$.
  \end{enumerate}
\end{lemma}
\begin{proof}[Sketch of proof]

  The full details of the proof are provided in Appendix~\ref{apx:phi}. We only describe the main ingredients here.

  \emph{Proof of (i) and (ii)} -- For a given configuration $\mm\in\Delta^d$, define $s(\mm)\in\{1\dots d\}$ to be the state such that $\sum_{i=1}^{s(\mm)-1}m_i\le \alpha < \sum_{i=1}^{s(\mm)}m_i$, with the convention that $\sum_{i=1}^{0}m_i=0$. WIP activates bandits by decreasing index order.  This means that when the system is in configuration $\mm$, WIP will activate all bandits that are in states $1$ to $s(\mm)-1$, and $N(\alpha-\sum_{i=1}^{s(\mm)-1}m_i)$ bandits that are in state $s(\mm)$. The other bandits will not be activated.  This implies that the map $\phi$ is defined as:
  \begin{align}
    \label{eq:phi}
    \phi_j(\mm) = \sum_{i=1}^{s(\mm)-1} m_i P^1_{ij} + (\alpha-\sum_{i=1}^{s(\mm)-1}m_i)P^1_{s(\mm)j} + (\sum_{i=1}^{s(\mm)}m_i - \alpha)P^0_{s(\mm)j} + \sum_{i=s(\mm)+1}^{d} m_i P^0_{ij}.
  \end{align}
  Let $\calZ_i := \{\mm\in\Delta^d \mid s(\mm)=i\}$. The above expression of $\phi$ implies that this map is affine on each zone $\calZ_i$, and there are $d$ such zones. Moreover, the value of $\phi$ coincides on the intersection of zones, hence $\phi$ is continuous.

  \emph{Proof of (iii)} -- This part of the proof is more involved, and it relies on indexability. By \eqref{eq:phi}, for each $i\in\{1\dots d\}$, there exist a matrix $\kk_i$ and a vector $\mathbf{b}_i$ such that for all $\mm\in\calZ_i$, we have:
  \begin{equation}
    \label{eq:phi_k}
    \phi(\mm) = \mm \cdot \kk_i  + \mathbf{b}_i.
  \end{equation}
  The assumption that each bandit is unichain implies that the linear equation $\phi(\mm) = \mm$ has a unique solution for each couple $(\kk_i, \mathbf{b}_i)$. Hence, $\phi$ has at most one fixed point inside each zone $\calZ_i$. To conclude, we prove in Appendix~\ref{apx:phi} that the indexability of bandits implies a monotonic property of $\phi$ that we use to obtain uniqueness.
\end{proof}

In what follows, we will denote by $\mm^*$ the unique fixed point of $\phi$. As we will see in Theorem \ref{th:expo_optimal}, the rate at which WIP becomes asymptotically optimal depends on (1) whether the iterations of $\phi$ converge to $\mm^*$ and (2) whether $\mm^*$ lies strictly inside a zone $\calZ_i$. Concerning the second property, we will call a point $\mm$ \emph{singular} if there exists $i\in\{1\dots d\}$ such that $\sum_{j=1}^i m_j=\alpha$. Said otherwise, a fixed point is singular if it is on the boundary of two zones.

\begin{figure}[h!]
  \centering
  \begin{subfigure}[b]{0.45\linewidth}
    \includegraphics[width=\linewidth]{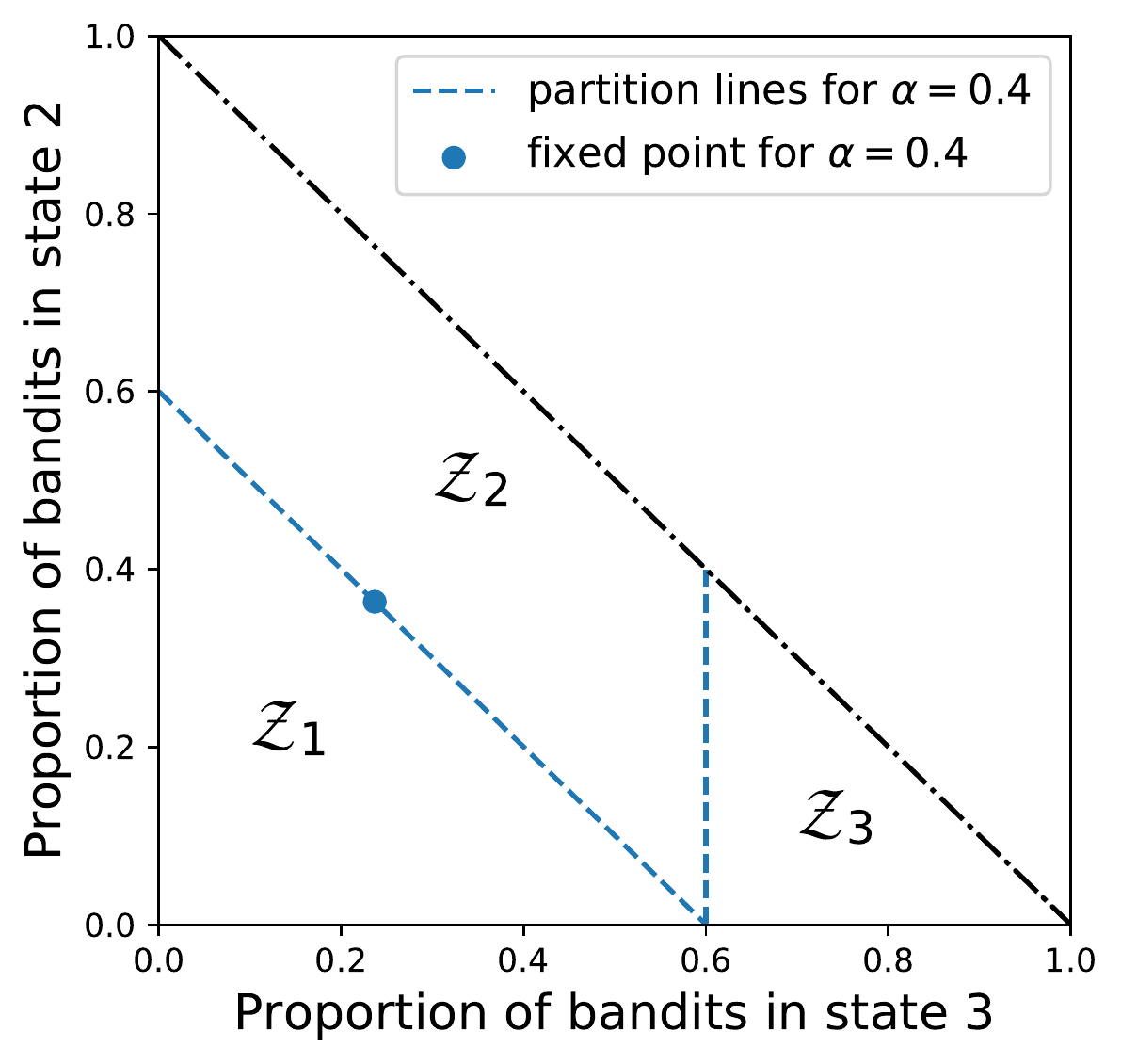}
    \caption{\label{fig1left}Singular fixed point.}
  \end{subfigure}
  \begin{subfigure}[b]{0.45\linewidth}
    \includegraphics[width=\linewidth]{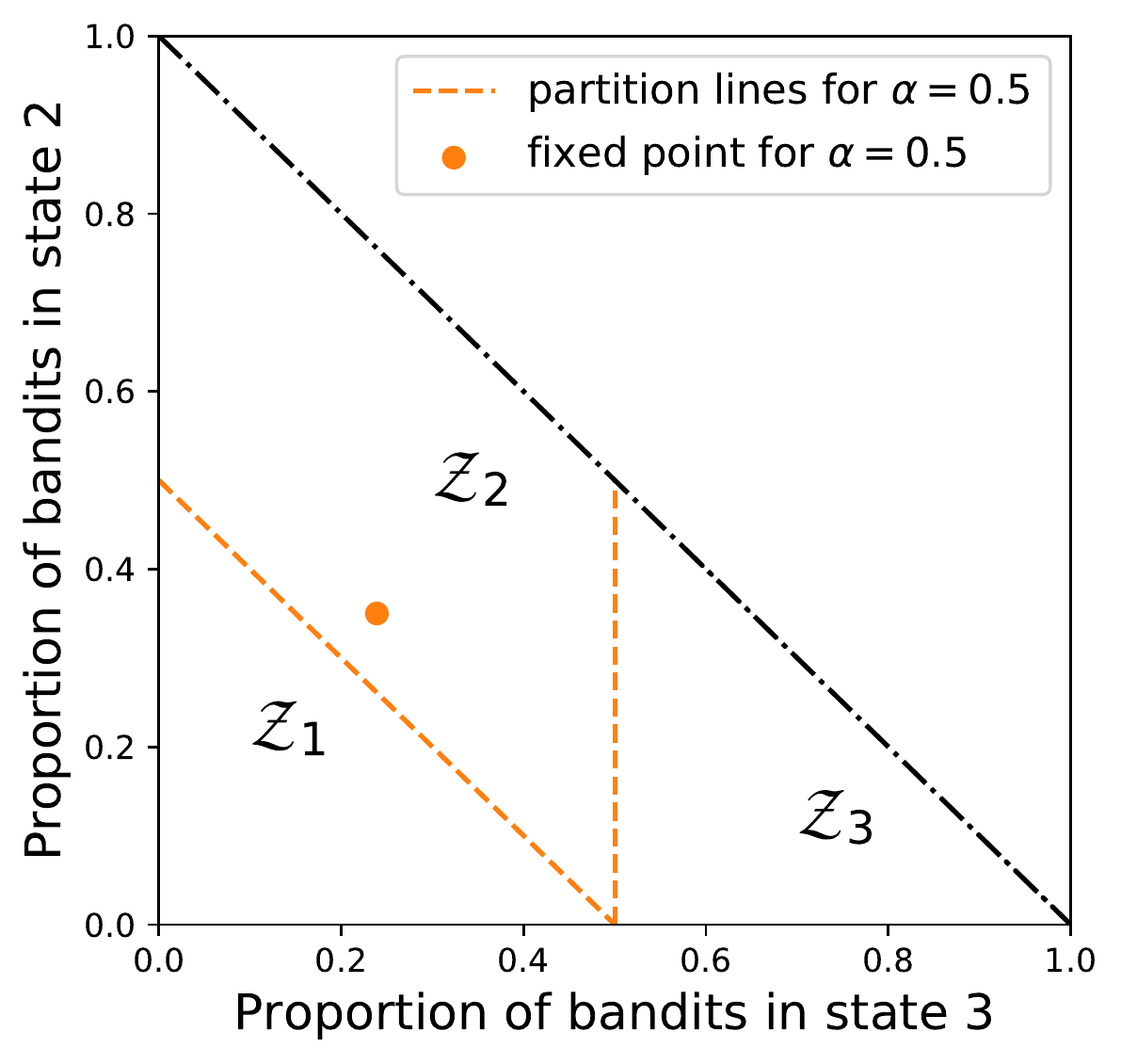}
    \caption{\label{fig1right}Non-singular fixed point.}
  \end{subfigure}
  \caption{An example with $d=3$. When  $\alpha= 0.4$ (Figure \ref{fig1left}) the fixed point is singular, while
    for $\alpha= 0.5$ (Figure \ref{fig1right}) it is not singular.}
  \label{fig1}
\end{figure}

In Figure~\ref{fig1}, we illustrate the notion of singular fixed point by an example in dimension $d=3$. As $m_1+m_2+m_3=1$, the simplex $\Delta^3$ can be represented in a $2$-dimensional space as $\Delta^2_c$, where $\Delta^d_c$ is the unit $d$-simplex and its interior. Our convention is that the $x$-coordinate of a point corresponds to $m_3$ (the proportion of bandits in state 3), and the $y$-coordinate corresponds to $m_2$ (the proportion of bandits in state 2). The colored dotted lines of Figures~\ref{fig1left} and \ref{fig1right} correspond to the set of singular points. These lines partition the different zones $\calZ_i$. The partition of zones, as well as the position of the unique fixed point depend on $\alpha$. For this example, when $\alpha=0.4$ (Figure~\ref{fig1left}), the fixed point is singular, while for $\alpha=0.5$ (Figure~\ref{fig1right}), it is non-singular.

\subsection{Exponential convergence rate}
\label{ssec:expo}

We are now ready to state our main theorem. Assume that bandits are indexable, at a given time $t$, WIP sorts all bandits according to the Whittle indices $\nu_{S_n(t)}$ and activates the $\alpha N$ bandits that have the highest indices. We denote the long-term average expected reward of WIP as $\ind$:
\begin{align*}
  &\ind :=  \lim_{T \rightarrow \infty} \frac{1}{T} \mathbb{E} \Big[ \sum_{t=0}^{T-1} \sum_{n=1}^N R^{a_n(t)}_{S_n(t)} \Big], &\quad \text{ where for all $t$, $\ba(t)$ is chosen according to WIP.}
\end{align*}

Let $\Phi_t$ be defined as the $t$-th iteration of the map $\phi$, \emph{i.e.} $\Phi_t : \Delta^d \rightarrow \Delta^d$ is $\Phi_0 (\mm) := \mm, \text{ and } \Phi_{t+1}(\mm) := \phi \big( \Phi_t (\mm) \big)$.
Recall that $\mm^*$ is the unique fixed point of $\phi$. As stated in the next theorem, the asymptotic optimality of WIP is guaranteed when $\mm^*$ attracts all trajectories of $\Phi_{t\geq0} (\cdot)$. In the rest of the paper, unless otherwise specified, we use $\| \cdot \|$ to denote the $\mathcal{L}^\infty$-norm of a vector.

\begin{theorem}[Exponential convergence rate theorem-synchronous case]
  \label{th:expo_optimal}
  Consider a synchronous restless bandit problem $\big\{ (\pp^0, \pp^1, \rr^0, \rr^1); \alpha \big\}$ such that:
  \begin{enumerate}[(i)]
    \item  Bandits are unichain and indexable.
    \item $\mm^*$ is a uniform global attractor of $\Phi_{t\geq0} (\cdot)$, \emph{i.e.} for all $\epsilon>0$, there exists $T(\epsilon)>0$ such that for all $t\ge T(\epsilon)$ and all $\mm\in\Delta^d$, one has $\norm{\Phi_t(\mm)-\mm^*}\le\epsilon$.
    \item $\mm^*$ is not singular.
  \end{enumerate}
  Then there exists two constants $b,c>0$ that depend only on $\pp^0, \pp^1, \rr^0, \rr^1$ and $\alpha$, such that for any $N$ with $\alpha N$ being an integer,
  \begin{equation}
    \label{eq:th_expo}
    0 \le \rel - \ind \le b \cdot e^{-cN}.
  \end{equation}
  Recall that $\rel$ is the value of the relaxed problem \eqref{eq3}.
\end{theorem}

\begin{proof}[Sketch of proof]
The full details of the proof are given in Appendix \ref{apx:proof_sync}. We first transform the evaluation of the performance to the analysis of the configuration of the bandits system. We then show that in stationary regime the expectation of $\bMN (0)$ concentrates exponentially fast on the fixed point $\mm^*$. More precisely, there exists constants $b',c>0$ such that $\norm{\mathbb{E}[\mathbf{M}^{(N)} (0)] - \mm^*} \leq b' \cdot e^{-c N}$.
In order to show this:
  \begin{itemize}
    \item We first use Hoeffding's inequality in Lemma~\ref{lem:Hoeffding} to show that for any configuration $\mm$: $\proba{\norm{\bMN(1)-\phi\big(\bMN(0)\big)}\ge \delta \mid \bMN(0)=\mm}\le e^{-2N \delta^2}$.
    \item By Lipschitz continuity of $\phi$, for a time $t$, we apply Lemma \ref{lem:Hoeffding} to prove Lemma \ref{lem:Hoeffding2}, which bounds $\proba{\norm{\bMN(t)-\Phi_t\big(\bMN(0)\big)}\ge\epsilon}$ by a term that depends on $t$ but decreases exponentially fast with $N$.
    \item We combine this with the uniformly global attractor property to show that when $t$ is large enough, $\bMN(t)$ is within a neighborhood $\mathcal{N}$ of $\mm^*$ with very high probability. As $\mm^*$ is non-singular, this neighborhood can be taken to be within a zone $\zz_i$ on which $\phi$ is affine. We will choose carefully this neighborhood $\mathcal{N}$ and make sure that its choice does not depend on $N$. We then deduce an exponentially small upper-bound for the probability of $\mmm^{N}(0)$ in stationary regime being outside $\mathcal{N}$ (see Subsection~\ref{apx:case1}), hence allows us to restrict our attention to a zone where $\phi$ is affine.
    \item The result then follows by using Stein's method on the process restricted to this affine zone, which shows that conditional on starting inside the neighborhood $\mathcal{N}$, the additive long-term distance between the large $N$ stochastic trajectory and the deterministic trajectory is exponentially small (see Subsection~\ref{apx:case2}).
  \end{itemize}
\end{proof}

Here are some comments on the assumptions of Theorem~\ref{th:expo_optimal}. These assumptions are very similar to the ones needed to prove the asymptotic optimality of WIP in the case of asynchronous bandits of \cite{WeberWeiss1990}. The indexability and unichain property of the bandit problem is a necessary condition for WIP to be well defined and was also assumed in \cite{WeberWeiss1990}. The most difficult assumption to verify is  point \emph{(ii)} that requires $\mm^*$ to be a global attractor.  Note that our assumption \emph{(ii)} is slightly stronger than the one used in \cite{WeberWeiss1990} as we assume that $\mm^*$ attracts the trajectories \emph{uniformly} in the initial condition. We use this to obtain our rate of convergence. We will see in Section~\ref{sec:numerical_cycles} that this condition is necessary in the sense that there exists examples that satisfy all assumptions of Theorem~\ref{th:expo_optimal} except this one and for which WIP is not asymptotically optimal (more comments on this in Remark \ref{rem:sub-optimal}). In \cite{WeberWeiss1990} the authors have proven that this is not possible when $d=3$ for the asynchronous case, and they give an example of dimension $d=4$ in \cite{weber1991addendum} for which the deterministic differential equation has an attracting limit cycle. Their proof of the impossibility in dimension 3 relies on Bendixson's Criteria to exclude limit cycles and Poincaré-Bendixson Theorem to exclude chaotic behaviors. The synchronous situation is different: by randomly generating examples, we are able to find attracting limit cycles (of period 2) for $d = 3$ and more complicated shape of cycles for $d=4$ (see Section~\ref{sec:numerical}).

\begin{rem} \label{rem:singular} {\bf The singular case.} The non-singularity of the fixed point $\mm^*$  is also necessary in the sense that the following  simple example  satisfies all the assumptions of Theorem~\ref{th:expo_optimal} except this one and does not satisfy  \eqref{eq:th_expo}. Consider the following 2 states bandit problem  with $\pp^0 = \pp^1 = $ $
\begin{pmatrix}
  0.5 & 0.5 \\
  0.5 & 0.5
\end{pmatrix}
$
, $\rr^0 = (0,0)$, $\rr^1 = (1,0)$, and $\alpha = 0.5$. The fixed point is $\mm^* = (0.5,0.5)$. It is singular.

It should be clear that $\rel[1]=0.5$. In stationary regime, the configuration $\bMN$ of the system of size $N$ is distributed independently from the policy employed. Moreover, WIP will activate in priority the bandits in state $1$. This implies that the reward of WIP will be $\ind=\expect{\min(M^{(N)}_1,0.5 \cdot N)}$. As $M^{(N)}_1$ follows a binomial distribution of parameter $(N,0.5)$, the central limit theorem shows that
\begin{align*}
  \lim_{N\to\infty}\frac{1}{\sqrt{N}}(\rel-\ind) = 0.5 \cdot \expect{\max(G,0)} = \frac{1}{\sqrt{2\pi}},
\end{align*}
where $G$ is a standard normal random variable.

This example shows that, in a case where $\mm^*$ is singular, the convergence in \eqref{eq:th_expo} may occur at rate $\mathcal{O}(1/\sqrt{N})$ and not at exponential rate. Note that if we take instead $\alpha \ne 0.5$, $\ind/N$ converges to $\rel[1] = \min(\alpha,0.5)$ at exponential rate due to the fact that almost all the mass of a Gaussian distribution is concentrated around its mean value $\alpha$ (which is different from $0.5$).
\end{rem}

\begin{rem}{\bf Cyclic and chaotic behaviors}.
\label{rem:sub-optimal}
Although the drift $\phi$ is piecewise affine and has a unique fixed point, the long run behavior of the deterministic dynamical system $\mm(t+1)  = \phi(\mm(t))$ can be cyclic or chaotic. In these cases, the fixed point is no longer a global attractor, and the performance of WIP is in general not asymptotically optimal.

More precisely, when the dynamical system admits a cycle as a global attractor for almost every initial configuration in the simplex, then as suggested in \cite{WeberWeiss1990}, one can infer a cyclic version of Theorem~\ref{th:rel}: The performance of WIP converges to the average reward on the cycle. This average reward is in general strictly smaller than $\rel[1]$, while $\opt / N$ always converge to $\rel[1]$, regardless to the behavior of the deterministic system (from Theorem \ref{th:rel}). Consequently, when cycles appear, the performance of WIP is asymptotically \emph{sub-optimal}. See also Sections~\ref{sec:numerical_cycles}.
\end{rem}

\begin{rem}{\bf What happens when $\alpha N$ is not an integer}.
  \label{rem:non-integer-alphaN}
  The exponential convergence rate in Theorem~\ref{th:expo_optimal} assumes that $\alpha N$ is an integer. When it is not the case, a decision maker cannot activate exactly $\alpha N$ bandits at each time step. There are three natural solutions to define the model in such cases: (1) activate $\lfloor \alpha N\rfloor$ bandits; (2) activate $\lceil N\alpha \rceil$ bandits; (3) activates $\lfloor \alpha N \rfloor$ bandits, plus one more bandit being activated with probability $\alpha N - \lfloor \alpha N \rfloor$. As we further discuss in Section~\ref{ssec:non-integer-alphaN}, the convergence rate in the first two solutions is much slower than in the third solution.
\end{rem}

\begin{rem}{\bf Finding optimal constants}.
  \label{rem:optimal-c}
  Theorem~\ref{th:expo_optimal} claims the existence of constants $b$ and $c$ for which the \emph{inequality} \eqref{eq:th_expo} holds true, but we do not emphasize on the optimality of the constant $c$, in the sense of finding constant $\tilde{c}$ such that
  \begin{equation*}
    \limsup_{N \rightarrow \infty} - \frac{1}{N} \log \big( \rel - \ind \big) = \tilde{c}.
  \end{equation*}
  Our choice of $c$ in the proof of Theorem \ref{th:expo_optimal} provided in Appendix \ref{apx:proof_sync} actually depends subtly on the given parameters, and we believe that finding $\tilde{c}$ is, if not impossible, a much more demanding task. Nevertheless, later on in Section \ref{sec:numerical_3states} we shall illustrate via numerical examples that the approximate value of $c$ is affected by the level of singularity of the fixed point, which in turn is affected by the value of $\alpha$, if all the other parameters $\big\{ \pp^0, \pp^1, \rr^0, \rr^1 \big\}$ are fixed.

\end{rem}

\section{Numerical Experiments}\label{sec:numerical}

In this section, we first provide statistical results to justify the conditions needed for Theorem \ref{th:expo_optimal}, and then verify numerically the exponential convergence rate for a general 3 states restless bandit model with non-singular fixed points. We also evaluate numerically the convergence rate for a singular fixed point example. We then investigate the situation when attracting limit cycle appears, as well as the solutions for $\alpha N $ not being integers. To ensure reproducibility\footnote{The code to reproduce all experiments and figures of the paper are available in a git repository \url{https://gitlab.inria.fr/phdchenyan/code_ap2021}.}, all parameters used in our numerical experiments are provided in Appendix~\ref{apx:paras}.

\subsection{How general is the general case?}
\label{sec:numerical_statistics}

The exponential convergence rate for the performance of WIP on a restless bandit problem is very desirable, however, several conditions have to be verified beforehand, listed in order as:
\begin{enumerate}[(C1)]
\item The restless bandit problem is indexable;
\item The unique fixed point is not singular;
\item The unique fixed point is a uniform global attractor.
\end{enumerate}

Condition (C1) is mostly verified through the specific structure of the restless bandit problem and by using various techniques that are  model dependent; a general method for the test of indexability is also presented in \cite{nino2007dynamic}. As its running time is exponential in the number of states, it is only applicable when this number is small (say $d<10$). For Condition (C2), being in an exact singular situation is improbable (for a given problem, the activation ratio $\alpha$ can only be singular if it satisfies an equality constraint). More generally, we will also check that the  closer  the fixed point to a singular situation, the smaller the coefficient $c$ in the estimation of the exponential rate could be. This point will be made more precise in the next subsection.

Condition (C3) is more complicated to verify as there is no general method to exclude cyclic or chaotic behaviors from a dynamical system. A necessary condition for (C3) to hold is that the unique fixed point is locally stable. Numerically, this is easy to verify when $\mm^*$ is not singular: indeed, in this case the dynamical system is affine in a neighborhood of $\mm^*$: $\phi(\mm)=(\mm-\mm^*) \cdot \kk_{s(\mm^*)} + \mm^*$, where $\kk_{s(\mm^*)}$ is defined as in \eqref{eq:phi_k}. The dynamical system is locally stable if $\kk_{s(m^*)}$ is a stable matrix, \emph{i.e.} if the norm of all eigenvalues of $\kk_{s(m^*)}$ is less than $1$ \footnote{Recall that $\phi$ is an application from $\Delta^d$ to $\Delta^d$. This means in particular that all the rows of all matrices $\kk_i$ sum to $1$. Therefore, each of these matrices have an eigenvalue $1$. When we write "the norm of all eigenvalues of $\kk_i$ is smaller than 1", we mean $1$ is an eigenvalue of $\kk_i$ and has multiplicity one; all other eigenvalues must be of norm strictly less than $1$.}. If $\kk_{s(m^*)}$ is not a stable matrix, the fixed point will not be a global attractor and an attracting cycle will appear in most cases, see Sections~\ref{sec:numerical_cycles}.

\begin{table}[ht]
\begin{tabular}{l|lllll}
\hline
\multicolumn{1}{r|}{Dimension $d$} & 3    & 4    & 5  & 6  & 7 \\ \hline
Nb. of non-indexable instances    & 653  & 81   & 5  & 0  & 0 \\
Nb. of indexable instances such that $\mm^*$ is not locally stable & 9878 & 1020 & 82 & 11 & 0 \\ \hline
\% of examples violating a condition of Theorem~\ref{th:expo_optimal} & $0.1\%$ & $0.01\%$ & $10^{-3}\%$
& $10^{-4}\%$ &  0.
\end{tabular}
\caption{\label{tab:random}%
  Number of randomly generated instances that violate any of the conditions of Theorem~\ref{th:expo_optimal} out of $10^7$ uniformly generated restless bandit models for each dimension $d\in\{3,4,5,6,7\}$. }
\end{table}

To check how general are these conditions, we generate a large number of synchronous restless bandit problems by choosing random parameter $(\pp^0,\pp^1,\rr^0,\rr^1)$ in dimensions $d\in\{3,4,5,6,7\}$. We estimate the rarity of violations of the above conditions. More precisely, for each $d$, we randomly generate $10^7$ instances of $(\pp^0,\pp^1,\rr^0,\rr^1)$, using a uniform distribution in $[0,1]$ for the rewards, and uniform distribution for probability vectors $\pp^0_i$ and $\pp^1_i$ over the simplex $\Delta^d$. We then count the number of instances that violate conditions (C1) or (C3), the results are reported in Table~\ref{tab:random}.  This table shows that the number of models that satisfy the conditions is more than $99.8\%$ for $d=3$; when $d=7$, all generated models (among $10^7$) satisfy our conditions. In our tests, what we mean by \emph{the number of indexable instances such that $\mm^*$ is not locally stable} is the number of models for which there exists $\alpha\in(0,1)$ such that $\mm^*$ is not locally stable. This can be done by testing each of the $d$ matrices $K_i$ (where $K_i$ is defined as in \eqref{eq:phi_k}). To test numerically if $\mm^*$ is a global attractor, we check if $\lim_{t\to\infty}\Phi_t(\mm)=\mm^*$ on $10^5$ randomly generated initial conditions $\mm$.  Note that we exclude $d=2$ from this table: it can be shown that when $d=2$, all models are indexable and the fixed point $\mm^*$ is always a globally stable attractor. Concerning indexability, a similar statistical test was already presented in \cite[(Table~2, left table, $\beta=1$)]{nino2007dynamic}. Our results are essentially equivalent (note however that \cite{nino2007dynamic} does not study the stability of $\mm^*$ but only tests indexability).

\subsection{The influence of how \emph{non-singular} is a fixed point}
\label{sec:numerical_3states}

\begin{figure}[hbtp]
  \centering
  \begin{subfigure}[b]{0.49\linewidth}
    \includegraphics[width=\linewidth]{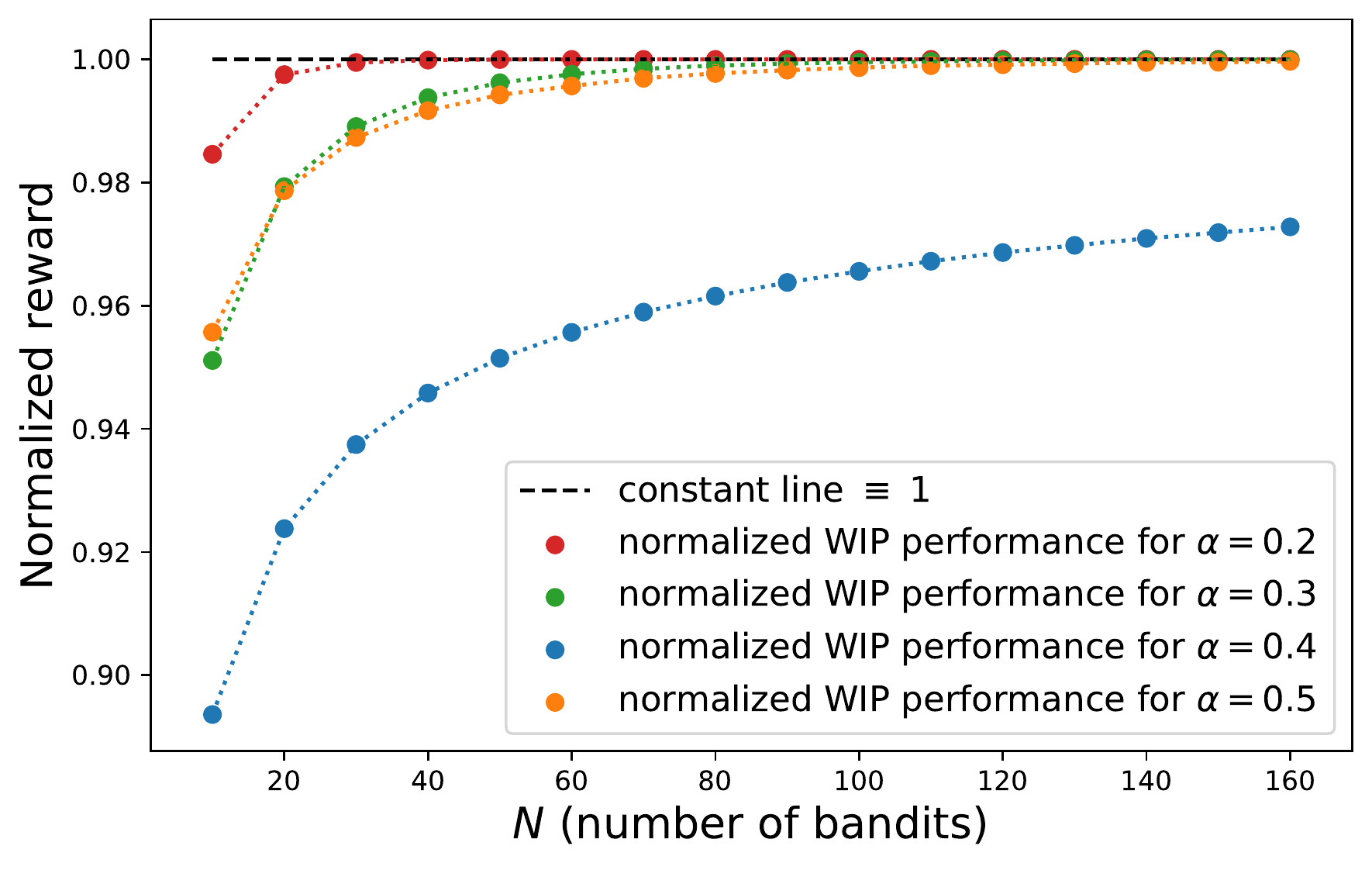}
    \caption{Varying $N$ for $\alpha= 0.2,0.3,0.4$ and 0.5. }
      \label{fig2left}
  \end{subfigure}
  \begin{subfigure}[b]{0.49\linewidth}
    \includegraphics[width=\linewidth]{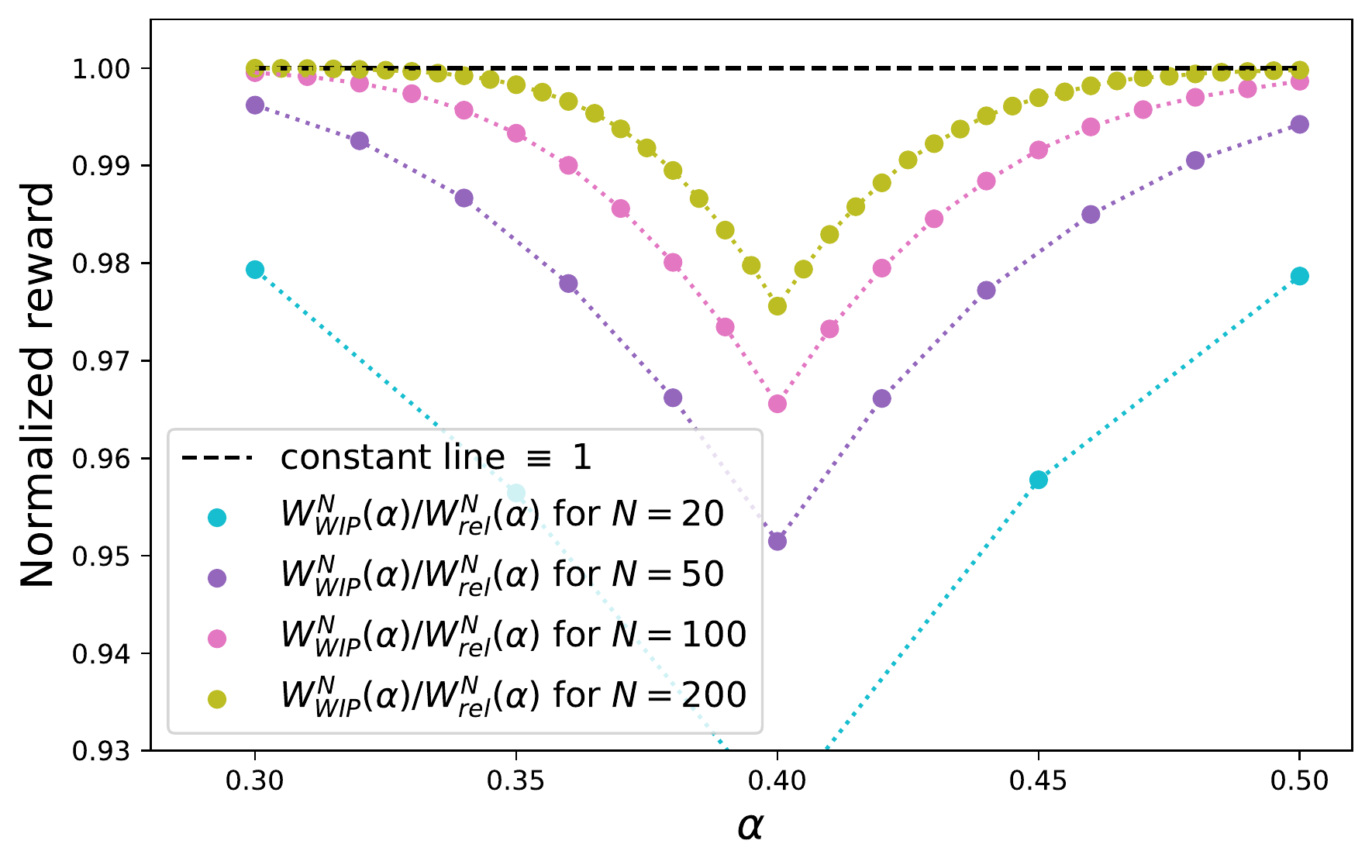}
    \caption{Varying $\alpha$ for $N=\in\{20,50,100,200\}$. }
      \label{fig2right}
  \end{subfigure}
  \caption{Normalized performance of WIP.}
  \label{fig2}
\end{figure}

To test how the "non-singularity" of the fixed point $\mm^*$ affects the convergence rate, we consider the example displayed in Figure~\ref{fig1} with varying values of $\alpha$ in the range between $0.20$ and $0.50$. Although we have no formal proof, we believe that the fixed points are uniform global attractors for two reasons:
\begin{itemize}
  \item All matrices $\kk_i$ are locally stable because the eigenvalues of $\kk_2$ are $\{1,-0.4\dots,0.08\dots\}$ while $\kk_1$ and $\kk_3=\kk_d$ are always stable matrices.
  \item For all tested values of $\alpha$, we simulated $\Phi_t(\mm)$ from random initial points $\mm$ and they all converge to the corresponding fixed point $\mm^*$.
\end{itemize}
The fixed point $\mm^*$ is singular when $\alpha=0.4$, and it is non-singular for any other values of $\alpha\in[0.2,0.5]$. This implies that all assumptions of Theorem~\ref{th:expo_optimal} are satisfied when $\alpha \neq 0.4$. As $\rel$ depends on the value of $\alpha$, to make better comparisons, we consider the quantity $\ind/\rel$, which is the normalized performance of WIP with respect to the relaxation upper-bound. In Figure \ref{fig2left}, we choose four values of $\alpha$ as 0.2, 0.3, 0.4 and 0.5, and plot the normalized performances as a function of the number of bandits $N$ that takes values on multiples of $10$, so that $\alpha N$ are always integers. The value of $\ind$ are cumputed by using simulations. We repeat each simulation so that 95$\%$ confidence intervals become negligible and hence can not be seen from the pictures. In Figure \ref{fig2right}, this time we fix the value of $N$ and plot the normalized performance as a function of $\alpha$ where $\alpha$ varies between $[0.3,0.5]$ with a stepsize of $1/N$: $\alpha\in\{0.3, 0.3+1/N, 0.3+2/N,\dots,0.5\}$ (so that again $\alpha N$ are always integers). These two figures suggest that the convergence rate is related to how far $\mm^*$ is away from the closest boundary of two zones (\emph{i.e.} how non-singular it is). Here is an intuitive explanation for this phenomenon:  the stochastic system in equilibrium will wander around the fixed point $\mm^*$ that gives the optimal reward, now if $\mm^*$ is near a boundary, it is more likely for the stochastic trajectory to jump into another neighboring polytope $\zz'$, in which case another affine drift applies and this may take the trajectory away from $\mm^*$.

\begin{figure}[hbtp]
  \centering
  \begin{subfigure}[b]{0.32\linewidth}
    \includegraphics[width=\linewidth]{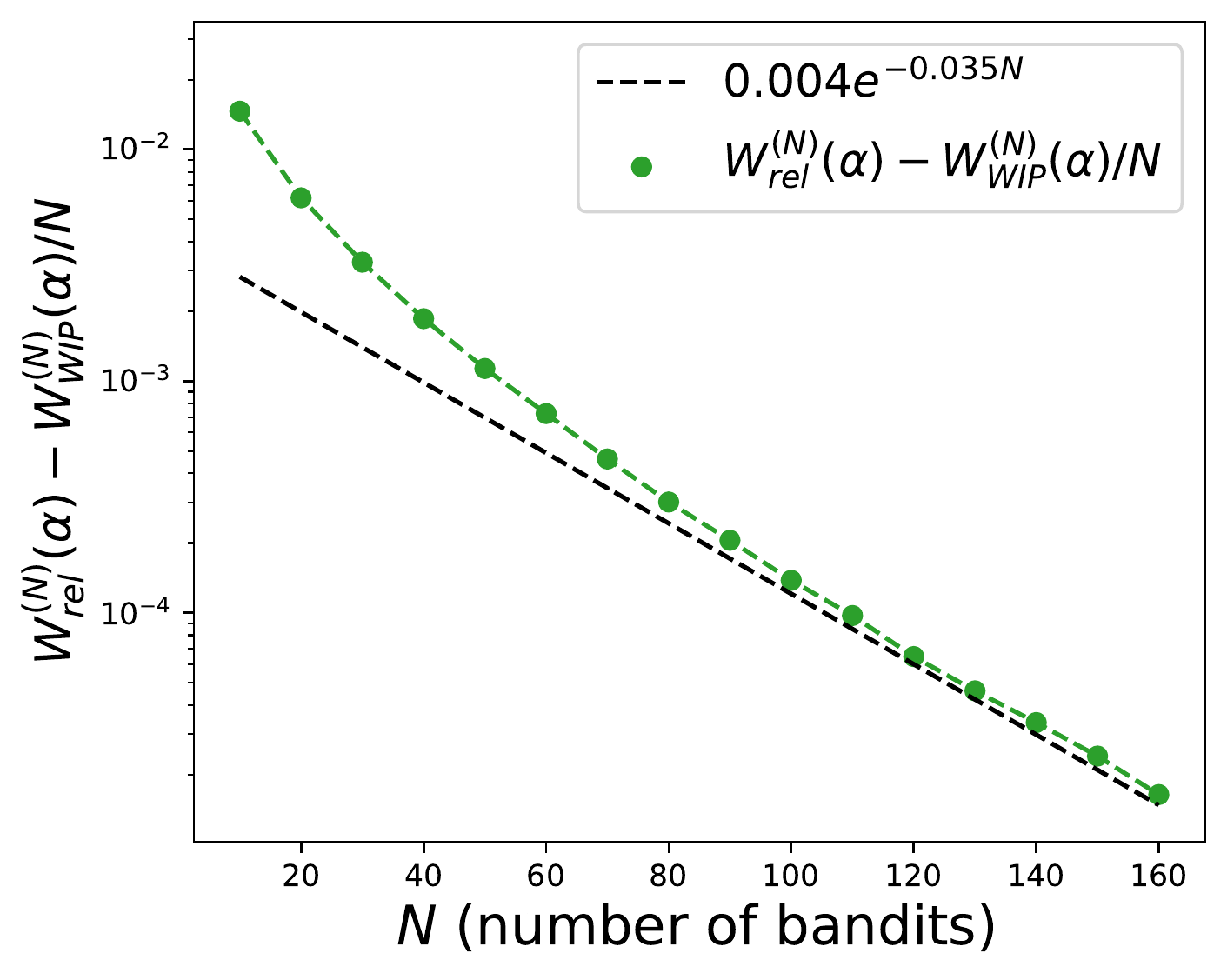}
    \caption{$c \approx 0.032$ for $\alpha=0.3$.}
  \end{subfigure}
  \begin{subfigure}[b]{0.32\linewidth}
    \includegraphics[width=\linewidth]{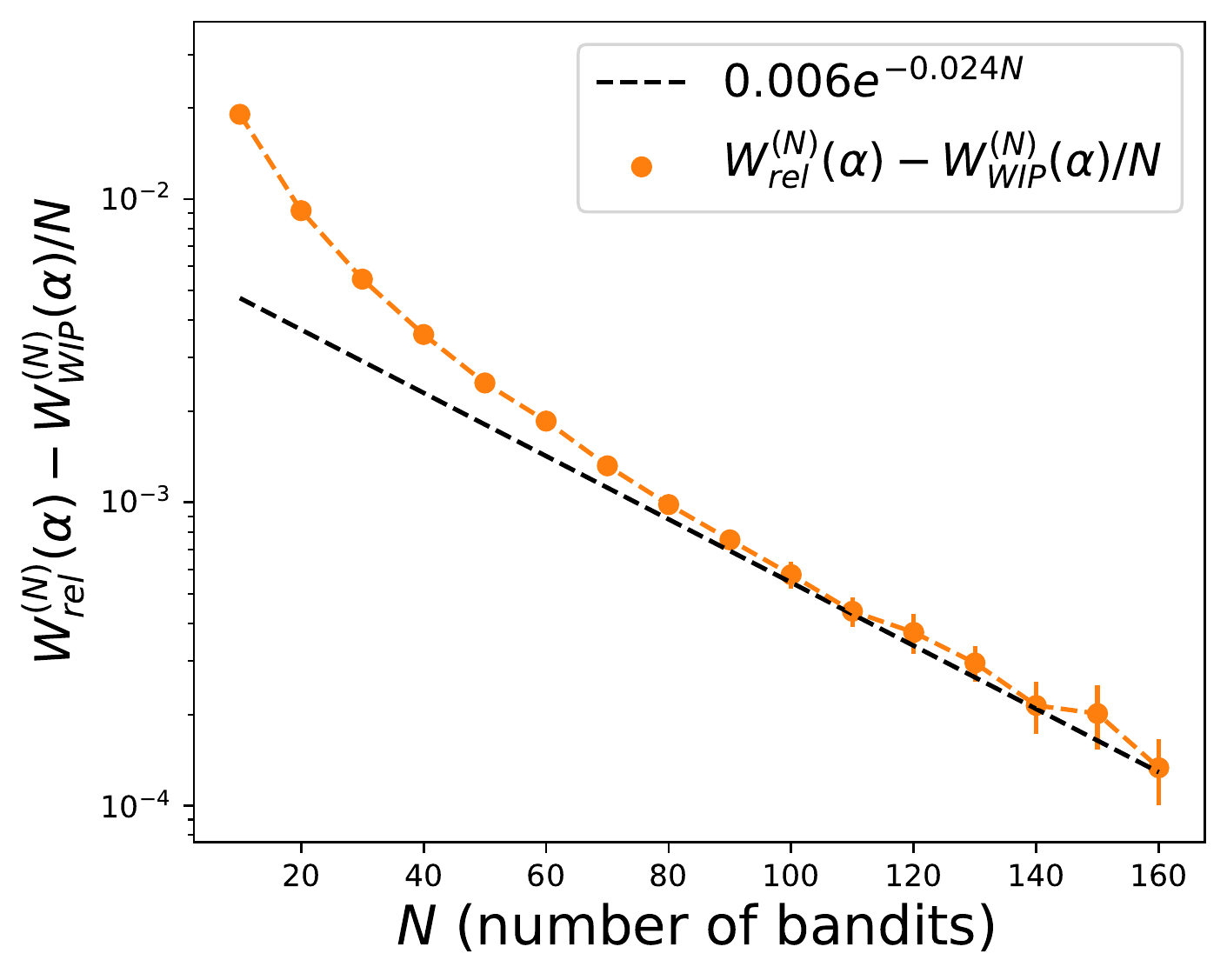}
    \caption{$c \approx 0.024$ for $\alpha=0.5$.}
  \end{subfigure}
  \begin{subfigure}[b]{0.32\linewidth}
    \includegraphics[width=\linewidth]{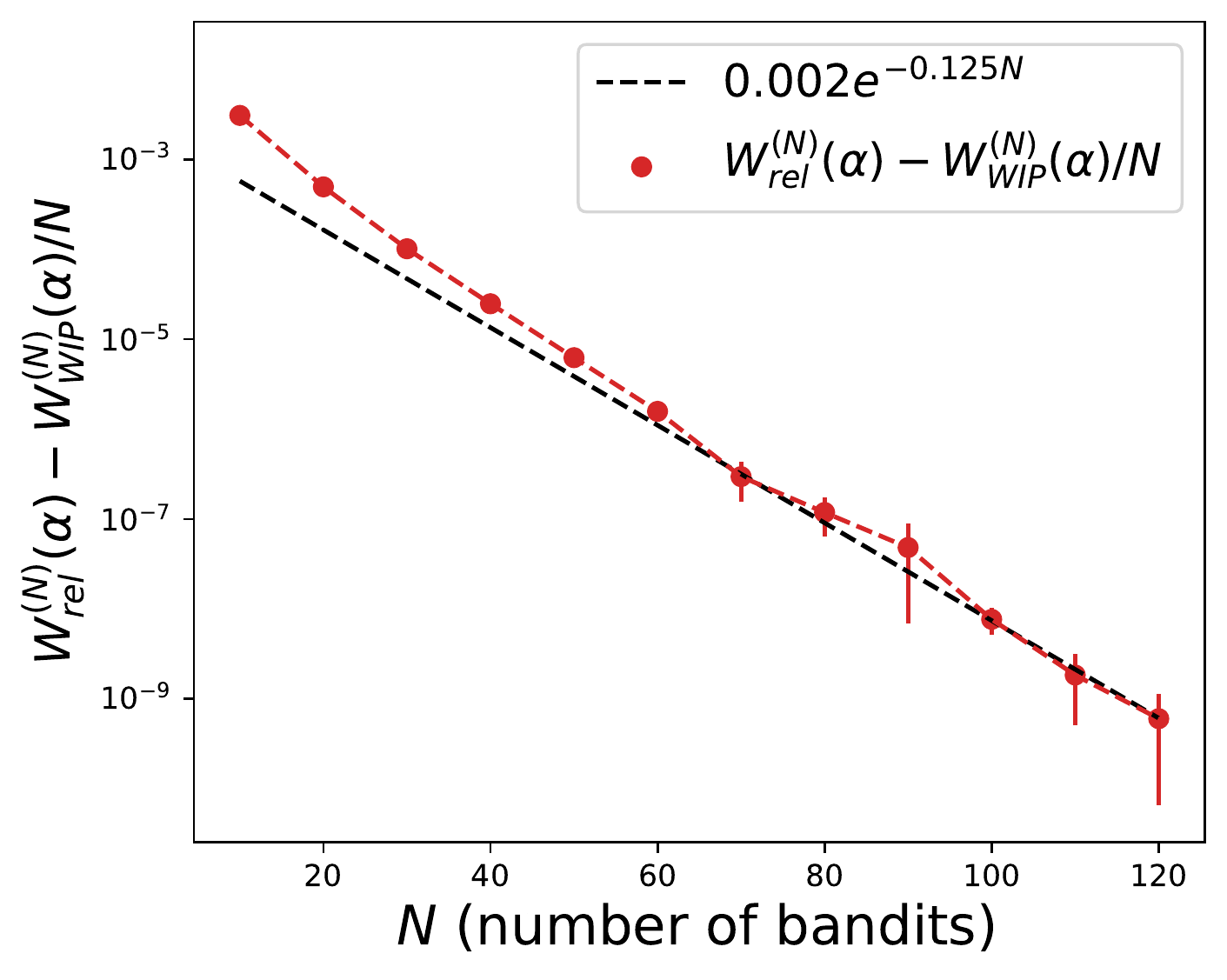}
    \caption{$c \approx 0.125$ for $\alpha=0.2$.}
  \end{subfigure}
  \caption{Fitting $\mbox{subgap}(N)$ with a line in log-scale.}
  \label{fig3}
\end{figure}

To examine more closely the convergence rate, let us consider the quantity
\begin{equation}
  \label{eq:subgap}
  \mathrm{subgap}(N) := \rel[1] - \frac{\ind}{N}.
\end{equation}
Theorem~\ref{th:expo_optimal} implies that $\mbox{subgap}(N)$ converges to 0 approximately as $b \cdot e^{-c \cdot N}$, for some constants $b,c>0$. In Figure~\ref{fig3}, we plot in log-scale the subgap \eqref{eq:subgap} as a function of $N$ for the same model as in Figure~\ref{fig2} and $\alpha = 0.2$, 0.3 and 0.5.  For each value of $\alpha$, we also plot the best-fit $b'e^{-c'N}$ which is a straight line in log-scale. The constant $c$ is around 0.03 for $\alpha = 0.3,0.5$, and it is around 0.125 for $\alpha=0.2$. However, in the singular case $\alpha=0.4$, we could not find a straight line to fit $\log \big( \mbox{subgap}(N) \big)$. But if we plot instead $\mbox{subgap}(N) \cdot \sqrt N$, the curve behaves like a constant. Moreover, this constant behavior is lost\footnote{We refer to our git repository for a more thorough numerical exploration of this case.} as soon as we plot $\mbox{subgap}(N) \cdot N^{\beta}$, with a power $\beta = 0.49$ or $\beta=0.51$. This gives numerical evidence for an $\mathcal{O}(\frac{1}{\sqrt N})$ convergence rate in this singular case, same as for the example given in Remark (\ref{rem:singular}). Actually, we believe that the convergence rate is $\mathcal{O}(\frac{1}{\sqrt N})$ for all singular global attractor situations, but a proof of this claim is still open to us.

\subsection{WIP is suboptimal in cases of cycle: examples of an attracting period-2 cycle with $d=3$}
\label{sec:numerical_cycles}

In the synchronous case, it is possible to have an attracting periodic cycle for the dynamics $\Phi_{t\geq0} (\cdot)$ as soon as $d=3$. Several new features that we have not encountered in the previous uniform global attracting fixed point situation will appear. To motivate our discussion, recall that under the assumptions of Theorem \ref{th:expo_optimal}, the quantity $\ind / N$ converges to $\rel[1] $ exponentially fast. As $N \cdot \rel[1] \geq \opt \geq \ind$, this guarantees that WIP becomes asymptotically optimal \emph{at least} exponentially fast in $N$. Such nice feature is no longer true, however, when the dynamical system has a periodic cycle.

\begin{figure}[hbtp]
  \centering
  \begin{subfigure}[b]{0.32\linewidth}
    \includegraphics[width=\linewidth]{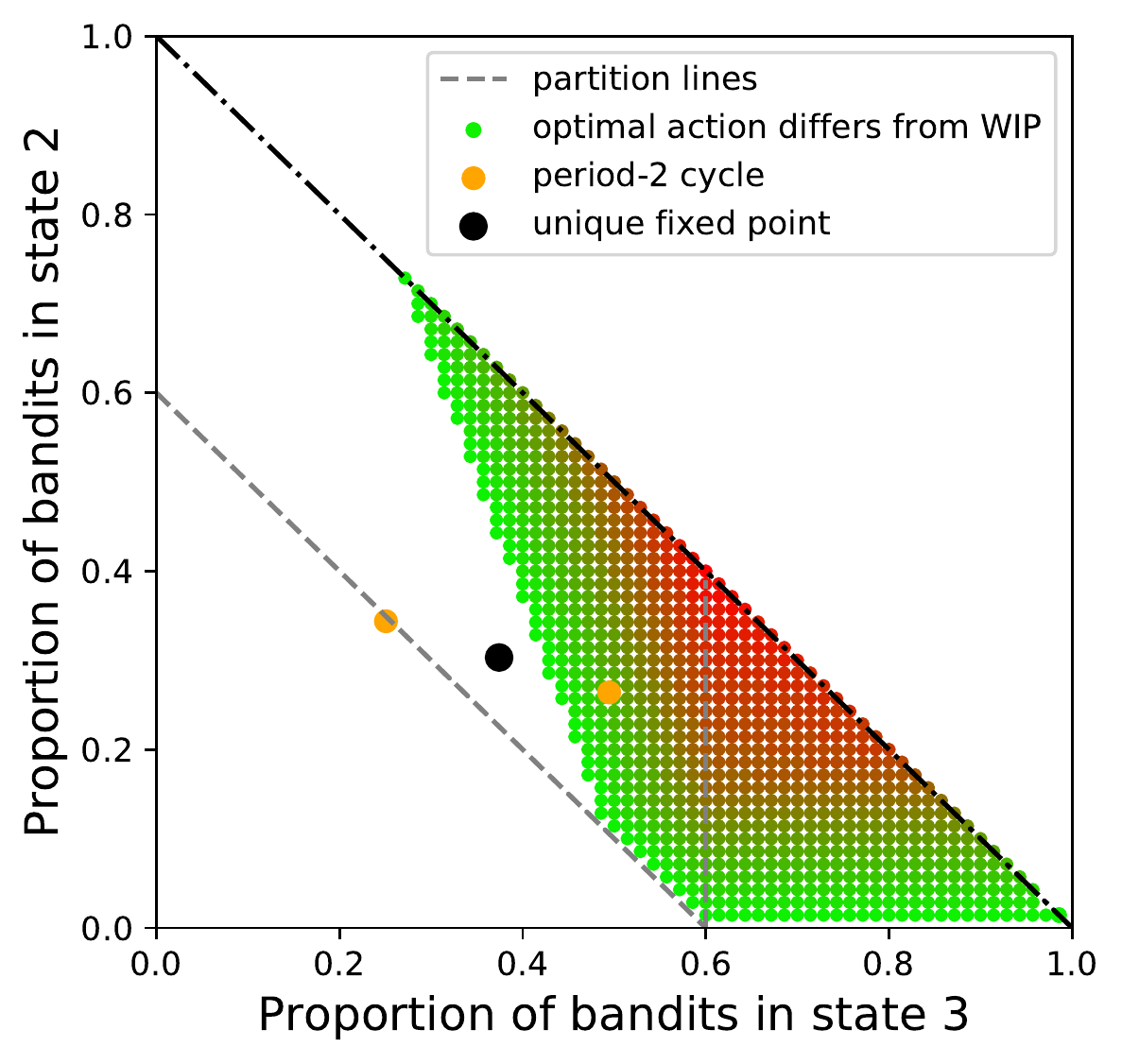}
    \caption{Example 1}
  \end{subfigure}
  \begin{subfigure}[b]{0.32\linewidth}
    \includegraphics[width=\linewidth]{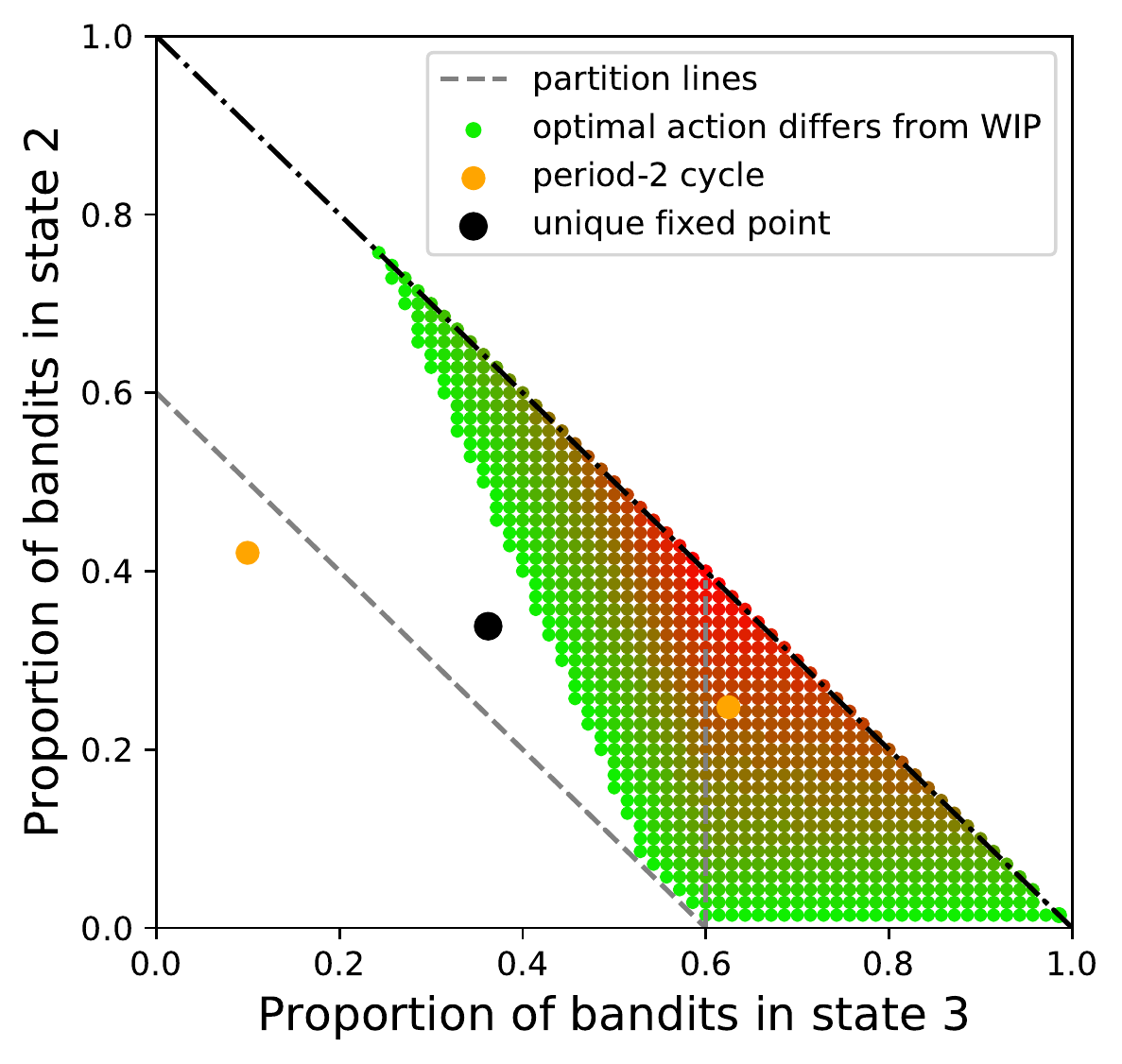}
    \caption{Example 2}
  \end{subfigure}
  \begin{subfigure}[b]{0.32\linewidth}
    \includegraphics[width=\linewidth]{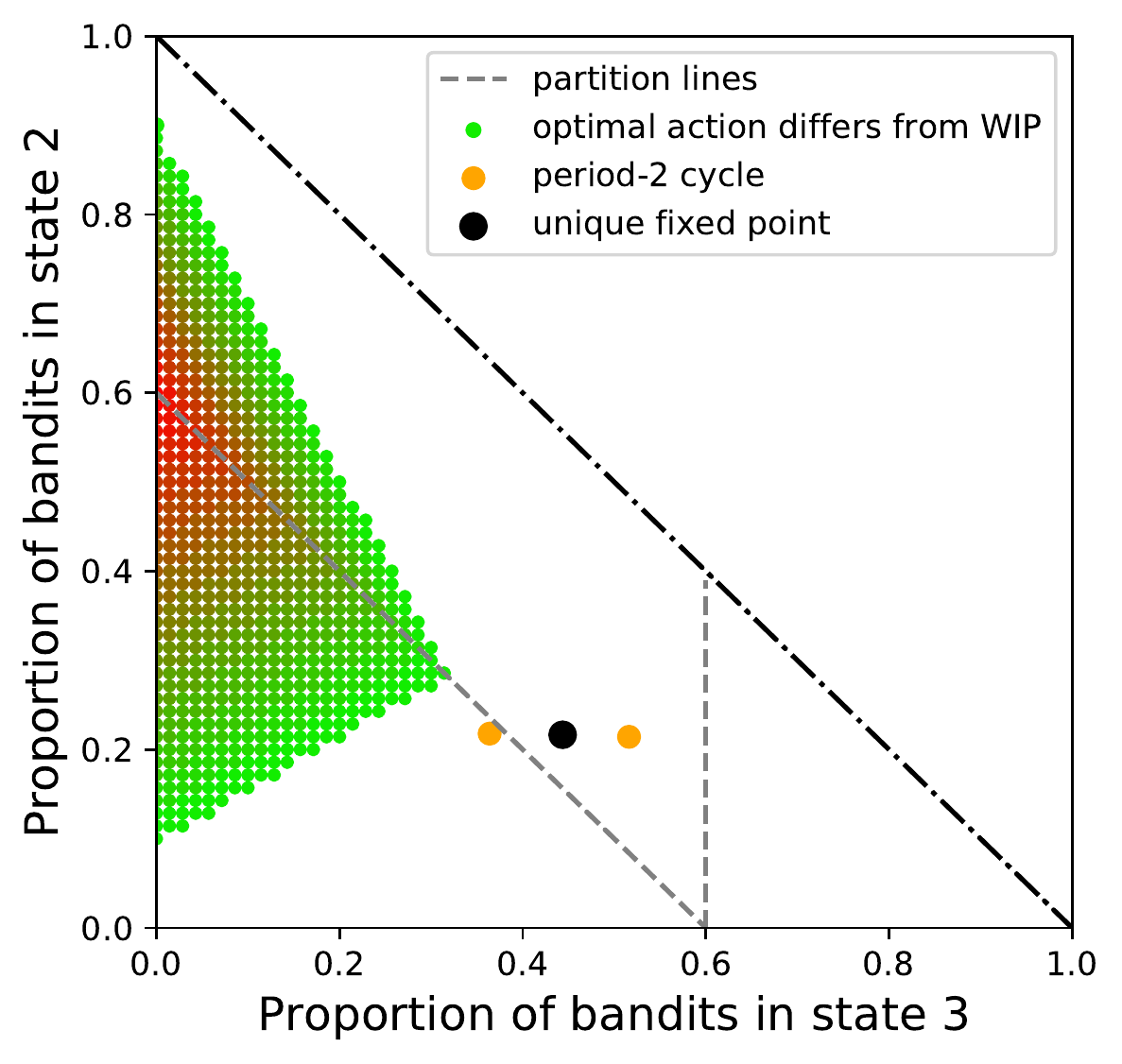}
    \caption{Example 3}
  \end{subfigure}
  \caption{Action differences plot for three period-2 cycle examples with $d=3$ and $N=70$.}
  \label{fig6}
\end{figure}

In all of the randomly generated examples presented in Table~\ref{tab:random}, only very few examples do not have a unique locally stable attractor. In all of those counter-examples, the limiting behavior is a cycle of period 2. In this subsection, we illustrate the behavior of restless bandits under WIP in three examples with $d=3$ and $\alpha=0.4$. For all of them, the dynamical system $\Phi_{t\geq0}(\cdot)$ has an attracting cycle of period 2. The fixed point and the two points of the attracting cycle for each example are shown in Figure \ref{fig6}. Note that for these three examples, the matrices $\kk_2$'s are not stable and they all have an eigenvalue smaller than $-1$.

In Figure~\ref{fig6}, we also highlight  the configurations in which the optimal policy takes a different action than  WIP  when the number of bandits is $N=70$. Such configurations are represented as colored dots, starting with the greenest color, the deeper the red, the more the optimal action deviates from WIP's action on this configuration. The blank area means that on these configurations WIP is an optimal action.

\begin{figure}[hbtp]
  \centering
  \begin{subfigure}[b]{0.32\linewidth}
    \includegraphics[width=\linewidth]{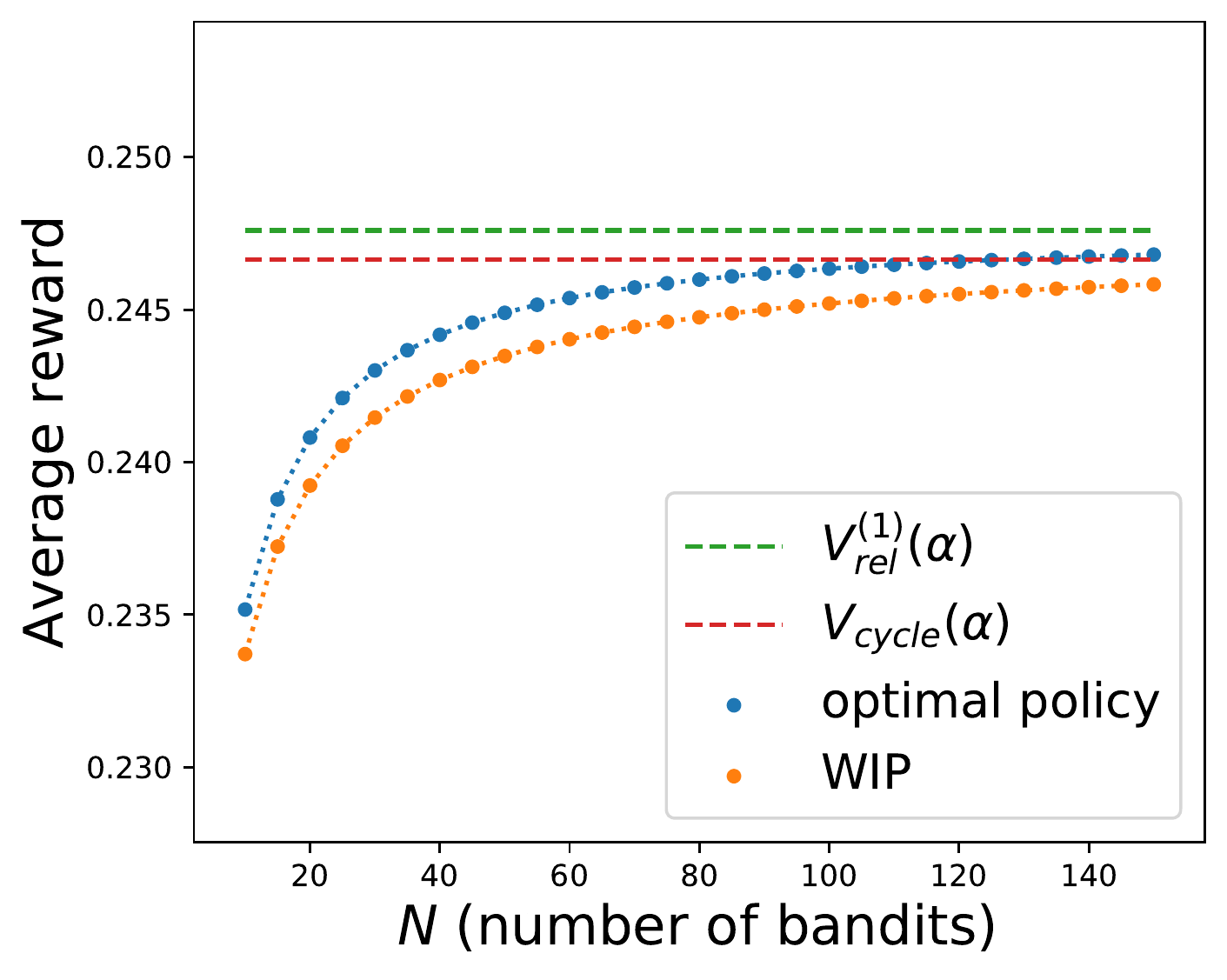}
    \caption{Performance for Example 1.}
  \end{subfigure}
  \begin{subfigure}[b]{0.32\linewidth}
    \includegraphics[width=\linewidth]{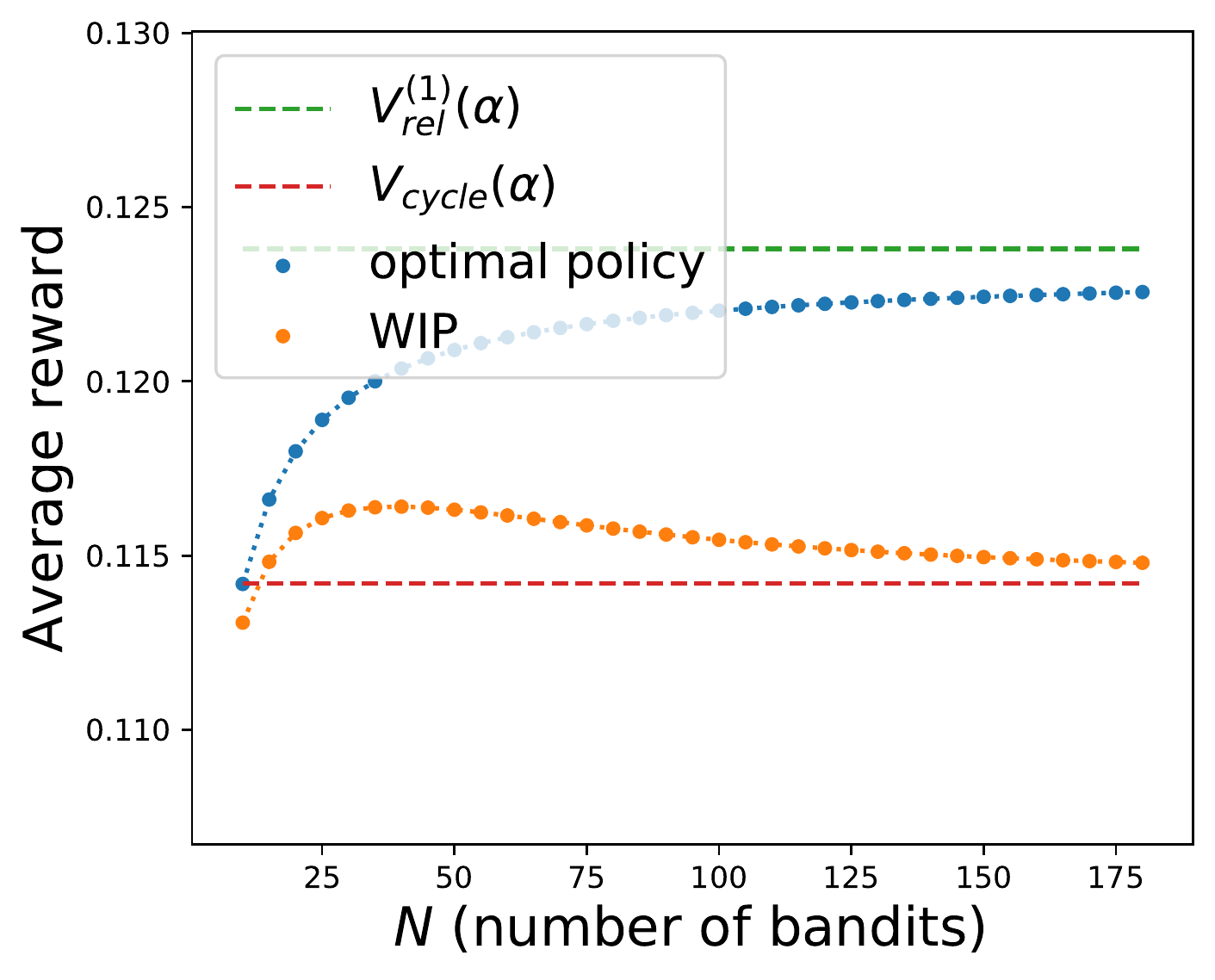}
    \caption{Performance for Example 2.}
  \end{subfigure}
  \begin{subfigure}[b]{0.32\linewidth}
    \includegraphics[width=\linewidth]{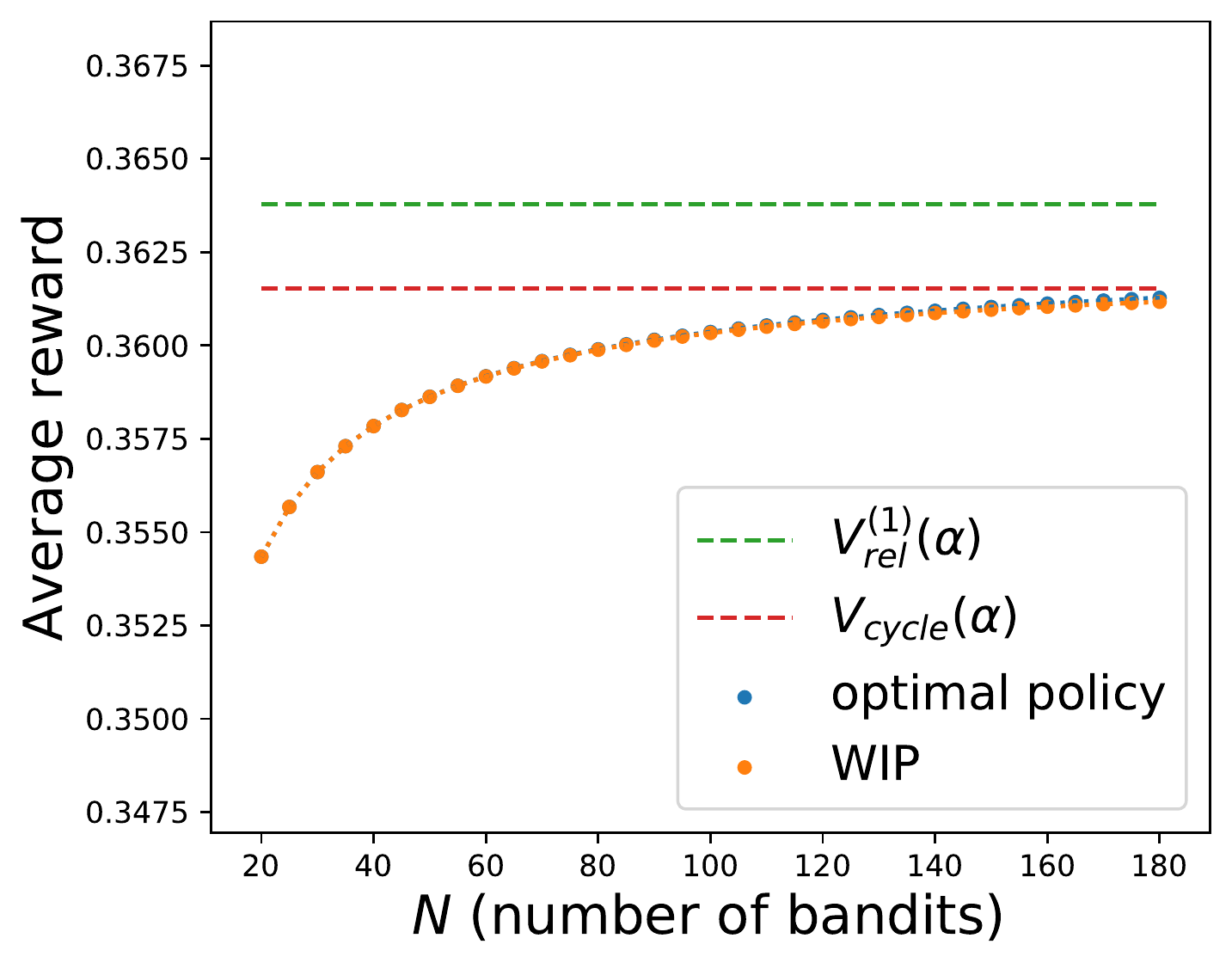}
    \caption{Performance for Example 3.}
  \end{subfigure}
  \caption{Performance of optimal policy and WIP for three period-2 cycle examples with $d=3$.}
  \label{fig7}
\end{figure}

We then plot in Figure~\ref{fig7} the value of the optimal decision rule, $\opt / N$, and of WIP, $\ind / N$, as a function of the number of bandits $N$. We take multiples of 5 for values of $N$ so that $\alpha N$ are always integers. Several comments are in order:
\begin{itemize}
    \item As mentioned in Remark \ref{rem:sub-optimal}, $\ind / N$ converges to the averaged reward on the cycle, denoted here by $\cycle$, instead of reward on the fixed point $\rel[1]$. Note that $\cycle$ is not an upper bound on $\ind/N$ and  sometimes, as in Example~2, $\ind / N$ becomes greater than $\cycle$ for $N\approx30$ before decreasing to this value from above.
    \item The quantity $\big(\opt - \ind\big) / N$ converges to $\rel[1] - \cycle$, which is strictly positive and might be relatively large, depending on the parameters.
    \item The gap $\big(\opt - \ind\big) / N$ can be increasing with $N$, as in Example 2 and 3. This violates the intuition that WIP should be closer to the optimal policy as $N$ grows. It should be contrasted with the uniform global attractor situation for which $ \big(\opt - \ind\big) / N \rightarrow 0$ exponentially fast.
 \end{itemize}

As a final remark, we would like to point out that it is possible to have more complicated shape of attracting limit cycles (of variant periods), as long as $d=4$, and an unstable matrix $\kk$ can have either a pair of conjugate complex eigenvalues or two real eigenvalues with norm bigger than 1.

\subsection{Non integer values of $\alpha N$}
\label{ssec:non-integer-alphaN}

Our previous analysis rely on the assumption that $\alpha N$ is an integer. Let us briefly discuss in this subsection how to deal with non integer values of $\alpha N$ for the optimization problem \eqref{eq1} under \eqref{eq2}. Consider the following three possible modifications for constraint \eqref{eq2}:
\begin{itemize}
  \item (\emph{floor}) At each decision epoch, we activate $\lfloor \alpha N \rfloor$ bandits;
  \item (\emph{ceil}) At each decision epoch, we activate $\lceil \alpha N \rceil$ bandits;
  \item (\emph{continue}) At each decision epoch, we activate $\lfloor \alpha N \rfloor$ bandits, and one more bandit is activated with probability $\{\alpha N\} := \alpha N - \lfloor \alpha N \rfloor$.
\end{itemize}
We denote by $\floor$, $\ceil$ and $\continue$ the reward of WIP under these three solutions. Note that these three values all coincide with our previous $\ind$ if $\alpha N$ is an integer, but otherwise are different in general. 
We claim  that under the assumptions of Theorem~\ref{th:expo_optimal}, the average reward when always activating $\lfloor\alpha N\rfloor$ bandits or always activating $\lceil\alpha N\rceil$ bandits will be at distance $\calO(1/N)$ from the relaxation $\rel$:
\begin{align}
   \liminf \Delta R \cdot \{\alpha N\} \leq \lim_{N\to\infty}\rel - \floor & \leq \limsup \Delta R \cdot \{\alpha N\}, \label{eq:non-integer1} \\
   \liminf -\Delta R \cdot (1-\{\alpha N\}) \leq \lim_{N\to\infty}\rel - \ceil & \leq \limsup -\Delta R \cdot (1-\{\alpha N\}), \label{eq:non-integer2}
\end{align}
 where $\Delta R := R^1_{s(\mm^*)} - R^0_{s(\mm^*)}$. On the contrary, $\continue$ converges at exponential rate to $\rel$. Indeed, recall the definition of $\phi(\mm)$, especially equation \eqref{eq:phi}. In Lemma \ref{lem:phi} we showed that $\phi$ does not depend on $N$ as long as $\alpha N$ is an integer. Now the modification \emph{(continue)} is actually a natural extension (and is the only one) on the definition of $\phi$ to all integers $N$, while keeping the property that its definition does not depend on $N$ for a given $\alpha$, in contrast with \emph{(floor)} and \emph{(ceil)}. Consequently a Hoeffding's equality holds in this extension for $\Phi_t$ on all $N$ (see Lemma \ref{lem:Hoeffding} and Lemma \ref{lem:Hoeffding2} in the appendix), regardless whether $\alpha N$ is an integer, and it gives an exponential bound in $N$ for the deviation. The rest of the proof then works exactly the same as in our original proof for Theorem \ref{th:expo_optimal} under this extension.

\begin{figure}[hbtp]
  \centering
  \begin{subfigure}[b]{0.48\linewidth}
    \includegraphics[width=\linewidth]{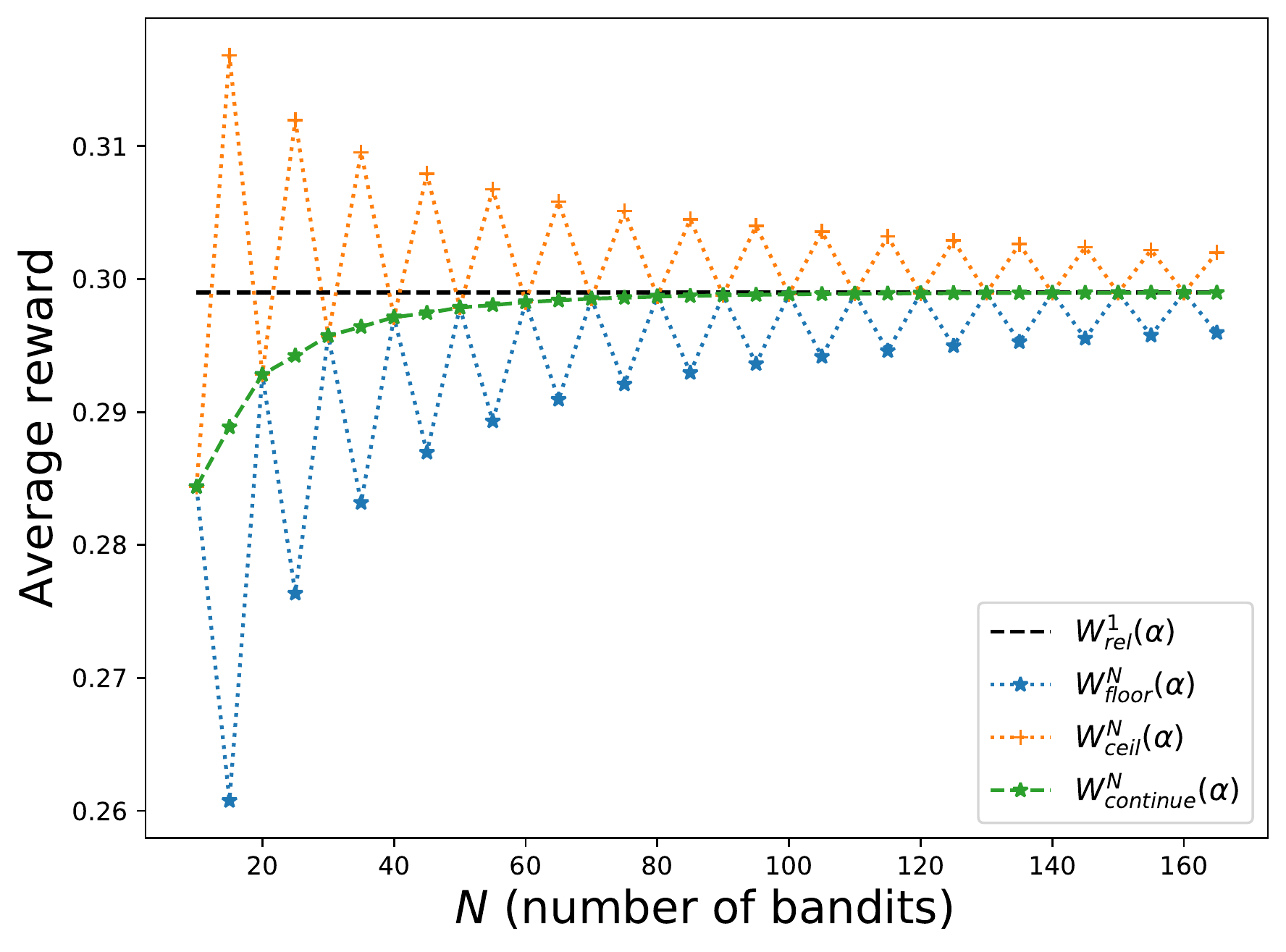}
    \caption{\label{fig10left}Performance of the three solutions}
  \end{subfigure}
  \begin{subfigure}[b]{0.48\linewidth}
    \includegraphics[width=\linewidth]{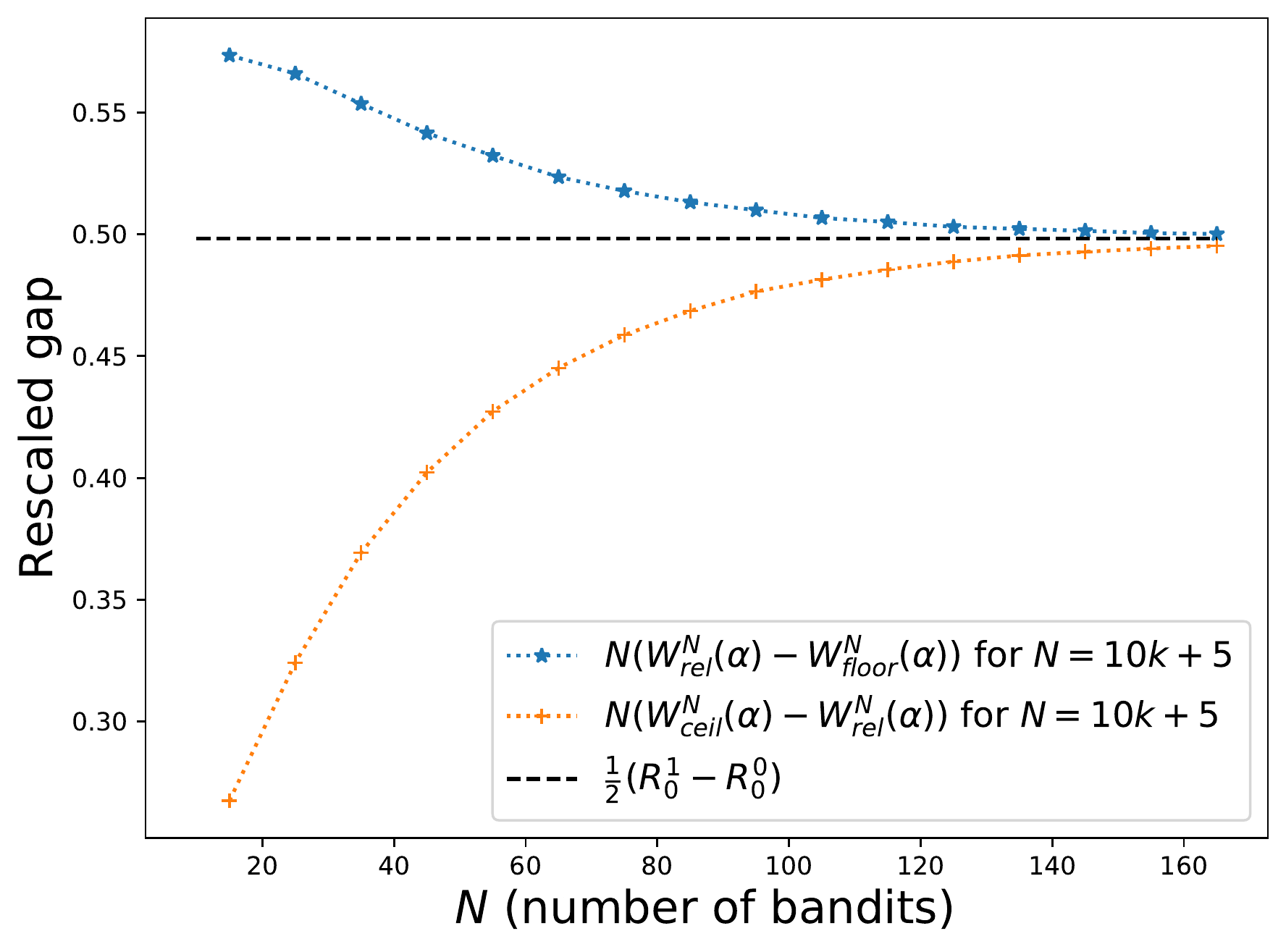}
    \caption{\label{fig10right} $\calO(1/N)$ convergence rate for $floor$ and $ceil$}
  \end{subfigure}
  \caption{Illustration for non integer values of $\alpha N$ on our previous example with $d=3$.}
  \label{fig:cycle3}
\end{figure}

To illustrate these points, we consider in Figure~\ref{fig:cycle3} the same example as in Section~\ref{sec:numerical_3states}, with $\alpha = 0.3$. As in Figure~\ref{fig2}, the green curve represents $\ind / N$ for $N$ being a multiple of $10$. Here, we extend this curve to all $N$ being a multiple of $5$, using the three modifications. The values of $\floor/N$, $\continue/N$ and $\ceil/N$ are plotted respectively in blue, green and red dots for $N\in\{25,35,45,\dots\}$, while their values coincide for $N$ being a multiple of $10$.  Now as claimed in \eqref{eq:non-integer1} and \eqref{eq:non-integer2}, the differences \emph{gap-floor} $\rel - \floor$ and \emph{gap-ceil} $\rel-\ceil$ converge to $\pm0.5 \cdot (R^1_1-R^0_1)$ when $N\to\infty$ and $\{N\alpha\} \equiv 0.5$, \emph{i.e.} $N=5\cdot(2k+1)$. This is indeed verified in Figure~\ref{fig10right}. However, the green dots in Figure~\ref{fig10left} have their positions along the green curve that is originally plotted only for $N = 10k$. This indicates that $\continue$ converges at exponential rate to $\rel$.

\section{Illustration: Markovian fading channels}
\label{sec:channels}

The Markovian fading channel is a typical synchronous restless bandit model. In \cite{ouyang2012asymptotically} a two-classes channel problem has been studied. By using the same scaling as here, the authors of \cite{ouyang2012asymptotically} have proven the asymptotic optimality of WIP for this model, after verifying the global attractor property of the deterministic system. In this section we take a step further, evaluate numerically the convergence rate of the performance, and verify if it is exponential, as claimed in Theorem \ref{th:expo_optimal}.

Let us first briefly review this two-classes channel model, more details could be referred from \cite{ouyang2012asymptotically}. A Gilbert-Elliott channel is modeled as a two-state Markov chain with a bad state $0$ and a good state $1$. Two classes of channels are available, with the transition probability matrix for class $k\in\{1,2\}$ being
$
\begin{pmatrix}
  p_k & 1-p_k \\
  r_k & 1-r_k
\end{pmatrix}
$
, where $p_k$ is the probability of a class $k$ channel being in good state at time $t+1$ if it was in good state at time $t$, and $r_k$ is the probability being in good state if one time step ago it was in bad state. We assume the channels are \emph{positively correlated}, namely $p_k > r_k$ for $k=1,2$.

We consider a total population of $N$ channels, a proportion $\beta$ of them are from class 1. Due to limited resource, each time we can only activate a proportion $\alpha$ of the channels, and only a channel in good state under activation can transmit data. We assume that we can observe the state of a channel only when it is activated. Otherwise, we keep track of the state of a channel by using a belief value $b^k_{s,t}$ where $k=1,2$, $s=0,1$ and $t \geq 1$. The value $b^k_{s,t}$ is the probability for a class $k$ channel to be in good state, provided that it was activated (hence observed) $t$ time steps ago and was observed to be in state $s$. The expression of $b^k_{s,t}$ is
\begin{equation*}
  b^k_{0,t} = \frac{r_k - (p_k-r_k)^t r_k}{1+r_k-p_k}, \ b^k_{1,t} = \frac{r_k + (1-p_k)(p_k-r_k)^l}{1+r_k-p_k}.
\end{equation*}
To cast this channel model into a synchronous restless bandit problem, we treat each channel as a bandit, and its state space is the whole set of possible values of $b^k_{s,t}$'s. The transition matrices $\pp^0$, $\pp^1$ can then be naturally written down, \emph{e.g.}
\begin{equation*}
  \mathbb{P}^0 (b^k_{s,t}, b^k_{s,t+1}) = 1,\qquad \mathbb{P}^1 (b^k_{s,t}, b^k_{1,1}) = b^k_{s,t},\qquad \ \mathbb{P}^1 (b^k_{s,t}, b^k_{0,1}) = 1-b^k_{s,t},
\end{equation*}
all other probabilities being $0$.

We evaluate the performance by the throughput of the system, hence we obtain a reward of 1 each time we activate a channel and it is in good state. Under the MDP framework, this is equivalent to assuming that state $b^k_{s,t}$ gives a reward $b^k_{s,t}$ under activation. It is shown in \cite{ouyang2012asymptotically} that this problem is indexable, and that Whittle index can be calculated explicitly (via techniques due to the specific structure of the model). The index of a state $b^k_{s,t}$ is denoted by $\nu (b^k_{s,t})$ and is equal to:
\begin{equation*}
  \nu (b^k_{s,t}) = \begin{cases}
                      \frac{(b^k_{0,t}-b^k_{0,t+1})(t+1)+b^k_{0,t+1}}{1-p_k+(b^k_{0,t}-b^k_{0,t+1})t + b^k_{0,t+1}}, & \mbox{if } s=0 \\
                      \frac{r_k}{(1-p_k)(1+r_k-p_k)+r_k}, & \mbox{otherwise}.
                    \end{cases}
\end{equation*}
We remark that for $k=1,2$, the index value $\nu (b^k_{0,t})$ is an increasing function of $t$, and furthermore $\nu (b^k_{0,t}) \xrightarrow{t \rightarrow \infty} \nu (b^k_{1,t'})$, where $\nu (b^k_{1,t'})$ is a constant for $t' \geq 1$. We shall also point out that the relative orders of the index values $\nu \big( b^k_{s,t} \big)$ between two classes $k=1$ and $k=2$ could be different from the orders of the belief values $b^k_{s,t}$. This indicates an interaction between classes and makes the Whittle indices for this model interesting.

The reader might have noticed that to apply Theorem~\ref{th:expo_optimal}, two assumptions are violated: first, the restless bandit model we consider here has a countable infinite state space; second, not all bandits are identical (there are two classes of bandits). The first point might raise some technical difficulties that we have not encountered on our previous finite state model. However, it can be shown that the states $b^k_{0,t}$ for $t$ large are extremely rarely visited, hence using a threshold $t^*$ and ignoring all states $b^k_{s,t}$ with $t > t^*$ (\emph{i.e.} treating them as $b^k_{s,t^*}$) makes a negligible difference. Concerning the two classes of bandit, we argue that having two classes of bandits can be represented by a single class of bandit by considering a larger state-space: the state of a bandit would be $(k,b^k_{s,t})$, where $k$ is its class and $b^k_{s,t}$ is its belief value. Compared to our model, in this new case, the bandits are no longer unichain as a bandit of class $k$ cannot become a bandit of class $k' \neq k$. This implies that the quantities $\ind$ and $\rel$ will depend on the initial condition of the system, \emph{i.e.} on the fraction $\beta$ of bandits that are in class $1$. Apart from that, our results apply mutatis mutandis to this case.

\begin{figure}[hbtp]
  \centering
  \begin{subfigure}[b]{0.48\linewidth}
    \includegraphics[width=\linewidth]{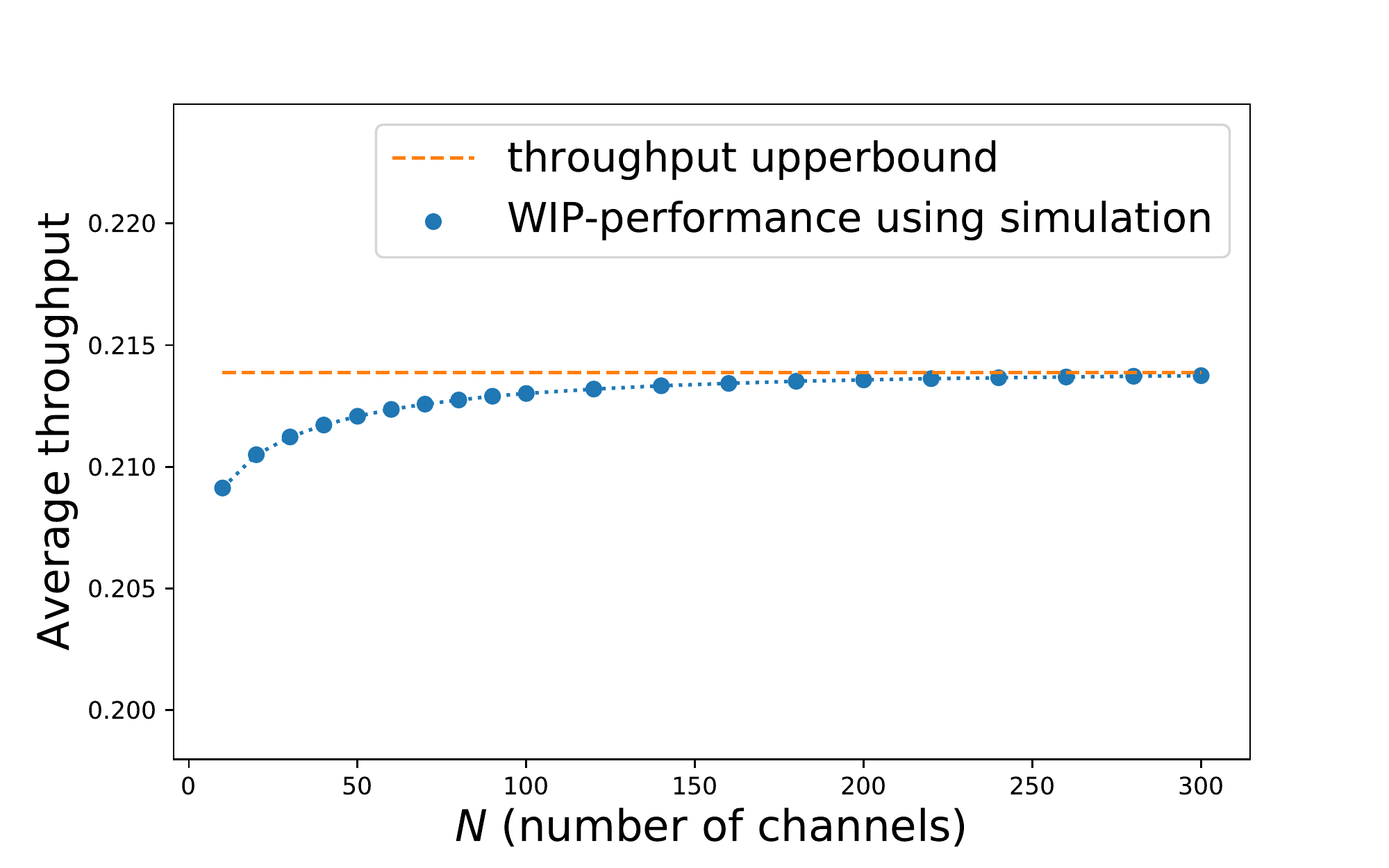}
    \caption{WIP for two-classes channel model.}
  \end{subfigure}
  \begin{subfigure}[b]{0.48\linewidth}
    \includegraphics[width=\linewidth]{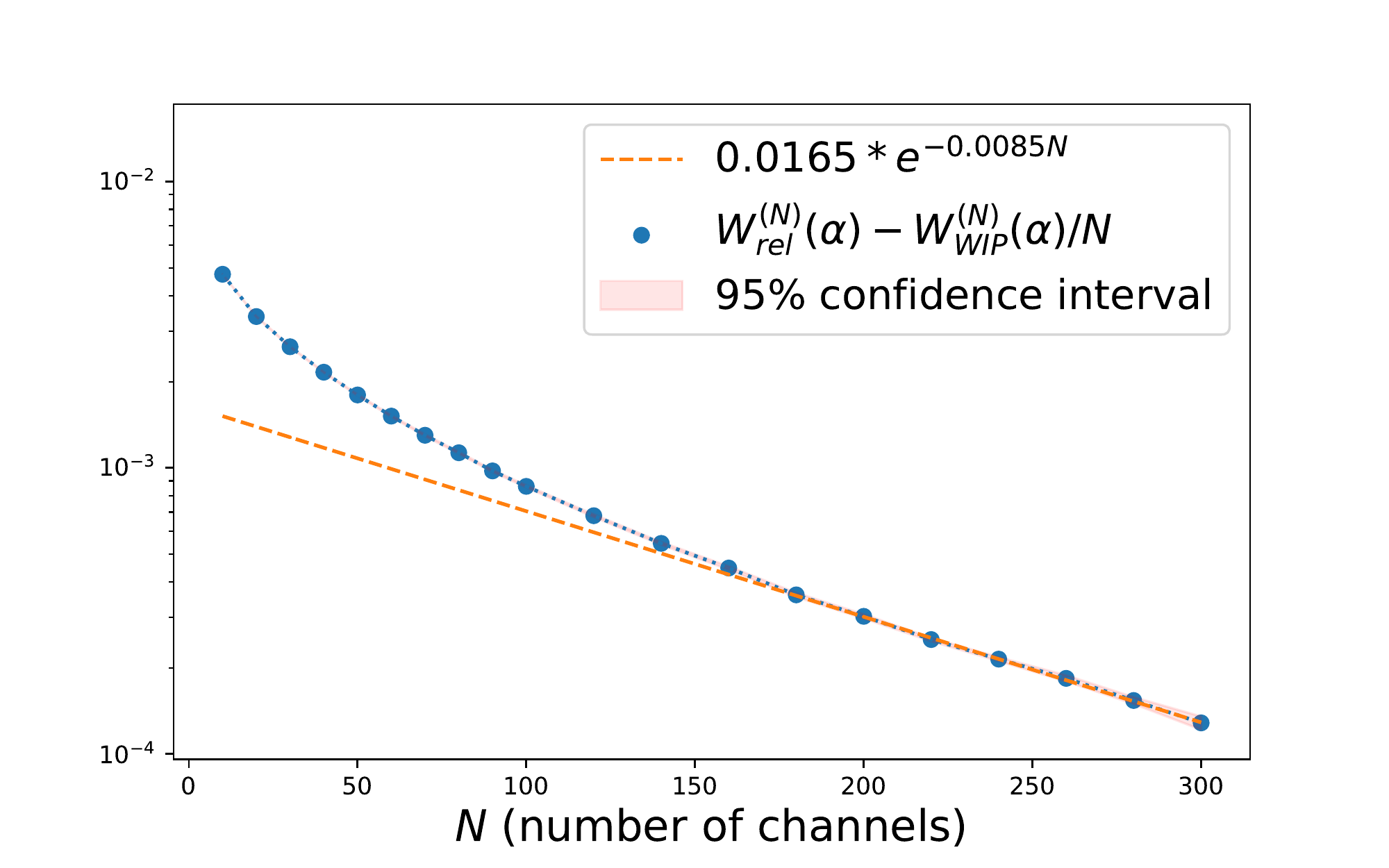}
    \caption{Verifying exponential convergence rate.}
  \end{subfigure}
  \caption{Convergence rate for two-classes channel model.}
  \label{fig5}
\end{figure}

We can now provide some numerical results. We shall choose a parameter set that is used in \cite{ouyang2012asymptotically}: $\beta = 0.6$, $\alpha = 0.3$, $(p_1,r_1) = (0.75,0.2)$, $(p_2,r_2) = (0.8,0.3)$.
It can be shown that using these parameters, a class 2 channel that has just been activated and has been observed in good state will have the highest priority, hence should always be activated. Also a class $2$ channel after $4$ time steps of being idle has higher priority than a class $1$ channel in any belief state. We can then characterize the fixed point $\mm^*$ by computing a threshold of activation of class 1 channels so that in steady-state, a proportion of $\alpha = 0.3$ of channels are activated. This gives that all class 1 channels in belief state $b^1_{0,t}$ with $t\leq 20$ will be kept idle, a fraction $0.89\dots$ of the class $1$ channels in belief state $b^1_{0,21}$ will be activated, and all class 1 channels in belief states $b^1_{0,t}$ with $t\ge22$ will be activated. As $0.89\dots\ne1$, the fixed point is \emph{not} singular.

Consequently, all conditions needed for Theorem \ref{th:expo_optimal} are satisfied for this model. We then use simulations to evaluate the average throughput, with $N$ ranging from 10 to 300. We see through Figure $\ref{fig5}$ that a similar convergence pattern as in the 3 states model occurs, and it suggests an exponential rate convergence as claimed, with a value of the constant $c \approx 0.0085$.

\section{Extension to the asynchronous model}
\label{sec:continuous}

Throughout the paper, we studied a \emph{synchronous} restless bandit problem in which all bandits synchronously make a transition. In this section, we explain how to adapt the proofs done in Section~\ref{sec:main} to the asynchronous model studied for example in \cite{WeberWeiss1990}. We start by recalling the model of \cite{WeberWeiss1990} in Section~\ref{ssec:asynchronous:model}. We show how the synchronous and asynchronous models are related in Section~\ref{ssec:asynchronous:equiv}. Finally, we state the equivalent of our main Theorem~\ref{th:expo_optimal} in Section~\ref{ssec:asynchronous:expo}.

\subsection{The asynchronous bandit model}
\label{ssec:asynchronous:model}

Similarly to Section~\ref{model}, an asynchronous restless bandit problem with parameters $\big\{ (\bQ^0, \bQ^1, \rr^0, \rr^1); \alpha, N\big\}$ is a Markov decision process defined as follows:
\begin{enumerate}
  \item As before, the model is composed of $N$ bandits each evolves in a finite state space.  The state space of the process at time $t\in\R^+$ is the vector $\bS(t)$.
  \item In continuous time, the decision maker chooses an action $\ba(t)\in\{0,1\}^N$. Decisions can be modified only when the process $\bS(t)$ changes state: At each jump of the process $\bS(t)$, the decision maker observes $\bS(t)$ and chooses a new action vector $\ba(t)$ that will be kept until the next jump of the process. The action vector must satisfy $\sum_{n=1}^N a_n(t)=\alpha N$ for all $t \in \R^+$.
  \item Bandit $n$ evolves as a continuous-time Markov chain of kernel $\bQ^{a_n(t)}$, \emph{i.e.} $\bQ^0$ and $\bQ^1$ are $d\times d$ matrices such that $Q^a_{ij}\ge0$ for $i\ne j$ and $\sum_{j}Q_{ij}^a=0$. For $i\ne j$, bandit $n$ jumps from state $i$ to state $j$ at rate $Q^{a_n(t)}_{ij}$. Given $\ba(t)$, the evolutions of the $N$ bandits are independent.
  \item The gain per unit time of the decision maker is $\sum_{n=1}^N R^{a_n(t)}_{S_n(t)}$.
\end{enumerate}
As before, the goal of the decision maker is to compute a decision rule in order to maximize the long-term average reward. Using our notation, this problem can be written as
\begin{align}
  \opt & := \sup_{\ba \in \Pi} \lim_{T \rightarrow \infty} \frac{1}{T} \mathbb{E} \Big[ \int_{t=0}^{T} \sum_{n=1}^N R^{a_n(t)}_{S_n(t)} dt \Big] \label{eq:optim_asynchro} \\
  & \mbox{subject to} \sum_{n=1}^N a_n(t) = \alpha N, \ \mbox{for all} \  t \in \mathbb{R}_{+}. \label{eq5}
\end{align}
This problem is the asynchronous (continuous-time) version of the synchronous (discrete-time) problem \eqref{eq1}--\eqref{eq2}. As before, we assume that the matrices $\bQ^0$ and $\bQ^1$ are such that bandit is unichain regardless of the policy employed.

\subsection{Whittle index, relaxation and equivalence with the synchronous model}
\label{ssec:asynchronous:equiv}

In this subsection, we recall briefly the definition of Whittle index and of the relaxation for the asynchronous bandit model. These definitions coincide with the ones of \cite{WeberWeiss1990}.

As for the synchronous case, Whittle index of asynchronous bandits is defined by considering a subsidized MDP for a single bandit $n$, in which a decision maker that takes the passive action $a_n(t)=0$ earns an extra reward $\nu$ per unit time. The definition of indexability is the same as the one in synchronous case and the index of a state $i$, denoted by $\nu_i$, is the smallest subsidy such that the passive action is optimal for state $i$.

Similarly to the synchronous problem, the definition of Whittle index in the asynchronous model can be justified by looking at the Lagrangian of the optimization problem \eqref{eq:optim_asynchro} where the constraint \eqref{eq5} is replaced by the constraint \eqref{eq5_relaxed} below. We again denote the value of this relaxed  problem as $\rel$.
\begin{align}
  \rel & := \sup_{\ba \in \Pi} \lim_{T \rightarrow \infty} \frac{1}{T} \mathbb{E} \Big[ \int_{t=0}^{T} \sum_{n=1}^N R^{a_n(t)}_{S_n(t)} dt \Big] \label{eq:relax_asynchronous} \\
  & \mbox{subject to} \lim_{T \rightarrow \infty} \int_{0}^{T} \sum_{n=1}^{N} \frac{a_n(t)}{T} dt = \alpha N.\label{eq5_relaxed}
\end{align}

As we show below, when considering bandits in isolation, using a synchronous or an asynchronous model is equivalent. In particular, neither the definition of Whittle index nor the value of the relaxation depend on the synchronization nature of a bandit.
\begin{definition}
  Let $(\bQ^0$, $\bQ^1,\rr^0,\rr^1)$ be the parameters of an asynchronous bandit. Let $\tau := \max_i\max_a|Q^a_{ii}|$. Let $(\pp^0,\pp^1,\tilde{\rr}^0,\tilde{\rr}^1)$ be the matrices defined as follows: for all states $i\ne j$ and all action $a\in\{0,1\}$:
  \begin{align}
    \label{eq:equiv_P_Q}
    P^a_{ij} := \frac{1}{\tau}Q^a_{ij}; \qquad P^a_{ii} := 1-\sum_{k\ne i}P^a_{ik}; \qquad \tilde{R}^a_{i} := \tau R^a_i.
  \end{align}
  We call $(\pp^0,\pp^1,\tilde{\rr}^0,\tilde{\rr}^1)$ the synchronous version of our asynchronous bandit model.
\end{definition}

The following lemma states the equivalence of Whittle relaxation between the synchronous and the asynchronous problems:
\begin{lemma}\label{lem:equiv}
  Let $(\bQ^0$, $\bQ^1,\rr^0,\rr^1)$ be an asynchronous bandit and let $(\pp^0,\pp^1,\tilde{\rr}^0,\tilde{\rr}^1)$ be its synchronous version \eqref{eq:equiv_P_Q}. Then:
  \begin{enumerate}[(i)]
    \item The matrices $\pp^0$ and $\pp^1$ are probability matrices.
    \item The synchronous bandit $(\pp^0,\pp^1,\tilde{\rr}^0,\tilde{\rr}^1)$ is indexable if and only if the asynchronous bandit $(\bQ^0$, $\bQ^1,\rr^0,\rr^1)$ is indexable. In such a case, the indices of both bandits coincide.
    \item The synchronous relaxed optimization problem \eqref{eq:relax_synchro} has the same value as its asynchronous counterpart \eqref{eq:relax_asynchronous}.
  \end{enumerate}
\end{lemma}
The proof of Lemma~\ref{lem:equiv} is a direct consequence of uniformization: the results rely on analysis of a bandit in isolation; when focus on a single bandit, Bellman's equation is identical for the synchronous and asynchronous version of the MDP.

\subsection{Exponential convergence in the case of asynchronous model}
\label{ssec:asynchronous:expo}

Lemma~\ref{lem:equiv} uses the fact that the Whittle relaxation is defined for bandit in isolation. Hence, considering synchronous or asynchronous bandits is equivalent. For the $N$ bandits model, however, the situation is different: in the synchronous model of Section~\ref{sec:model} all bandits change states at the same time, while in asynchronous situation, the probability that two bandits makes a jump at the exact same time is $0$. This implies that the reward of WIP for the $N$ bandits problem does depend on whether the model is synchronous or not. We denote the later by $\indcontinuous$.

It is shown in \cite{WeberWeiss1990} that the asymptotic optimality depends on the ergodic property of the solution of an ordinary differential equation (ODE) defined in Equation~(10) of \cite{WeberWeiss1990}. Using our notation, this differential equation can be written as
\begin{equation}
  \label{eq:ODE}
  \dot{\mm} = \tau(\phi(\mm) - \mm),
\end{equation}
where $\phi$ is defined as in Lemma~\ref{lem:phi} for a synchronous bandit problem $(\pp^0,\pp^1,\tilde{\rr}^0,\tilde{\rr}^1)$. By Lemma~\ref{lem:phi}, this equation has a unique fixed point, $\mm^*$. It is shown in \cite{WeberWeiss1990} that if all the solutions of the differential equation \eqref{eq:ODE} converge to $\mm^*$, then $\lim_{N\to\infty}\indcontinuous/N\to \rel[1]$. In the next theorem, we show that we can extend the result of Theorem~\ref{th:expo_optimal} to the asynchronous model.

\begin{theorem}[Exponential convergence rate theorem-asynchronous case]
\label{th:expo_optimal_async}
  Consider an asynchronous restless bandit problem $\{ (\bQ^0$, $\bQ^1,\rr^0,\rr^1); \alpha \} $ such that:
  \begin{enumerate}[(i)]
    \item Bandits are unichain and indexable.
    \item The unique fixed point $\mm^*$ of the ODE $\dot{\mm} = \tau(\phi(\mm)-\mm)$ is a uniform global attractor of the trajectories of the ODE.
    \item $\mm^*$ is not singular.
  \end{enumerate}
  Then there exists two constants $b,c >0$ that depend only on $\bQ^0$, $\bQ^1,\rr^0,\rr^1$ and $\alpha$, such that for any $N$ with $\alpha N$ being an integer,
  \begin{equation*}
    0\le \rel - \indcontinuous \le b \cdot e^{-c N}.
  \end{equation*}
\end{theorem}

\begin{proof}[Sketch of proof]
  The proof of this result follows the same structure as the proof of Theorem~\ref{th:expo_optimal} but needs substantial adaptation, the full details are given in Appendix \ref{apx:proof_async}. The main ingredients are:
  \begin{itemize}
    \item We first use a result from \cite{darling2008} to obtain an analogue of Hoeffding's inequality. This proves that the behavior of the $N$ bandits model is close to the dynamic of the ODE \eqref{eq:ODE}.
    \item Using this and the fact that $\mm^*$ is non-singular, we show that the stochastic system lies with high probability in a neighborhood $\n$ of $\mm^*$ where $\phi$ is affine. We again use Stein's method to obtain the exponential convergence result (but this time applied to a continuous-time process).
  \end{itemize}
\end{proof}

This theorem is a refinement of the original asymptotic optimality result of \cite[Theorem~2]{WeberWeiss1990}, as it provides a bound on the rate of convergence for the performance of WIP to the optimal one. The applicability conditions are essentially similar: \cite[Theorem~2]{WeberWeiss1990} also needs the assumption that $\mm^*$ is an attractor of the ODE. We add in addition that the attractivity is uniform in the initial points and that $\mm^*$ is not singular. Those conditions are also similar to the conditions of Theorem~\ref{th:expo_optimal}. It would be tempting to say that an asynchronous bandit satisfies the assumptions of Theorem \ref{th:expo_optimal_async} if and only if its synchronous version satisfies the assumptions of Theorem~\ref{th:expo_optimal}. Yet, this is true for the indexability and the non-singularity of the fixed point, but does not hold in general, since the behavior of the discrete-time dynamical system $\mm(t+1)=\phi(\mm(t))$ can be quite different from its continuous-time counterpart $\dot{\mm}= \tau \big( \phi(\mm)-\mm \big)$.  Consider for instance the synchronous model of Section \ref{sec:numerical_cycles}: the discrete-time dynamical system has an attracting cycle of period $2$; in the asynchronous version of this model, $\mm^*$ is a uniform global attractor. \footnote{From \eqref{eq:ODE} we infer that, if $\lambda_{\mathrm{sync.}}$ is an eigenvalue of an affine factor $\kk$ and $\lambda_{\mathrm{async.}}$ is its asynchronous counterpart, then $\tau \cdot (\lambda_{\mathrm{sync.}}-1) = \lambda_{\mathrm{async.}}$. So $\lambda_{\mathrm{sync.}} < -1$ for the examples in Section \ref{sec:numerical_cycles} implies $\Re(\lambda_{\mathrm{async.}})<0$ for the asynchronous models, and $\lambda_{\mathrm{async.}}$ is a stable eigenvalue (for the continuous-time ODE).} This indicates that there are bandit models for which WIP is asymptotically optimal under an asynchronous scaling but is not for the synchronous scaling.

\section{Conclusion and future work}
\label{sec:conclusion}

In this paper, we studied the performance of Whittle index policy (WIP) when there is a large number of bandits. We showed that, when WIP becomes asymptotically optimal, it does so at exponential rate (unless the fixed point is singular, which occurs with probability $0$). This explains why WIP is very efficient in practice, even when the number of bandits remains moderate. Our results hold for the classical model of \cite{WeberWeiss1990} where bandits evolve asynchronously as well as for a synchronous model in which all bandits make their transitions simultaneously. Yet, while Whittle indices are the same for both models, we provided examples for which WIP is asymptotically optimal for the asynchronous model but is not for the synchronous model.

As for future research, we plan on investigating more closely the singular situations, as well as extending the exponential convergence rate result to those generalizations of Whittle index as in \cite{oatao22728,hodge2015,Ve2016.6}.

\bibliographystyle{imsart-number}
\bibliography{reference}

\appendix

\section{Proof of Theorem~\ref{th:rel}}
\label{apx:proof_rel}

\begin{proof}

Let $\mm^*$ be the fixed point of $\phi$. As $\pp^0$, $\pp^1$ are rational, each coordinate of $\mm^*$ is a rational number. Note that the proof could be adapted to non rational $\pp^0,\pp^1$ by using a continuity argument.

Let $\{ N_k \}_{k \geq 0}$ be a sequence of increasing integers that goes to $\infty$, such that for all $k \geq 0$ and all $1 \leq i \leq d$, $m^*_i N_k$ and $\alpha N_k$ are integers. We then fix an $N$ from this sequence $\{ N_k \}_{k \geq 0}$. Recall that $m_i N$ is the number of bandits in state $i$ in configuration $\mm$ and that $S_n(t)$ is the state of bandit $n$ at time $t$. We use $\sss (t)$ to denote the state vector of the $N$ bandits system at time $t$. Let $\sss^*$ be a state vector corresponds to configuration $\mm^*$ with $N$ bandits. This is possible as $m^*_i N$ is an integer for all $i\in\{1\dots d\}$.

Note that in configuration $\mm^*$ (\emph{i.e.} state vector $\sss^*$), an optimal action $\ba^*$ under the relaxed constraint \eqref{eq3} will activate \emph{exactly} $\alpha N$ bandits. As $\ba^*$ is sub-optimal compared to an optimal policy for the original $N$ bandit problem \eqref{eq1}-\eqref{eq2}, we have
\begin{align*}
  \opt + V(\sss^*) & = \max_{\ba \in \{ 0,1 \}^N} \bigg \{ \sum_{n=1}^{N} R^{a_n}_{S^*_n} + \mathbb{E}_{\ba} \Big[ V\big(\sss(1)\big) \ \big| \ \sss(0) = \sss^* \Big] \bigg \}  \\
  & \ge \sum_{n=1}^{N} R^{a^*_n}_{S^*_n} + \mathbb{E}_{\ba^*} \Big[ V\big(\sss(1)\big) \ \big| \ \sss(0) = \sss^* \Big]\\
  & = N \rel[1] + \mathbb{E}_{\ba^*}\big[V(\sss(1))\mid \sss(0)=\sss^* \big],
\end{align*}
where in the above equation the function $V:\sss\to\R$ is the bias of the MDP. The first line corresponds to Bellman's equation (see \cite[Equation~(8.4.2) in Chapter~8]{Puterman:1994:MDP:528623}), the second line is because $\ba^*$ is a valid action for the $N$-bandits MDP but might not be the optimal action, and the last line is because $\sum_{n=1}^{N} R^{a^*_n}_{S^*_n}=\rel=N \rel[1]$.

We hence obtain
\begin{equation*}
  \rel[1] \geq \frac{\opt}{N} \geq \rel[1] + \frac{\mathbb{E}_{\ba^*}V\big(\sss(1)\big) - V(\sss^*)}{N}.
\end{equation*}
In the following, we bound $\mathbb{E}_{\ba^*}[V\big(\sss(1)\big) - V(\sss^*)]$. This will be achieved in two steps.

\paragraph*{Step One} We define for two state vectors $\by$, $\bz$ the distance
\begin{equation*}
  \delta (\by,\bz) :=  \sum_{n=1}^{N} \mathds{1}_{ \big\{ y_n \neq z_n \big\} },
\end{equation*}
which counts the number (among the $N$ bandits) of bandits that are in different states between those two vectors. Such distance satisfies the property that for all $\by$ and $\bz$ such that $\delta (\by,\bz) = k$, we can find a sequence of state vectors $\bz_1, \bz_2, ..., \bz_{k-1}$  that verify $\delta(\by,\bz_1) = \delta(\bz_1, \bz_2)= ... = \delta(\bz_{k-2},\bz_{k-1}) = \delta(\bz_{k-1}, \bz) = 1$. In what follows, we show that there exists $C>0$ independent of $N$ such that for all state vectors $\by$ and $\bz$,
$$
| V(\by) - V(\bz) | \leq C \cdot \delta(\by,\bz).
$$
In view of the above property of $\delta$, we only need to prove this for $\delta(\by,\bz)=1$, \emph{i.e.}
$$
| V(\by) - V(\bz) | \leq C.
$$

Let $\by$, $\bz$ be two state vectors such that $\delta (\by,\bz)=1$, and assume without loss of generality that it is bandit $1$ that are in different states: $y_1\ne z_1$ and $y_n=z_n$ for $n\in\{2\dots N\}$. We use a coupling argument as follows: We consider two trajectories of the $N$ bandits system, $\bY$ and $\bZ$, that start respectively in state vectors $\bY(0)=\by$ and $\bZ(0)=\bz$. Let $\pi^*$ be the optimal policy of the $N$ bandits MDP, and suppose that we apply $\pi^*$ to the trajectory $\bZ$. At time $t$, the action vector will be $ \pi^*(\bZ(t))$. We couple the trajectories $\bY$ and $\bZ$ by applying the same action vectors $\pi^*(\bZ(t))$ for $\bY$ and keeping $Y_n(t)=Z_n(t)$ for bandits $n\in\{2\dots N\}$.
The $\bZ$ trajectory follows an optimal trajectory, hence Bellman's equation is satisfied: for any $T>0$, we have:
\begin{equation}\label{eq:optimality1}
T \cdot \opt + V(\bz) = \sum_{n=1}^{N} R^{\pi^*_n (\bz)}_{z_n} + \mathbb{E}_{\pi^*} \bigg[ \sum_{n=1}^{N} \sum_{t=1}^{T-1} R^{\pi^*_n (\bZ(t))}_{Z_n(t)} + V\big( \bZ(T) \big) \ \Big| \ \bZ(0) = \bz \bigg].
\end{equation}
Since $\bY$ follows a possibly sub-optimal trajectory, we have:
\begin{equation}\label{eq:optimality2}
T \cdot \opt + V(\by) \geq \sum_{n=1}^{N} R^{\pi^*_n (\by)}_{y_n} + \mathbb{E}_{\pi^*} \bigg[ \sum_{n=1}^{N} \sum_{t=1}^{T-1} R^{\pi^*_n (\bZ(t))}_{Y_n(t)} + V\big( \bY(T) \big) \ \Big| \ \bY(0) = \by \bigg],
\end{equation}
Recall that the matrices $\pp^0,\pp^1$ are such that a bandit is unichain and aperiodic. This shows that the mixing time of a single bandit is bounded (independently of $N$): for any policy $\pi \in \Pi$
$$
\max_{i,j} \argmin_{t} \bigg\{ \mathbb{P}_{\pi} \Big[ Y_1(t) = Z_1(t) \ \Big| \ Y_1(0) = i, Z_1 (0) = j \Big] > 0 \bigg\} < \infty.
$$
Because of the coupling, for $0 \leq t \leq T$ and $\ 1 \leq n \leq N$, $Y_n(t) \neq Z_n(t)$ is only possible for $n=1$. Furthermore, as the mixing time of a bandit is bounded, for $T$ large enough, there is a positive probability, say at least $p > 0$, that $Y_1 (T) = Z_1 (T)$. Hence with probability smaller than $1-p$ we have $\delta \big(\by(T), \bz(T)\big) = 1$, conditional on $\bY(0)= \by$ and $\bZ(0)= \bz$.

Let $r := 2 \max_{1 \leq i \leq d, a\in \{0,1\}} | R^a_i |$. Subtracting \eqref{eq:optimality1} in \eqref{eq:optimality2} gives
\begin{align*}
  |V(\by) - V(\bz)| & \le T \cdot r + \Big|\mathbb{E}_{\pi^*} \Big[ V\big(\bY(T)\big)-V\big(\bZ(T)\big) \ \Big| \ \bY(0) = \by, \bZ(0)=\bz \Big] \Big| \\
    & \le T \cdot r + (1-p)\max_{\bu,\bv: \ \delta(\bu,\bv)=1} \big\{ |V(\bu)-V(\bv)| \big\}.
\end{align*}
This being true for all $\by, \bz$ with $\delta(\by,\bz)=1$, it implies that $\max_{\bu,\bv: \ \delta(\bu,\bv)=1} \big\{ |V(\bu)-V(\bv)| \big\} \le T \cdot r / p$, and we can take the constant $C:=T \cdot r/p$.

\paragraph*{Step Two} We now prove that
\begin{equation*}
  \mathbb{E}_{\ba^*}[\delta (\sss^*,\sss(1)) \mid \sss(0)=\sss^*] \leq D \sqrt{N},
\end{equation*}
with a constant $D$ independent of $N$, where $ \sss(1) $ is the random vector conditional on $\sss(0) = \sss^*$ under action vector $\ba^*$.

Indeed, let $\xx^* := \mm^* N$, and denote $\xxx := \mm(1) N$ to be the random $d$-vector, with $\mm(1)$ the random configuration corresponds to $\sss(1)$. For each $1 \leq i \leq d$, we may write
$$
X_i = (B_{i,1}^0+B_{i,1}^1)+ (B_{i,2}^0+B_{i,2}^1) + ... + (B_{i,d}^0 + B_{i,d}^1)
$$
where $B_{i,j}^a \sim Binomial (x^*_{j,a}, P^{a}_{ji})$ for $1 \leq j \leq d$, $a \in \{ 0,1 \}$; and $x^*_{j,0} + x^*_{j,1} = x^*_j$, with $x^*_{j,a}$ representing the number of bandits in state $j$ taking action $a$, when optimal action vector $\ba^*$ is applied to state vector $\sss^*$.

By stationarity, we have
$$
\mathbb{E}_{\ba^*} (X_i) = \sum_{j=1}^d \sum_{a=0,1} x^*_{j,a} \cdot P^{a}_{ji} = x^*_i,
$$
and
$$
\mbox{Var} (X_i) = \sum_{j=1}^d \sum_{a=0,1}  x^*_{j,a} \cdot P^{a}_{ji} (1-P^{a}_{ji}) = \mathcal{O}(N).
$$
Consequently, we can bound
$$
  \mathbb{E}_{\ba^*}[\delta (\sss^*,\sss(1))] \leq \sum_{i=1}^d \mathbb{E}_{\ba^*} \big| x^*_i - X_i \big| \leq D \sqrt{N},
$$
with a constant $D$ independent of $N$.

\

To summarize, we have
$$
\mathbb{E}_{\ba^*} \big[ |V(\sss(1)) - V(\sss^*)| \big]  \le \mathbb{E}_{\ba^*} \big[ C \cdot \delta \big(\sss(1), \sss^*\big) \big] \leq CD \cdot \sqrt{N},
$$
hence
\begin{equation}\label{eq:opt_rate}
\rel[1] \geq \frac{\opt}{N} = \rel[1] + \frac{\mathbb{E}_{\ba^*}V(\sss(1)) - V(\sss^*)}{N} \geq \rel[1] - \frac{CD}{\sqrt{N}},
\end{equation}
which implies that $\opt / N \rightarrow \rel[1]$ when $N$ goes to $+\infty$. Moreover, from \eqref{eq:opt_rate}, the convergence rate is at least as fast as $\calO(1/ \sqrt{N})$.
\end{proof}

\section{Proof of Lemma \ref{lem:phi}}
\label{apx:phi}

In this appendix we prove Lemma \ref{lem:phi}. We first show the piecewise affine property in Lemma \ref{lemma1}, which gives (i) and (ii). We then show the uniqueness of fixed point from a bijective property in Lemma \ref{lemma2}, from which we conclude (iii).

\begin{lemma}[Piecewise affine] \label{lemma1}
   $\phi$ is a piecewise affine continuous function, with $d$ affine pieces.
 \end{lemma}

 \begin{proof}
  Let $\mm\in\Delta^d$ be a configuration and recall $s(\mm)\in\{1\dots d\}$ is the state such that $\sum_{i=1}^{s(\mm)-1}m_i\le \alpha < \sum_{i=1}^{s(\mm)}m_i$. When the system is in configuration $\mm$ at time $t$, WIP will activate all bandits that are in states $1$ to $s(\mm)-1$ and not activate any bandit in states $s(\mm)+1$ to $d$. Among the $Nm_{s(\mm)}$ bandits in state $s(\mm)$, $N(\alpha-\sum_{i=1}^{s(\mm)-1}m_i)$  of them will be activated and the rest will not be activated.

  This implies that the expected number of bandits in state $j$ at time $t+1$ will be equal to
  \begin{align}
    \label{eq:phi_apx}
    \sum_{i=1}^{s(\mm)-1} N m_i P^1_{ij} + N(\alpha-\sum_{i=1}^{s(\mm)-1}m_i)P^1_{s(\mm)j} + N(\sum_{i=1}^{s(\mm)}m_i - \alpha)P^0_{s(\mm)j} + \sum_{i=s(\mm)+1}^{d} N m_i P^0_{ij}.
  \end{align}
  It justifies the expression \eqref{eq:phi}. Note that \eqref{eq:phi} can be reorganized to
  \begin{align*}
    \phi_j(\mm) = \sum_{i=1}^{s(\mm)-1} m_i (P^1_{ij}-P^1_{s(\mm)j}+P^0_{s(\mm)j}) + \sum_{i=s(\mm)}^{d} m_i P^0_{ij} + \alpha(P^1_{s(\mm)j}-P^0_{s(\mm)j}).
  \end{align*}
  Consequently
  $\phi(\mm)  = \mm \cdot \kk_{s(\mm)} + \mathbf{b}_{s(\mm)}$, where

  $\mathbf{b}_{s(\mm)} = \alpha (\pp^1_{s(\mm)} - \pp^{0}_{s(\mm)})$, and $\kk_{s(\mm)} = $
   $
   \begin{pmatrix}
           \pp^1_1 - \pp^1_{s(\mm)} + \pp_{s(\mm)}^0 \\
           \pp^1_2 - \pp^1_{s(\mm)} + \pp_{s(\mm)}^0 \\
           ... \\
           \pp^1_{s(\mm)-1} - \pp^1_{s(\mm)} + \pp^0_{s(\mm)} \\
           \pp^0_{s(\mm)} \\
           \pp^0_{s(\mm)+1} \\
           ... \\
           \pp^0_{d}
   \end{pmatrix}
   $.

   Let $\calZ_i := \{\mm\in\Delta^d \mid s(\mm)=i\}$. The above expression of $\phi$ implies that this map is affine on each zone $\calZ_i$. There are $d$ such zones with $1 \leq i \leq d$. It is clear from the expression that $\phi(\mm)$ is continuous on $\mm$.

 \end{proof}

\begin{lemma}[Bijectivity] \label{lemma2}
  Let $\pi(s,\theta) \in \Pi$ be the policy that activates all bandits in states $1,\dots,s-1$, does not activate bandits in states $s+1, s+2, \dots, d$, and that activates bandits in state $s$ with probability $\theta$. Denote by $\newalpha (s,\theta)$ the proportion of time that the active action is taken using policy $\pi (s,\theta)$. Then, the function $(s,\theta)\mapsto\newalpha(s,\theta)$ is a bijective map from $\{1 \dots d \} \times [0,1)$ to $[ 0,1 )$.
\end{lemma}

\begin{proof}
  The following proof is partially adapted from the proof of \cite[Lemma~1]{WeberWeiss1990}. For a given $\nu \in \mathbb{R}$, denote by $\gamma(\nu)$ the value of the subsidy-$\nu$ problem, \emph{i.e.}
\begin{equation}\label{4}
    \gamma(\nu) := \sup_{\pi \in \Pi} \lim_{T \rightarrow \infty} \frac{1}{T} \mathbb{E} \Big[ \sum_{t=0}^{T-1} \Big( R^{\pi(S(t))}_{S(t)}+\nu \big(1-\pi(S(t))\big) \Big) \Big]. \\
\end{equation}
We defined similarly $\gamma_{\pi} (\nu)$ as the value under policy $\pi$ for a such subsidy-$\nu$ problem. Note that for fixed $\pi$, the function $\gamma_{\pi} (\nu)$ is affine and increasing in $\nu$.

By definition of indexability, $\gamma(\nu) = \max_{\pi \in \Pi} \gamma_{\pi} (\nu)$ is a piecewise affine, continuous and convex function of $\nu$: it is affine on $(-\infty;\nu_d]$, on $[\nu_1;+\infty)$ and on all $[\nu_s;\nu_{s-1}]$ for $s\in\{2\dots d\}$.

Moreover, for $s\in\{2\dots d-1\}$ and $\nu\in[\nu_s;\nu_{s-1}]$, the optimal policy of \eqref{4} is to activate all bandits up to state $s-1$. Hence,
\begin{align*}
    \gamma(\nu) &= \gamma_{\pi(s,0)} (\nu) = \gamma(\nu_{s-1}) + \big(1-\newalpha(s,0)\big) \cdot (\nu - \nu_{s-1}).
\end{align*}
Similarly, and as $\newalpha(s+1,0)=\newalpha(s,1)$, for $\nu\in[\nu_{s+1};\nu_{s}]$ we have:
\begin{align*}
    \gamma(\nu) &= \gamma(\nu_{s}) + \big(1-\newalpha(s+1,0)\big) \cdot (\nu - \nu_{s}) \\
    &= \gamma(\nu_{s}) + \big(1-\newalpha(s,1)\big)\cdot (\nu - \nu_{s}).
  \end{align*}
  Consequently
  \begin{equation*}
   \frac{\partial \gamma}{\partial \nu} (\nu) = \begin{cases}
          1 - \newalpha(s,0), & \mbox{if } \nu_s < \nu < \nu_{s-1} \\
          1 - \newalpha(s,1), & \mbox{if } \nu_{s+1} < \nu < \nu_{s}.
          \end{cases}
  \end{equation*}
The convexity of $\gamma(\nu)$ implies that $1-\newalpha(s,0) > 1-\newalpha(s,1)$, hence $\newalpha(s,1) > \newalpha(s,0)$.

Now suppose that $\mm^0$ and $\mm^1$ are the equilibrium distributions of policies $\pi(s,0)$ and $\pi(s,1)$. Let $0 < \theta < 1$. The equilibrium distribution $\mm^{\theta}$ induced by $\pi(s,\theta)$ is then a linear combination of $\mm^0$ and $\mm^1$, namely $\mm^{\theta} = p\cdot \mm^0 + (1-p)\cdot \mm^1$, with
  $$
  p = \frac{(1- \theta)m_s^{1}}{\theta m_s^0 + (1-\theta)m_s^1}.
  $$
  Hence
  \begin{align*}
    m_s^{\theta} & = p m^0_s + (1-p)m^1_s \\
     & = \frac{m^1_s m^0_s}{\theta m^0_s + (1-\theta)m^1_s},
  \end{align*}
  and
  \begin{align*}
    \newalpha (s,\theta)  & = \bigg( \sum_{k=1}^{s-1}m_k^{\theta} \bigg) + \theta m_s^{\theta} \\
     & = \ \sum_{k=1}^{s-1} \big( (1-p) m^1_k + p m^0_k \big)  + \frac{\theta \cdot m^1_s m^0_s}{\theta m^0_s + (1-\theta)m^1_s} \\
     & = \ \frac{\sum_{k=1}^{s-1} \big( \theta \cdot m_s^0 m_k^1 + (1-\theta) m_s^1 m_k^0 \big) \ + \theta \cdot m^1_s m^0_s}{\theta m^0_s + (1-\theta)m^1_s}.
  \end{align*}
  Observe that $\newalpha(s,\theta)$ is the ratio of two affine functions of $\theta$, hence is monotone as $\theta$ ranges from 0 to 1; but as $\newalpha(s,1) > \newalpha(s,0)$, it is monotonically \emph{increasing}. We hence obtain
  $$
  1 = \newalpha(d,1) > \newalpha(d,0) = \newalpha(d-1,1) > \dots > \newalpha(2,0) = \newalpha(1,1) > \newalpha(1,0) = 0,
  $$
  which concludes the proof.
\end{proof}

We are now ready to finish the proof of Lemma~\ref{lem:phi}(iii). Let $\mm$ be a fixed point of $\phi$. Under configuration $\mm$, all bandits that are in states from $1$ to $s(\mm)-1$ are activated, and a fraction $\theta(\mm)=(\alpha-\sum_{i=1}^{s(\mm)-1} m_i)/m_{s(\mm)}$ of the bandits that are in state $s(\mm)$ are activated. This shows that $\mm$ also corresponds to the stationary distribution of the policy $\pi(s(\mm),\theta(\mm))$. The proportion of activated bandits of this policy is $\newalpha(s(\mm),\theta(\mm))=\alpha$. Consequently, if $\mm'$ is another fixed point of $\phi$, then $\mm'$ would have to be the stationary distribution of some other policy of the form $\pi(s',\theta')$, with $\newalpha(s',\theta') = \alpha$. As the function $(s,\theta)\mapsto\newalpha(s,\theta)$ is a bijection, this implies that $s' = s(\mm)$ and $\theta'=\theta(\mm)$. Hence the fixed point of $\phi$ is unique.

\section{Proof of Theorem~\ref{th:expo_optimal}}
\label{apx:proof_sync}

In this appendix, we explain technical details of the proof of our main result Theorem \ref{th:expo_optimal}.
In the following, we denote by $\mathcal{B}(\mm^*, r)$ the ball centered at $\mm^*$ with radius $r$.

\begin{theorem}
  \label{theorem2}
  Under the same assumptions as in Theorem \ref{th:expo_optimal}, and assume that $\mathbf{M}^{(N)}(0)$ is already in stationary regime. Then there exists two constants $b,c >0$ such that
  \begin{enumerate}[(i)]
    \item $\| \mathbb{E}[\mathbf{M}^{(N)} (0)] - \mm^* \| \leq b \cdot e^{-cN}$;
    \item $\proba{\bMN(0)\not\in\zzm} \le b \cdot e^{-cN}$.
  \end{enumerate}
\end{theorem}

Let us first explain how Theorem \ref{theorem2} implies Theorem~\ref{th:expo_optimal}. We prove below that:
\begin{lemma}
  \label{lem:rho(m)}
  Assume that bandits are indexable and unichain, and let $\rho(\mm)$ be the instantaneous bandit-averaged reward of WIP when the system is in configuration $\mm$. Then:
  \begin{enumerate}[(i)]
    \item $\rho$ is piecewise affine on each of the zone $\zz_i$ and for all $\mm\in\Delta^d$:
    \begin{align}
      \label{eq:rho}
      \rho(\mm)=\sum_{i=1}^{s(\mm)-1} m_i R^1_{i} + (\alpha-\sum_{i=1}^{s(\mm)-1}m_i)R^1_{s(\mm)} + (\sum_{i=1}^{s(\mm)}m_i - \alpha)R^0_{s(\mm)} + \sum_{i=s(\mm)+1}^{d} m_i R^0_{i}.
    \end{align}
    \item $\rho(\mm^*)=\rel[1]$.
  \end{enumerate}
\end{lemma}
By definition, the performance of WIP is $\ind= N \cdot \expect{\rho(\bMN(0))}$. Hence from Lemma~\ref{lem:rho(m)} we have
\begin{align*}
  \rel - \ind &= N \cdot \rel[1] - N \cdot \expect{\rho(\bMN(0))}\\
   &= N \cdot \e \bigg[ \big( \rho(\mm^*) - \rho(\bMN(0)) \big) \mathds{1}_{\{ \bMN(0)\in\zzm \} } \\
   &\qquad +\big( \rho(\mm^*) - \rho(\bMN(0)) \big) \mathds{1}_{\{ \bMN(0)\not\in\zzm\} } \bigg]
\end{align*}

By linearity of $\rho$ and Theorem~\ref{theorem2}(i), the first term inside the above expectation is exponentially small; by Theorem~\ref{theorem2}(ii) and since the rewards are bounded, the second term is also exponentially small.

Before proving Theorem~\ref{theorem2}, we start by proving a few technical lemmas.

\subsection{Relation between $\mm^*$ and $\rel[1]$ (Proof of Lemma~\ref{lem:rho(m)})}
\begin{proof}
Let $\mm\in\Delta^d$ be a configuration and recall $s(\mm)\in\{1\dots d\}$ is the state such that $\sum_{i=1}^{s(\mm)-1}m_i\le \alpha < \sum_{i=1}^{s(\mm)}m_i$. Similarly to our analysis of Lemma~\ref{lemma1}, when the system is in configuration $\mm$, WIP will activate all bandits that are in states $1$ to $s(\mm)-1$. This will lead an instantaneous reward of $\sum_{i=1}^{s(\mm)-1}Nm_iR^1_i$. WIP will not activate bandits that are in states $s(\mm)+1$ to $d$. This will lead an instantaneous reward of $\sum_{i=s(\mm)+1}^{d}Nm_iR^0_i$.  Among the $Nm_{s(\mm)}$ bandits in state $s(\mm)$, $N(\alpha-\sum_{i=1}^{s(\mm)-1}m_i)$  of them will be activated and the rest will not be activated. This shows that $\rho(\mm)$ is given by \eqref{eq:rho}.

For (ii), recall that $\mm^*$ is the unique fixed point, and consider a subsidy-$\nu_{s(\mm^*)}$ MDP, where $\nu_{s(\mm^*)}$ is the Whittle index of state $s(\mm^*)$. Denote by $L$ the value of this MDP:
\begin{align}
  L & := \max_{\ba \in \Pi} \lim_{T\to\infty}\frac1T \sum_{t=0}^{T-1} \expect{R^{a_n(t)}_{S_n(t)} + (\alpha-a_n(t))\nu_{s(\mm^*)}}\nonumber\\
  &= \max_{\ba \in \Pi} \lim_{T\to\infty}\frac1T \sum_{t=0}^{T-1} \expect{R^{a_n(t)}_{S_n(t)}} + \left(\alpha-\lim_{T\to\infty}\frac1T \sum_{t=0}^{T-1}\expect{a_n(t)}\right)\nu_{s(\mm^*)}.
  \label{eq:apx_relaxed}
\end{align}
By definition of Whittle index, any policy of the form $\pi(s(\mm^*),\theta)$ defined in Lemma~\ref{lemma2} is optimal for \eqref{eq:apx_relaxed}. Moreover, if $\theta^*$ is such that $\newalpha(s(\mm^*),\theta^*)=\alpha$, then such a policy satisfies the constraint \eqref{eq3}: $\lim_{T\to\infty}\frac1T \sum_{t=0}^{T-1}\expect{a_n(t)}= \alpha$. This shows that $L=\rel[1]$ and as all bandits are identical, we have $N \cdot \rel[1]=\rel$, and $\pi(s(\mm^*),\theta^*)$ is an optimal policy for the relaxed constraint \eqref{eq3}.

It remains to show that the reward of policy $\pi(s(\mm^*),\theta^*)$ is $\rho(\mm^*)$. This comes from the fact that the steady-state of the Markov chain induced by this policy is $\mm^*$, and $\pi (s(\mm^*), \theta^*)$ is such that $\alpha N$ bandits are activated on average. Indeed, the bandit-averaged reward of this policy is:
\begin{align}
  \label{eq:L}
  L = \sum_{i=1}^{s(\mm^*)-1} m^*_i R^1_i + \theta^* m^*_{s(\mm^*)}R^1_{s(\mm^*)} + (1-\theta^*) m^*_{s(\mm^*)}R^0_{s(\mm^*)} + \sum_{i=s(\mm^*)+1}^d m^*_i R^0_i
\end{align}
As the proportion of activated bandits is $\alpha$, we have $\sum_{i=1}^{s(\mm^*)-1} m^*_i + \theta^* m^*_{s(\mm^*)}=\alpha$. Hence \eqref{eq:L} coincides with the expression of $\rho(\mm^*)$ in \eqref{eq:rho}, and $\rho(\mm^*)= L = \rel[1]$. This concludes the proof of Lemma \ref{lem:rho(m)}.

\end{proof}

\subsection{Hoeffding's inequality (for one transition)}

\begin{lemma}[Hoeffding's inequality]
  \label{lem:Hoeffding}
  For all $t \in \mathbb{N}$, we have
  $$
  \mathbf{M}^{(N)}(t+1) = \phi \big( \mathbf{M}^{(N)}(t) \big) + \mathbf{E}^{(N)}(t+1)
  $$
  where the random vector $\mathbf{E}^{(N)} (t+1)$ is such that
  $$
  \mathbb{E} [ \mathbf{E}^{(N)} (t+1) \big| \mathbf{M}^{(N)}(t) ] = \mathbf{0},
  $$
  and for all $\delta>0$:
  $$
    \proba{\| \mathbf{E}^{(N)}(t+1) \| \geq \delta} \leq e^{-2N \delta^2}.
  $$
\end{lemma}

\begin{proof}
  Since the $N$ bandits evolve independently, we may apply the following form of Hoeffding's inequality:
  Let $X_1$, $X_2$, ..., $X_N$ be $N$ independent random variables bounded by the interval $[0,1]$, and define the empirical mean of these variables by $\overline{X} := \frac{1}{N} (X_1 + X_2 +...+ X_N)$, then
  $$
  \proba{ \overline{X} - \e [\overline{X}] \geq \delta } \leq e^{-2N\delta^2}.
  $$

  More precisely, for a fixed $1 \leq j \leq d$, we have
  $$
   \MN_j(t+1) = \frac{1}{N} \sum_{i=1}^{d} \sum_{k=1}^{N \cdot \MN_i(t) } \mathds{1}_{\{ U_{i,k} \leq P_{ij}(\mathbf{M}^{(N)}(t)) \}}
  $$
  where for $1 \leq i \leq d$, $\ 1 \leq k \leq N \cdot \MN_i(t)$, the $U_{i,k}$'s are in total $N$ independent and identically distributed uniform $(0,1)$ random variables, and $P_{ij}(\mm)$ is the probability for a bandit in state $i$ goes to state $j$ under WIP, when the $N$ bandits system is in configuration $\mm$.

  By definition, we have
  $$
   \phi_j (\bMN(t))  = \sum_{i=1}^{d} \MN_i(t) \cdot P_{ij}(\mathbf{M}^{(N)} (t)).
  $$
  Hence
  \begin{equation*}
    \e \big[ \MN_j(t+1) \big| \mathbf{M}^{(N)}(t) \big] = \sum_{i=1}^{d} \frac{1}{N} \cdot N \cdot \MN_i(t) \cdot P_{ij} (\mathbf{M}^{(N)} (t)) = \phi_j (\bMN(t)),
  \end{equation*}
  and
  \begin{align*}
      \proba{ \| \mathbf{M}^{(N)}(t+1) - \phi (\mathbf{M}^{(N)} (t)) \| \geq \delta }
    &= \proba{ \max_{1 \leq j \leq d} \big| \MN_j(t+1) - \phi_j (\bMN(t)) \big| \geq \delta } \\
    &\leq e^{-2N\delta^2} \ \ \ \ (\mathrm{By \ the \ above \ form \ of \ Hoeffding's \ inequality}).
  \end{align*}
\end{proof}

\subsection{Hoeffding's inequality (for $t$ transitions)}

\begin{lemma}
  \label{lem:Hoeffding2}
  There exists a positive constant $K$ such that for all $t \in \mathbb{N}$ and for all $\delta > 0$,
  $$
  \proba{ \| \mathbf{M}^{(N)}(t+1) - \Phi_{t+1}(\mm) \| \geq (1 + K + K^2 + ... + K^t)\delta \ \Big| \ \mmm^{(N)}(0) = \mm } \leq (t+1)e^{-2N\delta^2}.
  $$
\end{lemma}

\begin{proof}
  Since $\phi$ is a piecewise affine function with finite affine pieces, in particular $\phi$ is $K$-Lipschitz: there is a constant $K > 0$ such that for all $\mm_1, \mm_2 \in \Delta^d$:
  $$
  \| \phi(\mm_1) - \phi(\mm_2) \| \leq K \cdot \| \mm_1 - \mm_2 \|.
  $$
Let $t \in \mathbb{N}$ and $\mm \in $ be fixed, we have
\begin{align*}
   \| \mathbf{M}^{(N)} (t+1) - \Phi_{t+1} (\mm) \|
  &\leq \| \mathbf{M}^{(N)} (t+1) - \phi (\mathbf{M}^{(N)}(t)) \| + \| \phi (\mathbf{M}^{(N)} (t)) - \phi (\Phi_t (\mm)) \|  \\
  &\leq \| \mathbf{M}^{(N)} (t+1) - \phi (\mathbf{M}^{(N)}(t)) \| + K \cdot \| \mathbf{M}^{(N)} (t) - \Phi_{t} (\mm) \|.
\end{align*}

By iterating the above inequality, we obtain
\begin{align*}
   &\| \mathbf{M}^{(N)} (t+1) - \Phi_{t+1} (\mm) \| \\
   &\leq \| \mathbf{M}^{(N)} (t+1) - \phi (\mathbf{M}^{(N)}(t)) \| + K \cdot \| \mathbf{M}^{(N)} (t) - \phi (\mathbf{M}^{(N)}(t-1)) \| + K^2 \cdot \| \mathbf{M}^{(N)} (t-1) - \Phi_{t-1}(\mm) \|\\
   &\leq \sum_{s=0}^{t} K^s \cdot \| \mathbf{M}^{(N)} (t+1-s) - \phi (\mathbf{M}^{(N)}(t-s)) \|,
\end{align*}
where for each $0 \leq s \leq t$, we have by lemma \ref{lem:Hoeffding}: for all $\delta > 0$,
$$
\proba{ \| \mathbf{M}^{(N)} (t+1-s) - \phi (\mathbf{M}^{(N)}(t-s)) \| \geq \delta } \leq e^{-2N \delta^2}.
$$
Hence, using the union bound, we obtain
\begin{align*}
   & \proba{ \| \mathbf{\mathbf{M}}^{(N)}(t+1) - \Phi_{t+1}(\mm) \| \geq (1 + K + K^2 + ... + K^t)\delta \ \Big| \ \mmm^{(N)}(0) = \mm } \\
  &\leq \proba{ \sum_{s=0}^{t} K^s \cdot \| \mathbf{M}^{(N)} (t+1-s) - \phi (\mathbf{M}^{(N)}(t-s)) \| \geq (1 + K + K^2 + ... + K^t)\delta } \\
  &\leq \proba{ \bigcup_{s=0}^{t} \big\{ \| \mathbf{M}^{(N)} (t+1-s) - \phi (\mathbf{M}^{(N)}(t-s)) \| \geq \delta \big\} } \\
  &\leq \sum_{s=0}^{t} \proba{ \| \mathbf{M}^{(N)} (t+1-s) - \phi (\mathbf{M}^{(N)}(t-s)) \| \geq \delta } \\
  &\leq (t+1)\cdot e^{-2N \delta^2},
\end{align*}
and this ends the proof of Lemma \ref{lem:Hoeffding2}.
\end{proof}

\subsection{Exponential stability of $\mm^*$}

\begin{lemma}
  \label{lem:expo_stability}
  Under the assumptions of Theorem~\ref{th:expo_optimal}:
  \begin{enumerate}[(i)]
    \item The matrix $K_{s(\mm^*)}$ is a stable matrix.
    \item There exist constants $b_1,b_2>0$ such that for all $t\ge0$ and all $\mm\in\Delta^d$:
    \begin{equation}
      \label{eq:expo_stability}
       \| \Phi_t (\mm) - \mm^* \| \leq b_1 \cdot e^{-b_2 t} \cdot \| \mm - \mm^* \|.
    \end{equation}
  \end{enumerate}
\end{lemma}

\begin{proof}
  As $\mm^*$ is non-singular and is a uniform global attractor, there exists $T>0$ such that for all $\mm\in\Delta^d$ and $t\ge T$: $\Phi_t(\mm)\in\zz_{s(\mm^*)}$. Recall that for all $\mm\in\zz_{s(\mm^*)}$, we have $\phi(\mm)=(\mm-\mm^*) \cdot \kk_{s(\mm^*)}+\mm^*$. This shows that for all $\mm\in\Delta^d$ and $t\ge T$:
  \begin{align*}
    \Phi_t(\mm)= \big(\Phi_T(\mm) - \mm^*\big) \cdot \kk_{s(\mm^*)}^{t-T}+\mm^*.
  \end{align*}
It implies in particular that $\kk_{s(\mm^*)}$ is a stable matrix, \emph{i.e.} the norm of all of its eigenvalues is smaller than $1$. Consequently \eqref{eq:expo_stability} is true for all $\mm\in\zz_{s{(\mm^*)}}$. As $\zz_{s(\mm^*)}$ is reached in a finite time $T$ from any initial condition $\mm$, this implies \eqref{eq:expo_stability} for all $\mm\in\Delta^d$.
\end{proof}

\subsection{Proof of Theorem~\ref{theorem2}}

We are now ready to prove the main theorem.

\begin{proof}
   The proof consists of several parts.

\subsubsection{Choice of a neighborhood $\mathcal{N}$}
The fixed point $\mm^*$ is in zone $\zz_{s(\mm^*)}$ in which $\phi$ can be written as
$$
\phi (\mm) = (\mm - \mm^*) \cdot \kkm + \mm^*.
$$
As $\mm^*$ is not singular, let $\mathcal{N}_1$ be a neighborhood of $\mm^*$ included in $\zz_{s(\mm^*)}$. Since $\mm^*$ is a uniformly global attractor, $\kkm$ is a stable matrix. We can therefore choose a smaller neighborhood $\n_2 \subset \n_1$ so that $\Phi_t (\n_2) \subset \n_1$ for all $t\geq 0$. That is, the image of $\n_2$ under the maps $\Phi_{t\geq 0}$ remains inside $\n_1$. This is possible by stability of $\mm^*$. We next choose a neighborhood $\n_3 \subset \n_2$ and a $\delta > 0$ so that $(\phi(\n_3))^{\delta} \subset \n_2$, that is, the image of $\n_3$ under $\phi$ remains inside $\n_2$ and it is at least $\delta$ away from the boundary of $\n_2$. We finally fix $r > 0$ so that the intersection $\mathcal{B}(\mm^*, r) \cap \Delta^d \subset \n_3$, and we choose our neighborhood $\n$ as
$$
\mathcal{N} := \mathcal{B}(\mm^*, r) \cap \Delta^d.
$$
Note that the choice of $r$ and $\delta$ is independent of $N$. From $(ii)$ of Lemma \ref{lem:expo_stability}, we denote furthermore by $\tilde{T} := T(r/2)$ the finite time such that for all $\mm \in \Delta^d$, $\Phi_{\tilde{T}+1} (\mm) \in \mathcal{B}(\mm^*, r/2)$.

\subsubsection{Definition and properties of the function $G$.}
\label{ssec:G}

Following the generator approach used for instance in \cite{gast2018refined}. For $\mm \in \Delta^d$, define $G: \Delta^d \rightarrow \mathbb{R}^d$ as
\begin{equation*}
  G(\mm) := \sum_{t=0}^{\infty} \big( \Phi_t (\mm) - \mm^*\big).
\end{equation*}
By using Lemma~\ref{lem:expo_stability}(ii), for all $\mm \in \Delta^d$ we have
$\| G(\mm) \| \leq \sum_{t=0}^{\infty} b_1 \cdot e^{-b_2t} \cdot \| \mm - \mm^* \| < \infty$.  This shows that the function $G$ is well defined and bounded. Denote by $ \overline{G} := \sup_{\mm \in \Delta^d} \| G(\mm) \| <\infty$.

By our choice of $\n_2$ defined above, for all $t\ge0$ and $\mm \in \n_2$  we have:
\begin{equation}\label{5}
  \Phi_t (\mm) = (\mm - \mm^*) \cdot \kkm^t + \mm^*.
\end{equation}
Hence, for all $\mm\in\n_2$, we have
\begin{align*}
   G(\mm) = & \ \sum_{t=0}^{\infty} \big(\Phi_t (\mm) - \mm^*\big) \\
  = & \ \sum_{t=0}^{\infty} ( \mm - \mm^*) \cdot \kkm^t  \\
  = & \  (\mm - \mm^*) \cdot (\mathbf{I} - \kkm)^{-1},
\end{align*}
where the last equality holds because $\kkm$ is a stable matrix. Hence in $\n_2$, $G(\mm)$ is an \emph{affine} function of $\mm$.

From the definition of function $G$, we see that for all $\mm \in \Delta^d$:
\begin{align*}
  G(\mm) - G(\phi(\mm)) &= \sum_{t=0}^{\infty} \big( \Phi_t (\mm) - \mm^*\big)- \sum_{t=0}^{\infty} \big( \Phi_t (\phi(\mm)) - \mm^*\big) \\
  &= \sum_{t=0}^{\infty} \big( \Phi_t (\mm) - \mm^*\big)- \sum_{t=1}^{\infty} \big( \Phi_t (\mm) - \mm^*\big) \\
  &= \mm - \mm^*,
\end{align*}
Hence
\begin{align}
  \e [\mathbf{M}^{(N)} (0)] - \mm^* &= \e \big[ G(\mathbf{M}^{(N)} (0)) - G(\phi(\mathbf{M}^{(N)} (0))) \big] \ \ \ \ \text{(By  the above equality)}\nonumber\\
  &= \e \big[ G(\mathbf{M}^{(N)} (1)) - G(\phi(\mathbf{M}^{(N)} (0))) \big] \ \ \ \ \text{(Since $\mathbf{M}^{(N)}(0)$ is stationary)}\nonumber\\
  &= \e \bigg[ \e \big[ G(\mathbf{M}^{(N)} (1)) - G(\phi(\mm)) \mid \mathbf{M}^{(N)}(0) = \mm \big]\cdot \mathds{1}_{\{ \mm \notin \n\} }
  \label{eq:case1}\\
  &\qquad + \e \big[ G(\mathbf{M}^{(N)} (1)) - G(\phi(\mm)) \mid \mathbf{M}^{(N)}(0) = \mm \big]\cdot \mathds{1}_{ \{\mm \in \n \} } \bigg].
  \label{eq:case2}
\end{align}

In the following, we bound \eqref{eq:case1} and \eqref{eq:case2} separately.

\subsubsection{Bound on \eqref{eq:case1}}
\label{apx:case1}

As $G$ is bounded by $\overline{G}$, we have
\begin{align*}
   & \bigg| \bigg| \e \bigg[ \e \big[ G(\mathbf{M}^{(N)} (1)) - G(\phi(\mm)) \big| \mathbf{M}^{(N)}(0) = \mm \big]\cdot \mathds{1}_{\{ \mm \notin \n\} } \bigg] \bigg| \bigg| \le \ 2\overline{G} \cdot \proba{ \mathbf{M}^{(N)} (0) \notin \n }.
\end{align*}

We are left to bound $\proba{ \mathbf{M}^{(N)} (0) \notin \n }$. Let $u := \big( \frac{r}{2(1 + K + K^2 + ... + K^{\tilde{T})}} \big)^2$, where $K$ is the Lipschitz constant of $\phi$. We have by Lemma \ref{lem:Hoeffding2}:
\begin{align*}
  & \proba{ \| \mathbf{M}^{(N)}(\tilde{T}+1) - \Phi_{\tilde{T}+1} (\mm) \| \geq \frac{r}{2} \ \Big| \ \mmm^{(N)}(0) = \mm } \\
  &= \ \proba{ \| \mathbf{M}^{(N)}(\tilde{T}+1) - \Phi_{\tilde{T}+1} (\mm) \| \geq (1 + K + K^2 + ... + K^{\tilde{T}})\sqrt{u} \ \Big| \ \mmm^{(N)}(0) = \mm } \\
  &\leq \ (\tilde{T}+1) \cdot e^{-2uN}.
\end{align*}
This shows that
\begin{align}\label{eq:point(ii)}
  \proba{ \mathbf{M}^{(N)} (0) \notin \n } & = \ \proba{ \| \mathbf{M}^{(N)}(0) - \mm^* \| \geq r } \nonumber\\
  &  = \proba{ \| \mathbf{M}^{(N)}(\tilde{T}+1) - \mm^* \| \geq r } \ \ \ \ \mbox{(By stationarity)} \nonumber \\
  &\leq  \proba{ \| \mathbf{M}^{(N)} (\tilde{T}+1) - \Phi_{\tilde{T}+1} (\mm) \| \geq \frac{r}{2} \ \Big| \ \mmm^{(N)}(0) = \mm } + \proba{ \| \Phi_{\tilde{T}+1}(\mm) - \mm^* \| \geq \frac{r}{2} } \nonumber \\
  & \ (\mathrm{with \ } \mm \ \mathrm{being \ an \ arbitrary \ element \ of \ } \Delta^d ) \nonumber \\
  &= \proba{ \| \mathbf{M}^{(N)} (\tilde{T}+1) - \Phi_{\tilde{T}+1} (\mm) \| \geq \frac{r}{2} \ \Big| \ \mmm^{(N)}(0) = \mm }  \ \ (\mathrm{By \ our \ choice \ of \ } \tilde{T} = T(r/2)) \nonumber \\
  &\leq (\tilde{T}+1) \cdot e^{-2uN}
\end{align}

 \subsubsection{Bound on \eqref{eq:case2}}
 \label{apx:case2}
 By Lemma \ref{lem:Hoeffding}, we have
 \begin{align*}
  &\e \big[ G(\mathbf{M}^{(N)} (1)) - G(\phi(\mm)) \ \big| \ \mathbf{M}^{(N)}(0) = \mm \big]\cdot \mathds{1}_{ \{\mm \in \n \} } \\
   &= \ \e \big[ G(\phi(\mm) + \mathbf{E}^{(N)} (1)) - G(\phi(\mm)) \ \big| \ \mathbf{M}^{(N)}(0) = \mm \big]\cdot \mathds{1}_{ \{\mm \in \n \} } \\
   &= \ \e \bigg[ \big( G(\phi(\mm) + \mathbf{E}^{(N)} (1)) - G(\phi(\mm)) \big) \cdot \mathds{1}_{\{ \| \mathbf{E}^{(N)} (1) \| < \delta \}} \ \\
    & \qquad +\big( G(\phi(\mm) + \mathbf{E}^{(N)} (1)) - G(\phi(\mm)) \big) \cdot \mathds{1}_{\{ \| \mathbf{E}^{(N)} (1) \| \geq \delta \}} \ \bigg| \ \mathbf{M}^{(N)}(0) = \mm \bigg]\cdot \mathds{1}_{ \{\mm \in \n \} }
 \end{align*}

By our choice of $\n$ and $\delta$, for the first part of the above expectation, \emph{i.e.} when the event $\{ \| \mathbf{E}^{(N)} (1) \| < \delta \} $ occurs, $\phi(\mm) + \mathbf{E}^{(N)} (1)$ will remain in $\n_2$, hence $G\big( \phi (\mm) + \mathbf{E}^{(N)} (1) \big)$ takes the same affine form as $G(\phi(\mm))$. Consequently
\begin{align*}
   & \e \bigg[ \big( G(\phi(\mm) + \mathbf{E}^{(N)} (1)) - G(\phi(\mm)) \big) \cdot \mathds{1}_{\{ \| \mathbf{E}^{(N)} (1) \|  < \delta \}} \ \bigg| \ \mathbf{M}^{(N)}(0) = \mm \bigg]\cdot \mathds{1}_{ \{\mm \in \n \} } \\
   &= \bigg[ G \big( \e \big[\phi(\mm) + \mathbf{E}^{(N)} (1) \ \big| \ \mathbf{M}^{(N)} (0) = \mm \big] \big) - G \big( \e \big[\phi(\mm) \ \big| \ \mathbf{M}^{(N)} (0) = \mm\big] \big) \bigg] \mathbb{P} \big( \{ \| \mathbf{E}^{(N)}(1) \| < \delta \} \big)   \cdot \mathds{1}_{\{ \mm \in \n \}} \\
   & \ \mathrm{\big( Thanks \ to \ the \ affinity \ of \ } G \mbox{ in this case}, \ \mathrm{we \ can \ interchange \ } \e \ \mathrm{and} \ G \big) \\
  & = \ 0 \qquad\big( \text{By Lemma~\ref{lem:Hoeffding}} \big).
\end{align*}

For the second part of the above expectation,
\begin{align*}
  & \bigg| \bigg| \e \bigg[ \big( G(\phi(\mm) + \mathbf{E}^{(N)} (1)) - G(\phi(\mm)) \big) \cdot \mathds{1}_{\{ \| \mathbf{E}^{(N)} (1) \| \geq \delta \}} \bigg| \mathbf{M}^{(N)}(0) = \mm \bigg] \bigg| \bigg| \cdot \mathds{1}_{ \{\mm \in \n \} } \\
  &\leq \ 2\overline{G} \cdot \mathbb{P} \big( \| \mathbf{E}^{(N)} (1) \| \geq \delta \big) \\
  &\leq \ 2\overline{G} \cdot e^{-2N\delta^2} \ \ \ \ \big( \mathrm{By \ Lemma \ \ref{lem:Hoeffding}} \big).
\end{align*}
So finally
\begin{equation*}
  \big| \big| \e \big[ G(\mathbf{M}^{(N)} (1)) - G(\phi(\mm)) \big| \mathbf{M}^{(N)}(0) = \mm \big] \big| \big| \cdot \mathds{1}_{ \{\mm \in \n \} } \leq 0 + 2\overline{G} \cdot e^{-2N \delta^2} =  2\overline{G} \cdot e^{-2N \delta^2}.
\end{equation*}

\subsubsection{Conclusion of the proof} To summarize, we have obtained by \eqref{eq:point(ii)}:
\begin{align*}
  \proba{\bMN(0)\not\in\zzm} & \leq \proba{ \big( \mathbf{M}^{(N)} (0) \notin \n \big)} \\
   & \leq (\tilde{T}+1) \cdot e^{-2uN} \\
   & \leq b \cdot e^{-cN},
\end{align*}

and

\begin{align*}
   \| \mathbb{E} \big[ \mathbf{M}^{(N)}(0) \big] - \mm^* \|
  &\leq \ 2\overline{G} \cdot e^{-2N\delta^2} + 2\overline{G}(\tilde{T}+1)\cdot e^{-2Nu} \\
  &\leq \ b \cdot e^{-cN},
\end{align*}
where $b$, $c$ can be taken as $b := (2\overline{G}+1)(\tilde{T}+2)$, $c := \min(\delta^2, u)$, and this concludes the proof of Theorem \ref{theorem2}.

\end{proof}

\section{Proof of Theorem~\ref{th:expo_optimal_async}}
\label{apx:proof_async}

Recall that $\bMN(t)$ is the configuration of the system at time $t$, which means that $\MN_i(t)$ is the fraction of bandits that are in state $i$ at time $t$. Let $\ee_i$ be the $d$ dimensional vector that has all its component equal to $0$ except the $i$th one that equals $1$. The process $\bMN$ is a continuous-time Markov chain that jumps from a configuration $\mm$ to a configuration $\mm+\frac1N(\ee_j-\ee_i)$ when a bandit jumps from state $i$ to state $j$. For $i<s(\mm)$, this occurs at rate $Nm_iQ^1_{ij}$ as all of these bandits are activated. For $i>s(\mm)$, this occurs at rate $Nm_iQ^0_{ij}$ as these bandits are not activated. For $i=s(\mm)$, this occurs at rate $N\big((\alpha-\sum_{k=1}^{s(\mm)-1} m_k)Q^1_{ij} + (\sum_{k=1}^{s(\mm)} m_k-\alpha)Q^0_{ij}\big)$. Let us define:
\begin{align*}
  \lambda_{ij}(\mm) = \left\{
    \begin{array}{ll}
      m_iQ^1_{ij} & \text{ if $i<s(\mm)$}\\
      (\alpha-\sum_{k=1}^{s(\mm)-1} m_k)Q^1_{ij} + (\sum_{k=1}^{s(\mm)} m_k-\alpha)Q^0_{ij} & \text{ if $i=s(\mm)$}\\
      m_iQ^0_{ij} & \text{ if $i>s(\mm)$.}
    \end{array}
    \right.
\end{align*}
The process $\bMN$ jumps from $\mm$ to $\mm+(\ee_j-\ee_i)/N$ at rate $N \lambda_{ij}(\mm)$. This shows that $\bMN$ is a density dependent population process as defined in \cite{kurtz1978strong}. It is shown in \cite{kurtz1978strong} that, for any finite time $t$, the trajectories of $\bMN(t)$ converge to the solution of a differential equation $\dot{\mm}=f(\mm)$ as $N$ grows, with $f(\mm) := \sum_{i\ne j}\lambda_{ij}(\mm)(\ee_j-\ee_i)$. The function $f(\mm)$ is called the drift of the system. It should be clear that $f(\mm)=\tau(\phi(\mm)-\mm)$, where $\phi$ is defined for the synchronous version of our asynchronous bandit problem.

For $t \geq0$, denote by $\Phi_t \mm$ the value at time $t$ of the solution of the differential equation that starts in $\mm$ at time $0$, it satisfies
$$
\Phi_t \mm = \mm + \int_{0}^{t} f(\Phi_s \mm) ds.
$$
Following \cite{gast:hal-01622054,Ying2017}, we denote by $L^{(N)}$ the generator of the $N$ bandits system and by $\Lambda$ the generator of the differential equation. They associate to each almost-everywhere differentiable function $h$ two functions $L^{(N)}h$ and $\Lambda h$ that are defined as
\begin{align*}
  \big( L^{(N)}h \big)(\mm) &:= \sum_{i=1}^{d}\sum_{j \neq i} N \lambda_{ij}(\mm)\cdot \big( h(\mm+\frac{\ee_j-\ee_i}{N}) - h(\mm) \big),\\
  \big( \Lambda h \big)(\mm) &:= f(\mm) \cdot Dh(\mm),
\end{align*}
with $Dh$ being the differential of function $h$. The function $\Lambda h $ is defined only on points $\mm$ for which $h(\mm)$ is differentiable. Remark that if $h(\mm)$ is an affine function in $\mm$, \emph{i.e.} $h(\mm) = \mm\cdot \bb + \mathbf{b}$, with $\bb$ a $d$-dimensional matrix and $\mathbf{b}$ a $d$-dimensional vector, then $\big(L^{(N)}h\big)(\mm) = \big(\Lambda h \big) (\mm) = f(\mm) \cdot \bb$.

Now the analogue of Theorem \ref{theorem2}(i) in the asynchronous case is
\begin{thm}\label{theorem3}
  Under the same assumptions as in Theorem \ref{th:expo_optimal_async}, and assume that $\mmm^{(N)} (0)$ is already in stationary regime. Then there exists two constants $b,c >0$ such that
  $$
  \| \mathbb{E}[\mmm^{(N)} (0)] - \mm^{\ast} \| \leq b \cdot e^{-cN}.
  $$
\end{thm}

Note first that similarly, Theorem \ref{theorem3} implies Theorem \ref{th:expo_optimal_async}.

\begin{proof} Define the asynchronous version of function $G$ as
$$
G(\mm) := \int_{0}^{\infty} \big( \Phi_t \mm - \mm^{\ast} \big) dt.
$$
As for the synchronous case, our assumptions imply that the unique fixed point is an exponentially stable attractor and a result similar to Lemma~\ref{lem:expo_stability} can be obtained for the asynchronous case. This implies that the function $G$ is well-defined, continuous and bounded.

Recall that the function $f$ is affine in $\zzm$: since if $\mm \in \zzm$, then $\phi (\mm)=(\mm-\mm^*)\kk+\mm^*$ with the linear matrix $K$ as in \eqref{eq:phi_k}, and $f(\mm)= \tau(\phi(\mm)-\mm) = \tau(\mm-\mm^*)(\kk-\ii)$. Now suppose $\mm \in \Delta^d$ is such that $\Phi_t \mm$ remains inside $\zzm$ for all $t\geq0$, then
$$
\Phi_t \mm = (\mm-\mm^*)\cdot e^{t \cdot \tau(\kk-\ii)} + \mm^*, \mbox{ and } \ G(\mm) = \frac{1}{\tau} (\mm-\mm^*)(\kk-\ii)^{-1}.
$$
So as for the synchronous case, $G(\mm)$ is an affine function of $\mm$, with affine factor $\bb := \frac{1}{\tau}(\kk-\ii)^{-1}$.

As $\mm^*$ is non-singular, it is at a positive distance from the other zones $\zz_i \neq \zzm$ and we therefore define $\delta := \min_{i\ne s(\mm^*)} d(\mm^*,\zz_i)/2>0$, where $d(\cdot \ , \cdot)$ is the distance under $\| \cdot \|$-norm. We then choose a neighborhood $\mathcal{N}_1 := \mathcal{B}(\mm^*, \epsilon_1) \cap \Delta^d$ of $\mm^*$ such that for all $t \geq 0 $ and all initial condition $\mm \in \mathcal{N}_1$, $\Phi_t(\mm) \in \mathcal{B}(\mm^{\ast}, \delta)$. This is possible by the exponentially stable attractor property of $\mm^*$. Following Theorem 3.2 of \cite{gast:hal-01553133}, we have
\begin{align}
   \mm^* - \mathbb{E} \big[ \mmm^{(N)}(0) \big]  \nonumber
   = & \ \mathbb{E} \big[ \Lambda G \big( \mmm^{(N)} (0) \big) \big] \nonumber \\
   = & \ \mathbb{E} \big[ (\Lambda - L^{(N)}) G \big( \mmm^{(N)} (0) \big) \big] \nonumber \\
   = & \ \mathbb{E} \Big[ \Big( (\Lambda - L^{(N)}) G \big( \mmm^{(N)} (0) \big) \Big) \cdot \mathds{1}_{\{ \mmm^{(N)} (0) \in \mathcal{N}\} } 
   \label{async_eq1a}
   \\ &\qquad+ \Big( (\Lambda - L^{(N)}) G \big( \mmm^{(N)} (0) \big) \Big) \cdot \mathds{1}_{ \{ \mmm^{(N)} (0) \notin \mathcal{N} \} } \Big], \label{async_eq1b}
\end{align}
where $\mathcal{N}:= \mathcal{B}(\mm^*, \epsilon_1/2) \cap \Delta^d$.
Let $N_0 := \lceil 2/\epsilon_1 \rceil$. For $N\ge N_0$, $\mm \in \n$ verifies additionally that $\Phi_t \big( \mm + \frac{\ee_j-\ee_i}{N}\big) \in \zzm$ for all $1 \leq i \neq j \leq d$ and $t\geq0$. Hence, $G$ is locally affine and for all $m\in\n$ and $N \geq N_0$, we have:
\begin{equation}\label{asyc_eq2}
  \big( \Lambda G \big) (\mm) = \big( L^{(N)} G \big) (\mm) = f(\mm) \cdot \bb.
\end{equation}
This shows that the first term of \eqref{async_eq1a} is equal to zero.

For the second term, note that both $G$ and $\Lambda G$ are continuous functions defined on the compact region $\Delta^d$, hence they are both bounded, while $L^{(N)}G$ grows at most linearly with $N$. Hence we can choose constants $u,v > 0$ independent of $N$ such that:
$$
\sup_{\mm \in \Delta^d} \| \big( \Lambda G \big) (\mm) \| = u, \ \sup_{\mm \in \Delta^d}\| \big(L^{(N)} G \big) (\mm) \| \leq v N.
$$

We are left to bound $\mathbb{P} \big( \mmm^{(N)} (0) \notin \mathcal{N} \big)$ exponentially from above. This could be done by using the (unnamed) proposition on page 644 of~\cite{WeberWeiss1990}. Yet, we were not able to find the paper referenced for the proof of this proposition. Hence, we provide below a direct proof of this. To achieve this, we rely on an exponential martingale concentration inequality, borrowed from \cite{darling2008}, which in our situation can be stated as
\begin{lem}\label{lemma5}
  Fix $T > 0$. Let $K$ be the Lipschitz constant of drift $f$, denote $\lambda := \max_{i,j} \lambda_{ij}$, and $c_1 := e^{-2KT} / 18T$. If $\epsilon > 0$ is such that
  \begin{equation}\label{async_eq4}
    1 \geq \epsilon \lambda \cdot \mbox{exp} \big( \frac{\epsilon^2 e^{-KT}}{3T} \big),
  \end{equation}
  then we have
  \begin{equation}\label{async_eq5}
    \mathbb{P} \Big[ \sup_{t \leq T} \| \mmm^{(N)}(t) - \Phi_t \mm \| > \epsilon \Big| \ \mmm^{(N)}(0) = \mm \Big] \leq 2d \cdot e^{-c_1 N\epsilon^3}.
  \end{equation}
\end{lem}
The above lemma plays the role of Lemma \ref{lem:Hoeffding2} in synchronous case. Note that its original form stated as Theorem 4.2 in \cite{darling2008} is under a more general framework, which considered a continuous-time Markov chain with countable state-space evolves in $\mathbb{R}^d$, and discussed a differential equation approximation to the trajectories of such Markov chain. As such, the right hand side of \eqref{async_eq5} has an additional term $\mathbb{P} (\Omega_0^c \cup \Omega_1^c \cup \Omega_2^c)$, with $\Omega_i^c$ being the complementary of $\Omega_i$. In our case, $\Omega_0 = \Omega_1 = \Omega $ trivially holds; while the analysis of $\Omega_2$ is more involved. However, as remarked before the statement of  Theorem 4.2 in \cite{darling2008}, if the maximum jump rate (in our case $N \lambda $) and the maximum jump size (in our case $1/N$) of the Markov chain satisfy certain inequality, which in our situation can be sated as \eqref{async_eq4}, then $\Omega_2 = \Omega$. Note that the constraint \eqref{async_eq4} is satisfied as long as $\epsilon$ is sufficiently small, and consequently $\mathbb{P} (\Omega_0^c \cup \Omega_1^c \cup \Omega_2^c) = 0$.

Now let $\epsilon>0$ be such that $\mathcal{B}(\mm^*, 2\epsilon) \cap \Delta^d \subset \n$. The uniform global attractor assumption on $\mm^{\ast}$ ensures that there exists $T>0$ such that for all $\mm \in \Delta^d$  and $t\ge T$: $\Phi_t \mm \in \mathcal{B}(\mm^*,\epsilon)$. Let such $T$ and $\epsilon$ be as in Lemma \ref{lemma5} that verify additionally \eqref{async_eq4}. This is possible as the right hand side of \eqref{async_eq4} converges to $0$ when $\epsilon$ is small and $T$ is large.

We then have:
\begin{align*}
   \mathbb{P} \big[ \mmm^{(N)}(0) \notin \mathcal{N} \big]
  &= \mathbb{P} \big[ \mmm^{(N)}(T) \notin \mathcal{N} \big] \qquad (\mbox{By stationarity}) \\
  &\leq \mathbb{P} \big[ \| \mmm^{(N)}(T)-\mm^* \| \leq 2 \epsilon \big] \\
  &\leq \mathbb{P} \big[ \| \mmm^{(N)}(T) - \Phi_T (\mmm^{(N)}(0)) \| > \epsilon \big] \ + \ \mathbb{P} \big[ \| \Phi_T (\mmm^{(N)}(0)) - \mm^{\ast} \| > \epsilon \big] \\
  &= \mathbb{P} \big[ \| \mmm^{(N)}(T) - \Phi_T (\mmm^{(N)}(0)) \| > \epsilon \big] \qquad  \big( \mbox{By our choice of $T$}\big) \\
  &\leq 2d \cdot e^{-c_1 N\epsilon^3}  \qquad  \big( \mbox{We apply \eqref{async_eq5} of Lemma \ref{lemma5}} \big).
\end{align*}
So in summary, \eqref{async_eq1a}-\eqref{async_eq1b} gives
\begin{align}
  \label{eq:asyncrho_last}
  \| \mathbb{E} \big[ \mmm^{(N)}(0) \big] - \mm^{\ast} \|
  &\leq (u+vN)\cdot 2d \cdot e^{-c_1 N\epsilon^3}.
\end{align}
Moreover, for any $c'>0$ and $0<c<c'$, $N \cdot e^{-c'N}=\calO(e^{-cN})$, so the right hand side of \eqref{eq:asyncrho_last} can be bounded by a term of the form $b \cdot e^{-cN}$. This concludes the proof of Theorem~\ref{theorem3}.
\end{proof}

\section{Parameters used in the numerical experiments.}
\label{apx:paras}

For completeness, in this appendix we provide the parameters used in our numerical experiments in Section \ref{sec:numerical}. The numbers are recorded in $8$ digits of precision.

\subsection{Parameters for the example of Section~\ref{sec:numerical_3states}}

The parameters for the example  with $d=3$ in Section \ref{sec:numerical_3states} are

$$\pp^0 =
\begin{pmatrix}
  0.30368587 & 0.25184515 & 0.44446898 \\
  0.40839084 & 0.41334941 & 0.17825975 \\
  0.66146205 & 0.1840829 & 0.15445505
\end{pmatrix}, \ \
\pp^1 = \begin{pmatrix}
  0.23763148 & 0.42381178 & 0.33855674 \\
  0.54401527 & 0.27028947 & 0.18569526 \\
  0.06938943 & 0.38776507 & 0.54284550
\end{pmatrix},$$ $\rr^1 = (0.99663977, 0.22770951, 0.17300611)$ and $\rr^0 = \mathbf{0}$. This example has also been used in our discussion at the end of  Section \ref{ssec:phi} as well as in Section \ref{ssec:non-integer-alphaN}.

\subsection{Parameters for the examples of Section~\ref{sec:numerical_cycles}}

The parameters for the three period-$2$ cyclic examples in Section \ref{sec:numerical_cycles} are:

Example 1

$$\pp^0 =
\begin{pmatrix}
  0.5214073 & 0.40392496 & 0.07466774 \\
  0.0158415 & 0.21455666 & 0.76960184 \\
  0.53722329 & 0.37651148 & 0.08626522
\end{pmatrix}, \ \
\pp^1 = \begin{pmatrix}
  0.24639364 & 0.23402385 & 0.51958251 \\
  0.49681581 & 0.49509821 & 0.00808597 \\
  0.37826553 & 0.15469252 & 0.46704195
\end{pmatrix},$$ $\rr^1 = (0.72232506, 0.18844869, 0.25752477)$ and $\rr^0 = \mathbf{0}$.

Example 2

$$\pp^0 =
\begin{pmatrix}
  0.02232142 & 0.10229283 & 0.87538575 \\
  0.03426605 & 0.17175704 & 0.79397691 \\
  0.52324756 & 0.45523298 & 0.02151947
\end{pmatrix}, \ \
\pp^1 = \begin{pmatrix}
  0.14874601 & 0.30435809 & 0.54689589 \\
  0.56845754 & 0.41117331 & 0.02036915 \\
  0.25265570 & 0.27310439 & 0.47423991
\end{pmatrix},$$ $\rr^1 = (0.37401552, 0.11740814, 0.07866135)$ and $\rr^0 = \mathbf{0}$.

Example 3

$$\pp^0 =
\begin{pmatrix}
  0.47819592 & 0.02090623 & 0.50089785 \\
  0.08063373 & 0.15456935 & 0.76479692 \\
  0.66552514 & 0.08481946 & 0.24965540
\end{pmatrix}, \ \
\pp^1 = \begin{pmatrix}
  0.00279465 & 0.37327924 & 0.62392611 \\
  0.51582335 & 0.46333908 & 0.02083756 \\
  0.41875202 & 0.17776712 & 0.40348086
\end{pmatrix},$$ $\rr^1 = (0.97658608, 0.53014109, 0.40394919)$ and $\rr^0 = \mathbf{0}$.

\section*{Acknowledgements}

This work was supported by the ANR project REFINO (ANR-19-CE23-0015).

\end{document}